\documentclass[11pt,a4paper]{article}
\pdfoutput=1

\usepackage{jheppub}
\usepackage{epstopdf}

\def \be {\begin{equation}}
\def \ee {\end{equation}}
\def \bsp {\begin{split}}
\def \esp {\end{split}}
\def \bea {\begin{eqnarray}}
\def \eea {\end{eqnarray}}

\def\mc{\mathcal}

\def\mb{\mathbb}

\def \bp{\begin{pmatrix}}
\def\ep{\end{pmatrix}}

\DeclareMathOperator*{\argmin}{arg\,min}

\title{Learning non-Higgsable gauge groups in 4D F-theory}

\author[a]{Yi-Nan Wang}
\author[b]{Zhibai Zhang}

\affiliation[a]{Center for Theoretical Physics,\\Department of Physics\\Massachusetts Institute of Technology\\77 Massachusetts Avenue\\Cambridge, MA 02139, USA}

\affiliation[b]{Department of Finance and Risk Engineering \\
Tandon School of Engineering \\
New York University \\
12 Metro Tech Center\\Brooklyn NY 11201, USA}

\emailAdd{wangyn@mit.edu,z.zhibai@gmail.com}

\preprint{\today \hspace*{0.1in} MIT-CTP-5005}

\abstract{We apply machine learning techniques to solve a specific classification problem in 4D F-theory. For a divisor $D$ on a given complex threefold base, we want to read out the non-Higgsable gauge group on it using local geometric information near $D$. The input features are the triple intersection numbers among divisors near $D$ and the output label is the non-Higgsable gauge group. We use decision tree to solve this problem and achieved 85\%-98\% out-of-sample accuracies for different classes of divisors, where the data sets are generated from toric threefold bases without (4,6) curves. We have explicitly generated a large number of analytic rules directly from the decision tree and proved a small number of them. As a crosscheck, we applied these decision trees on bases with (4,6) curves as well and achieved high accuracies. Additionally, we have trained a decision tree to distinguish toric (4,6) curves as well. Finally, we present an application of these analytic rules to construct local base  configurations with interesting gauge groups such as SU(3).}


\begin{document}

\maketitle

\section{Introduction}

The existence of mutiple vacuum solutions is a central feature of string/M-theory paradigm of quantum gravity. This ensemble of string vacuum solutions is commonly denoted as the ``landscape of string vacua''. Specifically, one can choose a particular regime of string theory (such as IIB, heterotic or M-theory) and a class of geometries to probe a part of the landscape. 

In particular, F-theory~\cite{Vafa-F-theory, Morrison-Vafa-I, Morrison-Vafa-II} provides a geometric framework to describe the largest finite number of string vacua to date. In this geometric description of strongly coupled IIB superstring theory, we compactify on an elliptic fibered Calabi-Yau manifold $X$ of $(d+1)$ complex dimensions to get a low energy theory in the Minkowski space $\mathbb{R}^{9-2d,1}$. The base manifold $B$ of this elliptic fibration has $d$ complex dimensions, which is not Calabi-Yau. Hence F-theory can also be thought as a compactification of IIB string theory on a non-Ricci-flat space $B$, while the non-zero curvature is balanced by the inclusion of 7-branes.

The classification of F-theory landscape has the following three layers:

(1) Classify all the $d$-dimensional base manifolds $B$ up to isomorphism. For $d=2$, the base surfaces have been almost completely classified~\cite{mt-toric,Hodge,Martini-WT,non-toric}. For $d=3$, there are some partial classification and probing results in the subset of toric threefold bases~\cite{Halverson-WT,MC,Halverson:2017,skeleton}, but we do not have a global picture of non-toric and non-rational threefolds yet. 

(2) Classify all the distinct elliptic fibrations $X$ over $B$. Physically, different elliptic fibrations will give rise to different gauge groups and matter spectra ~\cite{Bershadsky-all, Katz-Vafa, Morrison-sn, Grassi-Morrison-2, mt-singularities, Johnson:2016qar}.

(3) For a given geometry, classify other non-geometric information relevant to the low energy physics, such as the $G_4$ flux in 4D F-theory~\cite{Grana:2005jc, Douglas:2006es, Denef-F-theory, dgkt, Acharya:2006zw, Braun-Watari, Watari}. 

As one can see, the classification and characterization of the base manifolds is the foundation of this program. In this paper, we will mostly consider the generic fibration $X_{\rm gen}$ over $B$. For most of the base manifolds, it turns out that $X_{\rm gen}$ has singularities corresponding to a stack of 7-branes carrying non-Abelian gauge groups $G_{\rm gen}$. For any other elliptic fibration $X$ over $B$, the gauge group $G$ always contains $G_{\rm gen}$ as a subgroup. Hence $G_{\rm gen}$ is minimal among all the elliptic fibrations over $B$ and it is called non-Higgsable gauge group~\cite{clusters,4d-NHC}, which is a physical characterization of the base manifold $B$\footnote{In 4D F-theory, these non-Higgsable gauge groups may be broken by the $G_4$ flux}. 

In 6D F-theory, the base $B$ is a complex surface and the non-Higgsable gauge groups are carried by the complex curves on $B$. Such curves form ``non-Higgsable clusters'' and they are well understood~\cite{clusters}. For example, a curve $C$ with self-intersection $(-3)$ always carries non-Higgsable SU(3) gauge group if it is not connected to any other curve with self-intersection $(-2)$ or lower. These non-Higgsable clusters are fundamental building blocks of the classification of compact 2D bases~\cite{mt-toric,Hodge,Martini-WT,non-toric} and the non-compact bases giving rise to 6D (1,0) SCFTs~\cite{Heckman:2013pva,6dCM,Heckman:2015bfa}.

\begin{figure}
\centering
\includegraphics[height=3cm]{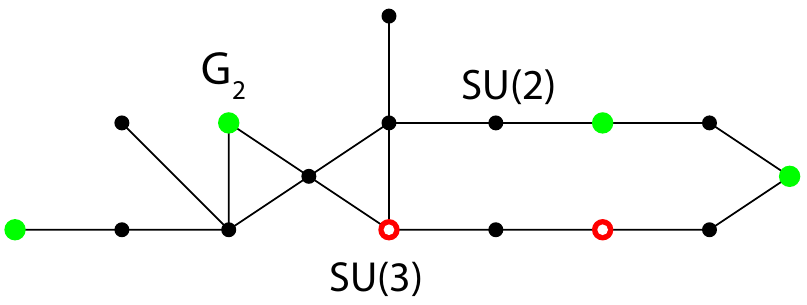}
\caption[E]{\footnotesize A typical non-Higgsable cluster with SU(2), SU(3) and $G_2$ gauge groups found on a generic base in the random walk approach~\cite{MC}.}\label{f:typical-NHC}
\end{figure}

In 4D F-theory, the base is a complex threefold and the non-Higgsable gauge groups locate on complex surfaces (divisors). The triple intersection structure among divisors on a complex threefold is highly involved, and the topology of non-Higgsable clusters seem to be arbitary~\cite{4d-NHC,MC}. In figure~\ref{f:typical-NHC}, we show a typical non-Higgsable cluster found in \cite{MC}. In fact, there does not even exist a dictionary between the local geometric information on the threefold base $B$ and the non-Higgsable gauge groups. 

Nonetheless, we have generated a large number of compact toric threefolds with various non-Higgsable gauge groups that can be easily computed with toric geometry techniques~\cite{MC,skeleton}. Thus we have a different approach: starting from the geometric data generated by Monte Carlo methods and try to find patterns and rules in it. Because of the large volume of data and potentially complex pattern, it is natural to use the recently flourishing machine learning techniques to simply the task. Various machine learning tools have been applied to data sets in string theory~\cite{He:2017aed,Krefl:2017yox,Ruehle:2017mzq,Carifio:2017bov,He:2017set,Hashimoto:2018ftp}. For the data in the string geometric landscape, a common feature is that they are mostly tuples of integers with no error. Hence it provides a brand new playground for data science. 

In this paper, we formulate a classification problem on the data and apply supervised machine learning techniques. Given local triple intersection numbers near a divisor $D$ as the input vector (the features)\footnote{The notion ``near'' involves the  neighbor divisors that intersect $D$, see section~\ref{s:features} for more details.}, we train a classifier to predict the non-Higgsable gauge group on $D$ (the label). There are 10 classes corresponding to the 10 possible non-Higgsable gauge groups: $\varnothing$, SU(2), SU(3), $G_2$, SO(7), SO(8), $F_4$, $E_6$, $E_7$ and $E_8$. We train the classifier with the set of divisors $D$ on toric threefold bases we have generated. Based on accuracy and model interpretability, we find that decision tree provides the best performance in this problem. Hence we will mostly use decision tree and generate a number of analytic rules which are inequalities on the features. 

To specify the features, we need to pick a set of local triple intersection numbers near a divisor $D$. Although one may expect the accuracy to increase with more features included, the decision tree structure and analytic rules will be more complicated. In this work, we only use the triple intersection number information among the divisor $D$ and its neighbor divisors $D_1,\dots,D_n$. On a toric threefold, the number of neighbors of $D$ is $n=h^{1,1}(D)+2$. Hence the input vector has different dimensions for different $h^{1,1}(D)$, and we need to train a different decision tree for this class of divisors with a specific $h^{1,1}(D)$. 

To choose the data set, we need to used the notion of ``resolvable bases'' and ``good bases'' introduced in~\cite{skeleton}. The resolvable bases have toric (4,6) curves which will give rise to a strongly coupled sector in the 4D low energy effective theory, while the good bases do not have these (4,6) curves. We choose our training data set to be the toric divisors on two classes of good bases generated by the random walk approach~\cite{MC} and the random blow up approach~\cite{skeleton} respectively. The out-of-sample accuracies we have achieved range from 85\% to 98\% depending on the $h^{1,1}(D)$, which is remarkably high considering that the task is a multiclass prediction problem \footnote{For some $h^{1,1}(D)$, not all of the 10 non-Higgsable gauge groups are found in the data, so there may be 7$\sim$10 different classes for different data sets.}. Since the decision trees typically have $\mc{O}(10^3\sim 10^4)$ leaves, the number of analytic rules the algorithm generates is too large to present. To circumvent this, we have selected a number of rules which apply to the largest number of samples or have small depth in the decision tree.

The decision trees are tested on a set of resolvable bases as well. It turns out that the accuracies are  usually slightly lower than accuracies on the good bases. Nonetheless, for the Hirzebruch surface $D=\mb{F}_n$, the accuracy is 98.04\%, which is even higher than the out-of-sample accuracies on the good bases. This shows that the decision trees and analytic rules generated from the good bases will apply to resolvable bases as well. 

The structure of this paper is as follows: in section~\ref{s:toric}, we introduce the fundamentals of toric threefolds and a useful diagrammatic representation of triple intersection numbers. In section~\ref{s:F}, we review the basic setups of 4D F-theory and the non-Higgsable gauge groups. In section~\ref{s:gendata}, we show how to generate the data sets in this paper for machine learning. First, we clarified a subtlety involving the codimension-two (4,6) singularity and review the definition of resolvable and good bases in~\cite{skeleton}. Then we show how to generate the good toric threefold bases and construct the input vector (features) from them. In section~\ref{s:ml}, we briefly review the basic definitions in machine learning and the methods used in this paper. In section~\ref{s:method}, we present the universal machine learning framework that will be applied to various data sets in section~\ref{s:divisors} and ~\ref{s:46curve}. Section~\ref{s:divisors} will be focusing on the classification of non-Higgsable gauge groups on divisors with different Picard rank, and we will list a number of analytic rules extracted from the decision tree explicitly. Section~\ref{s:46curve} will be focusing on distinguishing toric (4,6) curves. We then discuss two potential applications of the decision tree trained in section~\ref{s:divisors}: applying them to the resolvable bases and constructing local configurations reversely with the analytic rules. Finally, we summarize the results and discuss future directions in section~\ref{s:con}.

\section{Geometry of toric threefolds}
\label{s:toric}

Toric threefolds are the central geometric objects in this paper. A basic introduction to toric variety can be found in~\cite{Fulton,Danilov}. In this paper, we always assume that the toric threefold is smooth and compact, unless otherwise indicated.

A toric threefold $B$ is characterized by a simplicial fan $\Sigma$ with a set of rays 
\be
\Sigma(1)=\{v_i=(x_{i,1},x_{i,2},x_{i,3})\in\mathbb{Z}^3\}\ (i=1,\dots,n)
\ee
 and a set of 3D cones $\Sigma(3)$. The intersection of $\sigma\in\Sigma(3)$ forms the set of 2D cones $\Sigma(2)$ in the fan. From the compactness and smoothness conditions, the 3D cones span the whole $\mathbb{Z}^3$ and each of them has unit volume. For a smooth compact toric threefold, we always have
\bea
n&=&h^{1,1}(B)+3\\
|\Sigma(3)|&=&2h^{1,1}(B)+2\\
|\Sigma(2)|&=&3h^{1,1}(B)+3.
\eea

Geometrically, the 1D rays $v_i$ correspond to the toric divisors $D_i$, which generates the effective cone of $B$. The 2D cones $v_i v_j$ correspond to the toric curves $D_i\bigcap D_j$, which generates the Mori cone of $B$. The 3D cones $v_i v_j v_k$ are the intersection points of three toric divisors $D_i$, $D_j$ and $D_k$.

In terms of the local coordinates $z_i(i=1,\dots,n)$ on $B$, the toric divisors $D_i$ are given by hypersurface equations $z_i=0$. An important fact is that the global holomorphic section of a general line bundle
\be
L=\sum_{i=1}^n a_i D_i\ (a_i\in\mb{Z})
\ee
can be easily written out as a linear combination of monomials:
\be
s_L=\sum_{u\in \mc{L}}c_u\prod_{i=1}^n z_i^{\langle u,v_i\rangle+a_i}.\label{toricsec}
\ee
Here $\mc{L}$ is a lattice polytope defined by
\be
\mc{L}=\{u\in\mb{Z}^3|\forall v_i\in\Sigma(1),\langle u,v_i\rangle\geq -a_i\},
\ee
and $c_u$ is an arbitrary complex number. In contrast, we do not have a analogous expression for non-toric threefolds.

A very important class of line bundles on $B$ is the multiple of anticanonical line bundle $-mK_B(m\in\mb{Z}_+)$, where
\be
-K_B=\sum_{i=1}^n D_i
\ee
on a toric variety.

On a toric threefold $B$, there are three linear relations among the toric divisors $D_i$:
\be
\bsp
\sum_{i=1}^n x_{i,a} D_i=0\ (a=1,2,3)\label{linrel}
\end{split}
\ee

Now we can compute the triple intersection numbers among the divisors using the information of rays and 3D cones. First, the following equation holds for smooth toric threefolds:
\be
D_i D_j D_k(i\neq j\neq k)=\left\{\begin{array}{rl} 1 & \ v_i v_j v_k\in\Sigma(3) \\ 0 & \ v_i v_j v_k\notin\Sigma(3)\end{array}\right.\label{Dijk}
\ee

The other triple intersection numbers can be computed using the linear relations (\ref{linrel}). For all the triple intersection numbers in form of $D_i^2 D_j(i\neq j)$, they all vanish if $v_i v_j\notin\Sigma(2)$. Otherwise, suppose that the two 3D cones sharing the same 2D cone $v_i v_j$ are $v_i v_j v_k$ and $v_i v_j v_l$ ($i\neq j\neq k\neq l$), then with the equations
\be
\bsp
(D_i D_j)\sum_{m=1}^n x_{m,a} D_m=0\ (a=1,2,3)
\end{split}
\ee
or simply
\be
x_{i,a}D_i^2 D_j+x_{j,a}D_i D_j^2+x_{k,a}+x_{l,a}=0\ (a=1,2,3),
\ee
we can solve $D_i^2 D_j$ and $D_i D_j^2$.

Finally, with all the information of triple intersection numbers in form of $D_i^2 D_j$, we can solve $D_i^3$ by using the data $D_i^2 D_j$ for all the neighbors of $v_i$:
\be
D_i^2\left(\sum_{v_i v_j\in\Sigma(2)} x_{j,a} D_j+x_{i,a}D_i\right)=0\ (a=1,2,3),\label{tripint1}
\ee
Hence we can pick an arbitrary $x_{i,a}\neq 0$, and solve
\be
D_i^3=-\frac{1}{x_{i,a}}\sum_{v_i v_j\in\Sigma(2)} x_{j,a} D_i^2 D_j.\label{tripint2}
\ee

In the end, we are always able to solve all the triple intersection numbers on $B$ uniquely, and they are all integers.

\begin{figure}
\centering
\includegraphics[height=6cm]{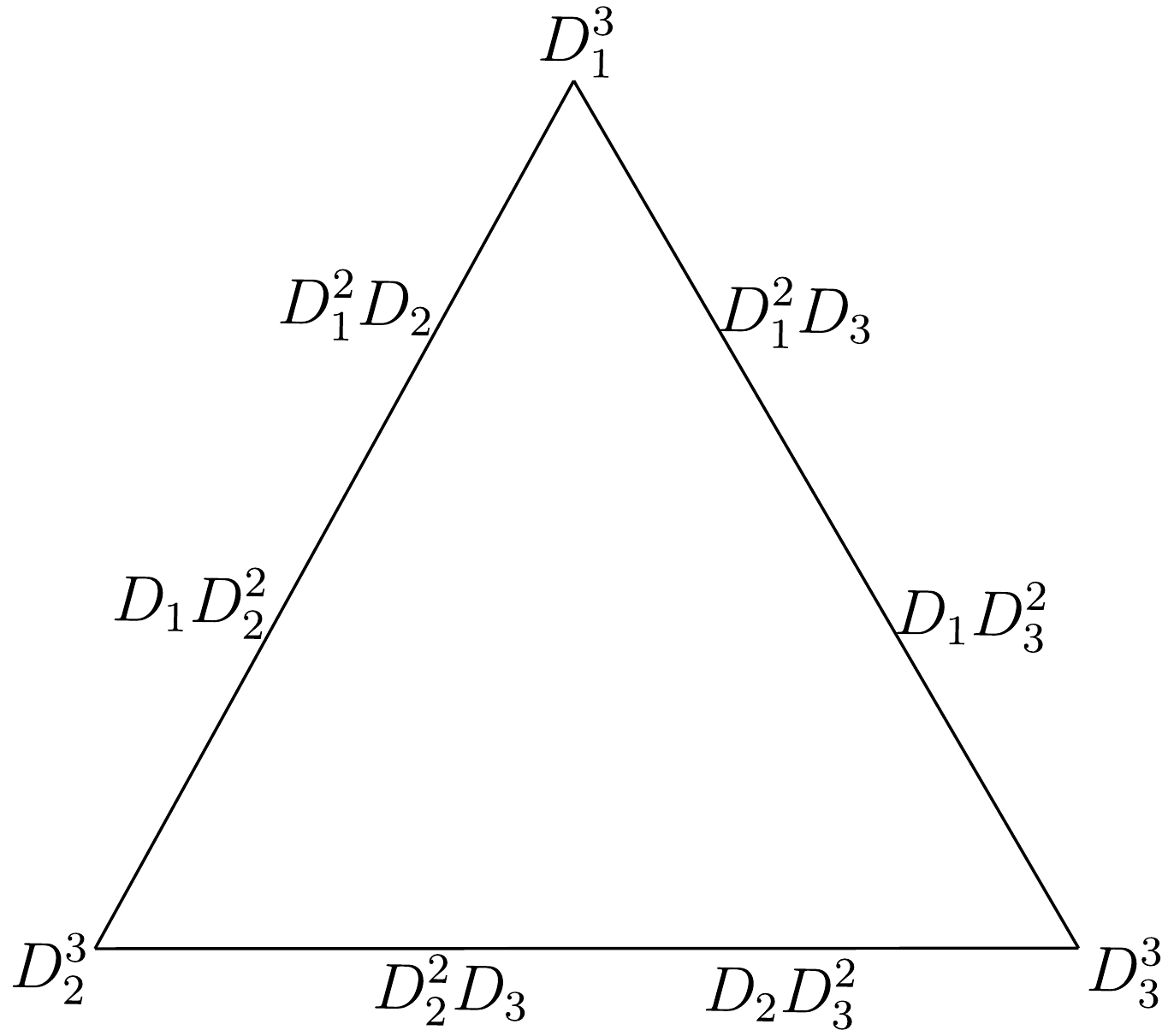}
\caption[E]{\footnotesize A diagrammatic way to show the triple intersection numbers between the divisors.}\label{f:diag}
\end{figure}

Next we introduce a diagrammatic way to present the triple intersection numbers of a toric variety in figure~\ref{f:diag}. On each vertex $v_i$, we label the triple self-intersection number $D_i^3$. On the edges $v_i v_j$, we label the triple intersection numbers $D_i^2 D_j$ and $D_i D_j^2$, where $D_i^2 D_j$ lies closer to the vertex $v_i$. We do not need to label the $D_i D_j D_k(i\neq j\neq k)$ since they are straight forward to read out from the triangulation structure and (\ref{Dijk}). 

\begin{figure}
\centering
\includegraphics[height=6cm]{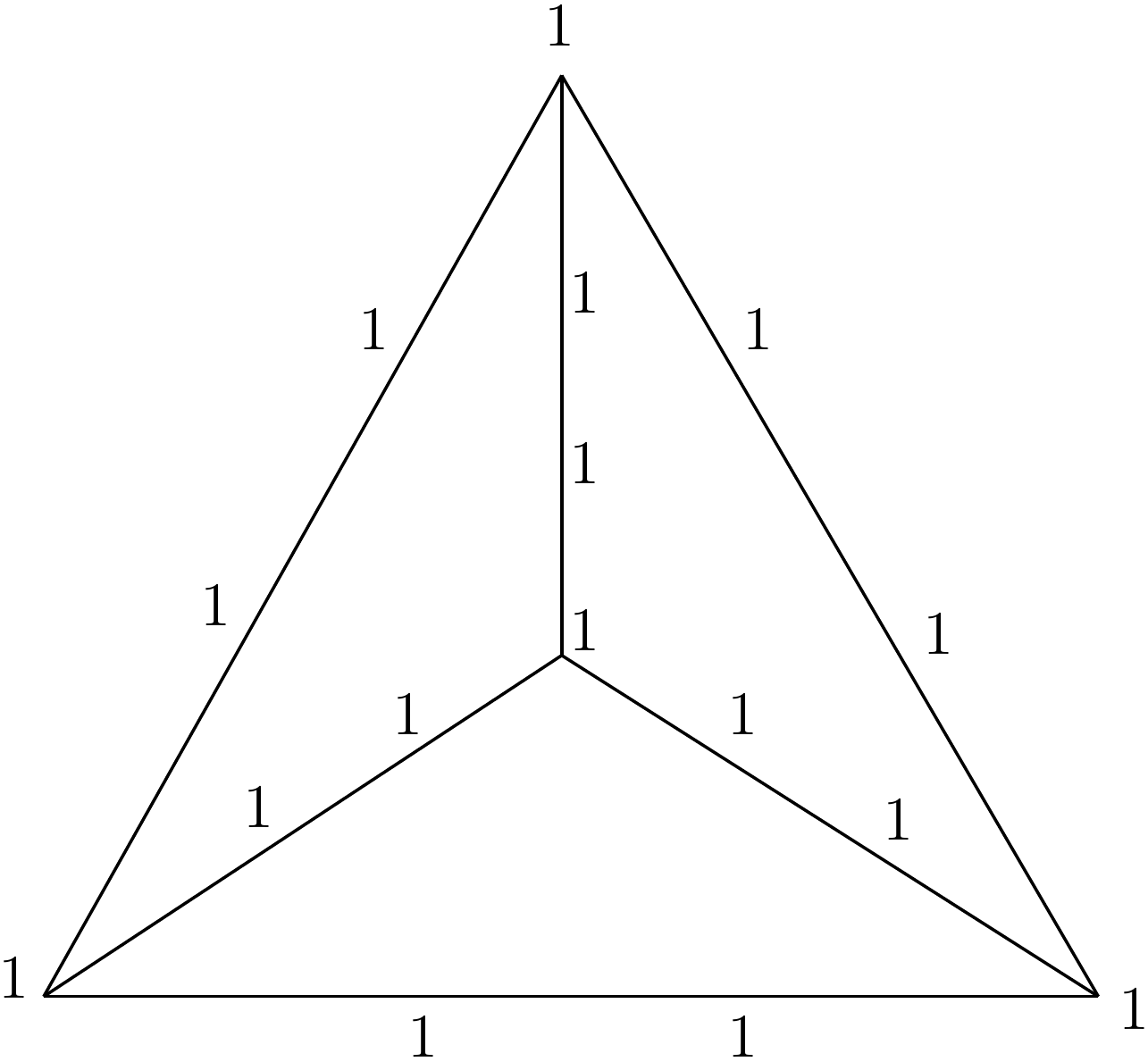}
\caption[E]{\footnotesize Diagrammatic representation of triple intersection numbers on $\mathbb{P}^3$.}\label{f:P3}
\end{figure}

For example, we show the diagrammatric presentation of $\mathbb{P}^3$ in figure~\ref{f:P3}. Clearly all the triple intersection numbers equal to 1.

\begin{figure}
\centering
\includegraphics[height=7cm]{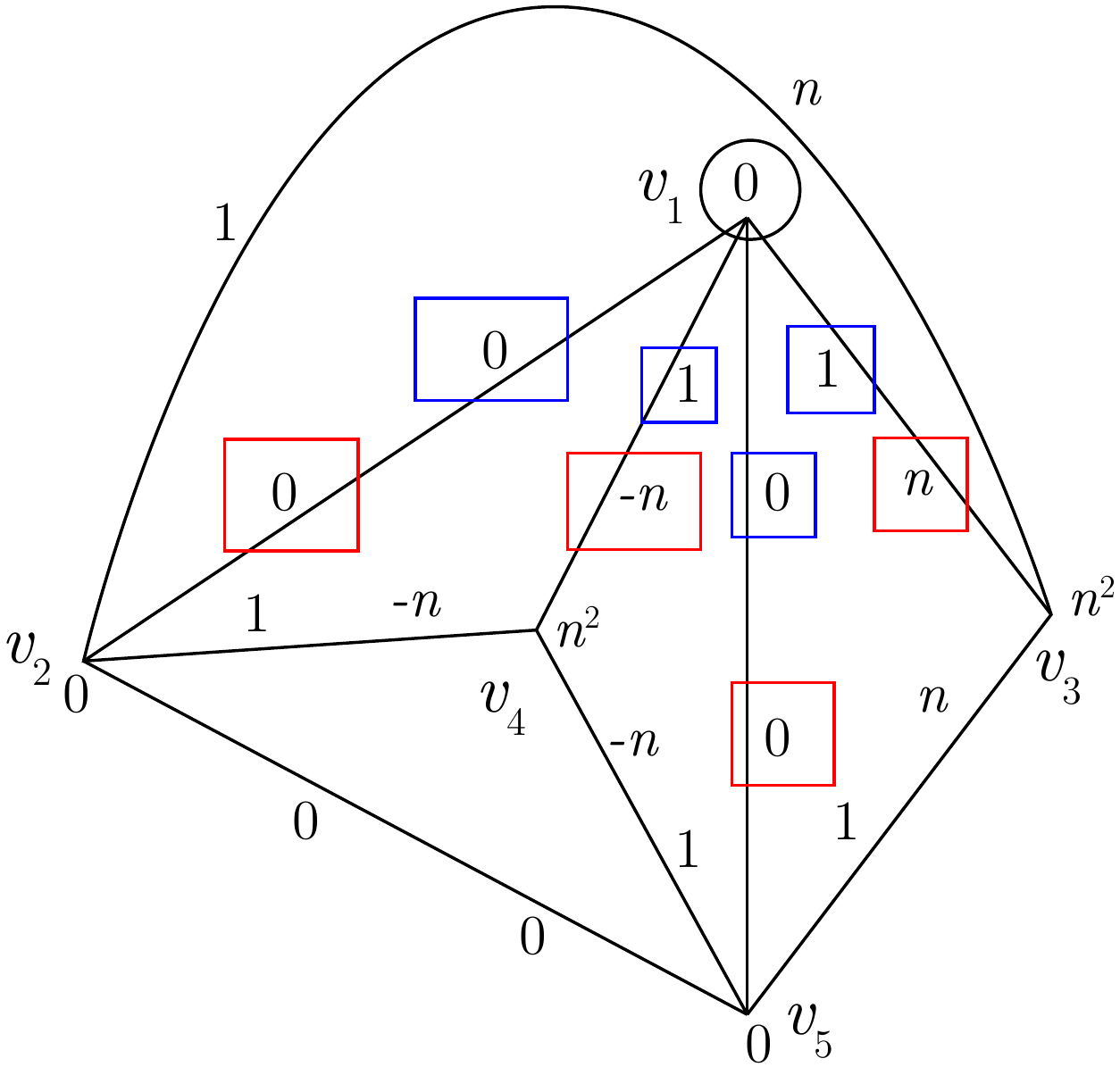}
\caption[E]{\footnotesize Diagrammatic representation of triple intersection numbers on $\tilde{\mathbb{F}}_n$.}\label{f:Fn}
\end{figure}

For generalized Hirzebruch threefold $\tilde{\mathbb{F}}_n$, the 1D rays are $v_1=(1,0,0)$, $v_2=(0,1,0)$, $v_3=(0,0,1)$, $v_4=(0,0,-1)$ and $v_5=(-1,-1,-n)$ and the 3D cones in the fan are $\{v_1 v_2 v_3$, $v_1 v_5 v_3$, $v_2 v_5 v_3$, $v_1 v_2 v_4$, $v_1 v_5 v_4$, $v_2 v_5 v_4\}$. We draw the diagrammatic representation in figure~\ref{f:Fn}.

These numbers on the edges encode the geometric information of toric divisors explicitly. For example, for the divisor $D_1$ in figure~\ref{f:Fn}, the numbers in red squares are $D_1 D_j^2 (j=2,3,4,5)$, which are actually the self intersection number of curves $C_{1j}=D_1 D_j$ on the surface $D_1$. Hence we can directly read off that the divisor $D_1$ is a Hirzebruch surface $F_n$ since the self-intersection of curves $C_{1j}$ are $(0,n,0,-n)$. The numbers in the blue squares are $a_j=D_1^2 D_j (j=2,3,4,5)$, which are the intersection number between the normal bundle $N_{D_1}$ and the curves $C_{1j}$. With the intersection form on the surface $D_1$, we can use $a_j$ to solve the normal bundle $N_{D_1}$, see section~\ref{s:features} for more detail.

The triple intersection numbers in the diagram are not entirely independent. In fact, the self triple intersection number $D^3$ of a divisor $D$ is uniquely fixed by the triple intersection numbers $D^2 D_i$ and $D_i D^2$ where $D_i$ is a toric divisor intersecting $D$. We present a derivation of $D^3$ in Appendix A for small values of $h^{1,1}(D)$.

\begin{figure}
\centering
\includegraphics[height=5.5cm]{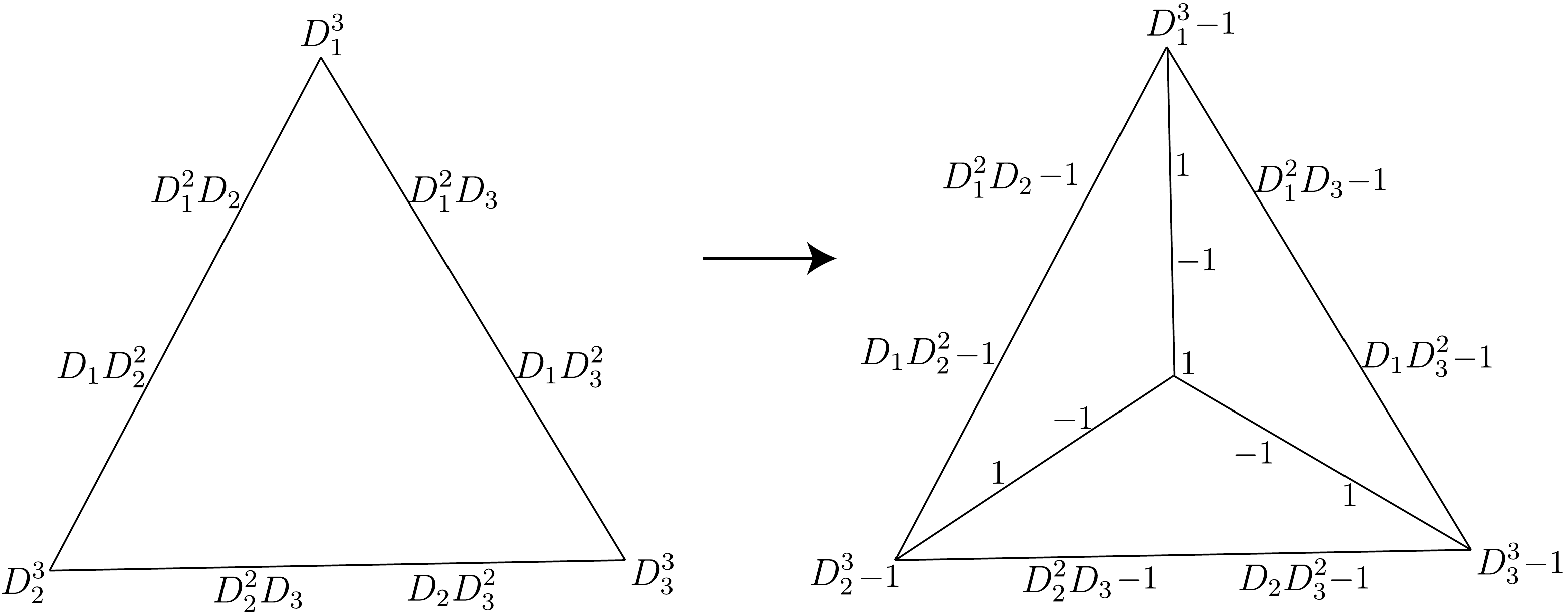}
\caption[E]{\footnotesize Diagrammatic representation of the change in triple intersection numbers after a point $D_1\bigcap D_2\bigcap D_3$ is blown up.}\label{f:blp3}
\end{figure}

The change in triple intersection numbers after a blow up can also be easily computed. If we blow up a 3D cone $v_1 v_2 v_3$ corresponding to a point $D_1\bigcap D_2\bigcap D_3$, we get a new divisor class: the exceptional divisor $E$. The divisors $D_1$, $D_2$ and $D_3$ are properly tranformed to $D_1'=D_1-E$, $D_2'=D_2-E$, $D_3'=D_3-E$. The new 3D cones containing the ray of exceptional divisor $v_E$ are $v_1 v_2 v_E$, $v_1 v_3 v_E$, $v_2 v_3 v_E$, where $v_i(i=1,2,3)$ corresponds to the transformed divisors $D_i'$. From (\ref{Dijk}), we have equations
\be
\bsp
(D_1-E)(D_2-E)E&=1\\
(D_1-E)(D_3-E)E&=1\\
(D_2-E)(D_3-E)E&=1\\
(D_1-E)(D_2-E)(D_3-E)&=0
\end{split}
\ee
Along with the fact that $D_i D_j E=0$ for all $i,j=1,2,3$, which follows from that the curve $D_i D_j$ does not contain the  Poincar\'{e} dual of $E$, we can solve all the triple intersection numbers:
\be
\bsp
D_i E^2&=0\ (i=1,2,3)\\
E^3&=1\\
D_i' D_j' D_k'&=D_i D_j D_k-1\ (i,j,k=1,2,3)\\
D_i'^2 E&=1\ (i=1,2,3)\\
D_i' E^2&=-1\ (i=1,2,3)
\end{split}
\ee
We show the change in triple intersection numbers after the blow up in figure~\ref{f:blp3}. Hence the exceptional divisor is a $\mathbb{P}^2$ with normal bundle $N_E=-D_i|_E=-H$, where $H$ is the hyperplane class on $\mb{P}^2$.

\begin{figure}
\centering
\includegraphics[height=6cm]{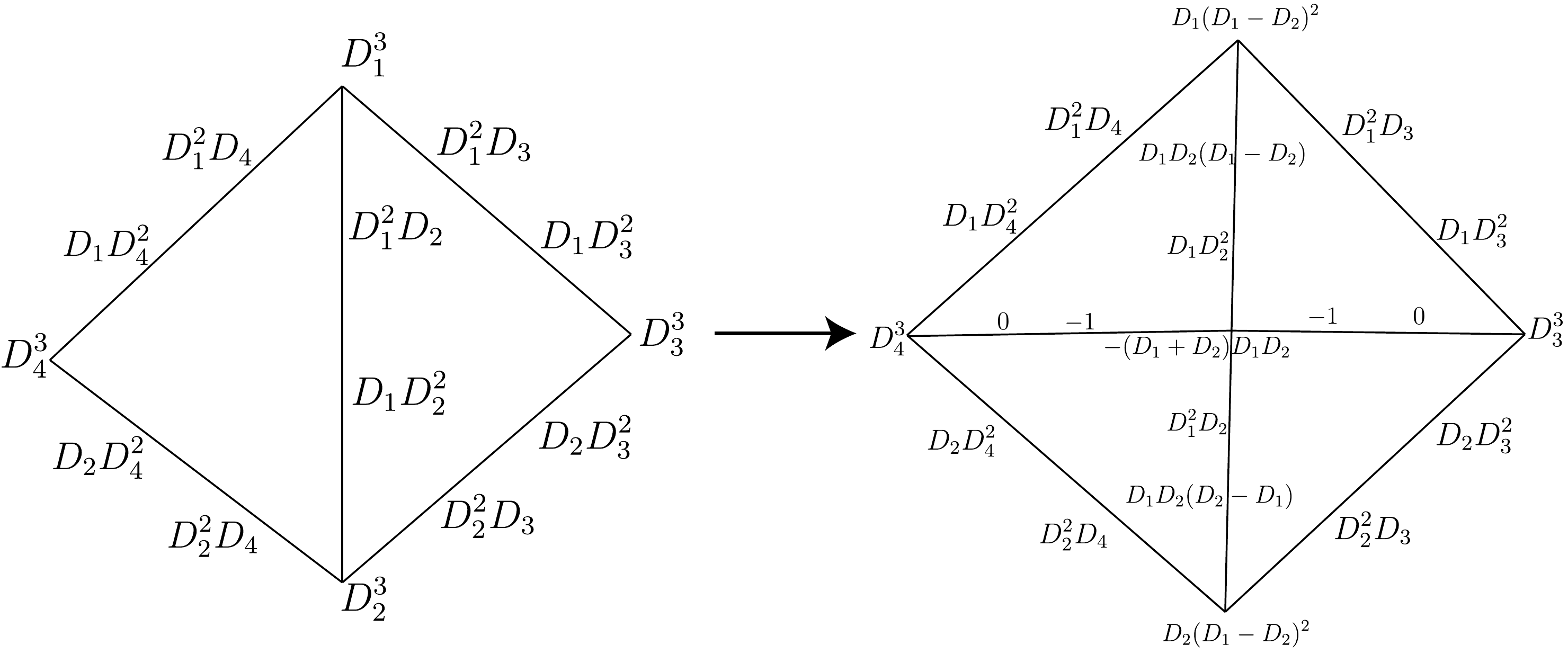}
\caption[E]{\footnotesize Diagrammatic representation of the change in triple intersection numbers after a curve $D_1\bigcap D_2$ is blown up.}\label{f:blp2}
\end{figure}

We can do the similar analysis for the case of blowing up a curve $D_1 D_2$, and the change in triple intersection numbers is shown in figure~\ref{f:blp2}. As we can see, the exceptional divisor is $\mathbb{F}_n$ in this case, where $n=|D_1 D_2(D_1-D_2)|$. 

\section{F-theory on toric threefold bases and the non-Higgsable gauge groups}
\label{s:F}

An introduction to F-theory can be found in~\cite{Weigand}, and we will only present the essential information for our setup.

In this paper, to get a 4D $\mc{N}=1$ effective field theory, we always consider an elliptic Calabi-Yau fourfold $X$ over the base manifold $B$ with a global section. It is described by the Weierstrass equation:
\be
y^2=x^3+fx+g,\label{Weierstrass}
\ee
where the Weierstrass polynomials $f$ and $g$ are holomorphic sections of line bundles $\mc{O}(-4K_B)$, $\mc{O}(-6K_B)$. $-K_B$ is the anticanonical divisor (line bundle) of the base $B$, which is always effective. When the discriminant $\Delta=4f^3+27g^2$ vanishes over a subset $L$ of $B$, the elliptic fiber is singular over $B$. If $L$ is complex codimension-one, 7-branes will locate on $L$ and the attached open string modes will give rise to gauge fields in the 4D low energy effective theory.

Hereafter we assume that $f$ and $g$ are generic sections, such that the order of vanishing of $\Delta$ over the codimension-one locus $L$ is minimal. Under this condition, the gauge groups from the codimension-one locus $L$ are the minimal non-Higgsable gauge groups~\cite{clusters,4d-NHC}. We list the possible non-Higgsable gauge groups with the order of vanishing of $(f,g,\Delta)$ in table~\ref{t:NHG}.

\begin{table}
\begin{center}
\begin{tabular}{|c |c |c |c |c |}
\hline
Kodaira type &
ord ($f$) &
ord ($g$) &
ord ($\Delta$) &
gauge group\\ \hline \hline
$III$ &1 & 2  &3 & SU(2) \\
$IV$ & $\geq 2$ & 2 & 4 & SU(3) or SU(2)\\
$I_0^*$&
$\geq 2$ & 3 & $6$ & SO(8) or SO(7) or $G_2$ \\
$IV^*$& $\geq 3$ & 4 & 8 & $E_6$ or $F_4$\\
$III^*$&3 & 5 & 9 & $E_7$\\
$II^*$& 4 & 5 & 10 & $E_8$ \\
\hline
non-min & 4 & 6 & 12 & - \\
\hline
\end{tabular}
\end{center}
\caption[x]{\footnotesize  table of
non-Higgsable non-Abelian gauge groups and their Kodaira singular fiber type.
For the Kodaira fibers $IV$, $I_0^*$ and $IV^*$, the gauge group is not uniquely determined by the orders
of vanishing of $f, g$. One need additional monodromy information in the Weierstrass polynomials to fix the precise gauge group. When $(f,g,\Delta)$ vanishes to order $(4,6,12)$ or higher on a codimension-one locus, the geometry does not describe any supersymmetric vacua.

}
\label{t:NHG}
\end{table}

For the fiber types $IV$, $I_0^*$ and $IV^*$, the gauge group is
specified by additional information encoded in the ``monodromy cover
polynomials'' $\mu(\psi)$ \cite{Grassi-Morrison-2}. Suppose that the
divisor is given by a local equation $w=0$, then for the case of type
$IV$,
 \be
 \mu(\psi)=\psi^2-(g/w^2)|_{w=0}=\psi^2-g_2.
 \ee
The gauge group is SU(3) if and only if
$g_2$ is a complete square. The case of type $IV^*$ is similar, where
the monodromy cover polynomial is 
\be
\mu(\psi)=\psi^2-(g/w^4)|_{w=0}=\psi^2-g_4.
\ee
When $g_4$ is a complete square, then the corresponding gauge group is $E_6$, otherwise it is $F_4$. 

For the case of type $I_0^*$, the monodromy cover polynomial is
\be
\mu(\psi)=\psi^3+(f/w^2)|_{w=0}\psi+(g/w^3)|_{w=0}=\psi^3+f_2\psi+g_3.\label{monocover}
\ee
When $\mu(\psi)$ can be decomposed into three factors:
\be
\mu(\psi)=(\psi+a)(\psi+b)(\psi-a-b),
\ee
the corresponding gauge group is SO(8). Otherwise, if it can be decomposed into two factors:
\be
\mu(\psi)=(\psi+a)(\psi^2-a\psi+b),
\ee
the gauge group is SO(7). If $\mu(\psi)$ is irreducible, then the gauge group is $G_2$.

On a general threefold base, suppose that the divisor $D$ is given by the hypersurface equation $w=0$ locally, and we expand
\bea
f&=&\sum_k f_{k,D} w^k\\
g&=&\sum_k g_{k,D}w^k.
\eea
Then the line bundle generators for $f_k$ and $g_k$ can be written down using the normal bundle $N_D$ and canonical line bundle $K_D$~\cite{4d-NHC}:
\be
f_{k,D}\in\mc{O}(-4K_{D}+(4-k)N_{D}-\sum_{D\bigcap D_i\neq\varnothing}\phi_i C_i),\label{normf}
\ee
\be
g_{k,D}\in\mc{O}(-6K_{D}+(6-k)N_{D}-\sum_{D\bigcap D_i\neq\varnothing}\gamma_i C_i).\label{normg}
\ee
Here $\mc{O}(\cdot)$ denotes the holomorphic section of a line bundle on the complex surface $D$. $\phi_i$ and $\gamma_i$ denote the order of vanishing of $f$ and $g$ on another divisor $D_i$ which intersects $D$. $C_i=D\bigcap D_i$ is the intersection of $D$ and $D_i$, which has the topology of $\mb{P}^1$. If $f_{k,D}\in\mc{O}(C_k)$ where $C_k$ is not an effective divisor on $D$  for all $k< k_0$, then $f$ vanishes to at least order $k_0$ on $D$. Similar statement holds for $g$.

The problem of this formula is that there may be other non-local constraints that are not encoded in the neighboring divisors of $D$. In reality, the order of vanishing of $(f,g)$ may be higher than the values given by (\ref{normf}, \ref{normg}). Hence we can only read out a subgroup of the actual non-Higgsable gauge group on $D$.

For toric threefold bases, the exact form of $f$ and $g$ can be easily computed with the holomorphic section formula (\ref{toricsec}). The sets of monomials in $f$ and $g$ are given by the following lattice polytopes:
\be
\mc{F}=\{u\in\mathbb{Z}^3|\forall v_i\in\Sigma(1),\langle u,v_i\rangle\geq -4\},\label{fset}
\ee
\be
\mc{G}=\{u\in\mathbb{Z}^3|\forall v_i\in\Sigma(1),\langle u,v_i\rangle\geq -6\},\label{gset}
\ee
where $v_i$ are the 1D rays in the fan of the toric base. 

The order of vanishing of $f$ and $g$ on a toric divisor $D$ corresponding to the ray $v\in\Sigma(1)$ are 
\be
\bsp
&\text{ord}_{D}(f)=\min(\langle u,v\rangle+4)|_{u\in\mc{F}},\\
&\text{ord}_{D}(g)=\min(\langle u,v\rangle+6)|_{u\in\mc{G}},
\end{split}
\ee
Now we are going to present the explicit monodromy criteria to distinguish the gauge groups for the cases of type $IV$, $IV^*$ and $I_0^*$ fiber in table~\ref{t:NHG}. We denote the toric ray of the neighboring divisors of $D$ by $v_1,\dots,v_p$.

When $\text{ord}_{D}(f)\geq 2$, $\text{ord}_{D}(g)=2$, the
singularity type is $IV$. In this case, when $g_2$ only contains one
monomial $u$ and $2|\langle u,v_i\rangle$ $(i=1,\dots,p)$, then the gauge group is SU(3). Otherwise the gauge group is SU(2).

When $\text{ord}_{D}(f)\geq 3$, $\text{ord}_{D}(g)=4$, the
singularity type is $IV^*$. In this case, when $g_4$ only contains one
monomial $u$ and $2|\langle u,v_i\rangle$ $(i=1,\dots,p)$, then the gauge
group is $E_6$. Otherwise the gauge group is $F_4$.

 When $\text{ord}_{D}(f)\geq 2$, $\text{ord}_{D_i}(g)=3$ or $\text{ord}_{D}(f)=2$, $\text{ord}_{D}(g)>3$, the singularity type is $I_0^*$. If $\text{ord}_{D}(f)=2$ and $\text{ord}_{D}(g)>3$,
  $\mu(\psi)=\psi^3+f_2\psi$, the gauge group is either SO(7) or
  SO(8). The gauge group is SO(8) only when $f_2$ only contains one
  monomial $u$ and $2|\langle u,v_i\rangle$ $(i=1,\dots,p)$. Otherwise the gauge group is SO(7).

For the other case, $\text{ord}_{D}(f)\geq 2$ and
$\text{ord}_{D}(g)=3$, if the following two conditions are satisfied:

 (1) $f_2$ only contains a single monomial $u$ and $2|\langle u,v_i\rangle$ $(i=1,\dots,p)$ or $f_2=0$;
 
 (2) $g_3$ only contains a single monomial $w$ and $3|\langle w,v_i\rangle$ $(i=1,\dots,p)$,
 
\noindent
then the gauge group is SO(8). Otherwise it is $G_2$.

We will always apply this method to compute the non-Higgsable non-Abelian gauge group on a toric divisor, which is the label of the data samples. It is worth pointing out, the determination of gauge groups involves several inequalities. Later we will see that this is coincidentally reflected in the machine learning algorithm selection.

For some particular class of bases, there exist non-Higgsable Abelian gauge groups from the Mordell-Weil group of the elliptic fibration~\cite{Martini-WT,Morrison:2016lix,NHU(1)s}. However, they do not appear on toric bases~\cite{NHU(1)s} and are not considered.

\section{Generation of toric data}
\label{s:gendata}

\subsection{Resolvable and good bases}
\label{s:resgood}

In \cite{skeleton}, we introduced the terminology of ``resolvable bases'' and ``good bases'' depending on the existence of complex codimension-two locus $L\subset B$ where $(f,g)$ vanishes to order $(4,6)$ or higher. 

If these codimension-two (4,6) loci exist, then we can try to blow up these loci and lower the order of vanishing of $(f,g)$ to be  under $(4,6)$. If this blow-up process can be done without introducing a codimension-one (4,6) locus in the process, then we call this base $B$ a ``resolvable base''. 

If $B$ is free of codimension-two (4,6) locus, then we call it a ``good base''. 

For a toric threefold base, we can write down the order of vanishing of $f$ and $g$ 
on a toric curve $D_i D_j$ corresponding to a 2D cone $v_i v_j$:
\be
\bsp
&\text{ord}_{D_i D_j}(f)=\min(\langle u,v_i+v_j\rangle+8)|_{u\in\mc{F}},\\
&\text{ord}_{D_i D_j}(g)=\min(\langle u,v_i+v_j\rangle+12)|_{u\in\mc{G}},\label{dovcurve}
\end{split}
\ee

If there is a toric curve $D_i D_j$ with $\text{ord}_{D_i D_j}(f)\geq 4,\text{ord}_{D_i 
  D_j}(g)\geq 6$, we can see that there does not exist $u\in\mc{F}$ where $\langle u,v_i+v_j\rangle<4$ or $u\in\mc{G}$ where $\langle u,v_i+v_j\rangle<6$. Hence if we try to blow up the toric curve by adding a new ray $\tilde{v}=v_i+v_j$, the sets of Weierstrass monomials will not change. 

To check whether a base is resolvable or not, one only needs to check whether the origin $(0,0,0)$ lies on the
boundary of $\mc{G}$. If $(0,0,0)$ does not lie on the boundary of
the lattice polytope $\mc{G}$, then after the resolution process
where all the (4,6) curves are blown up, there will not be a
codimension-one (4,6) locus on any divisor. The reason is that if
there exists such a divisor corresponding to the ray $v$, then all the
points $u\in\mc{G}$ satisfying $\langle u,v\rangle<0$ will vanish
 and the origin $(0,0,0)$ lies on the boundary plane $\langle
 u,v\rangle=0$ of $\mc{G}$. Since the polytope $\mc{G}$ does not change when we blow up (4,6) curves, this condition applies to the original resolvable base as well.

Now we clarify the physical difference of the non-resolvable, resolvable and good bases.

\begin{itemize}
\item{For the non-resolvable bases, they cannot support any elliptic Calabi-Yau manifold with only terminal singularities. For this reason, they do not describe any supersymmetric vacua in F-theory. Hence we never include these bases in the classification program of F-theory geometries.}

\item{For the resolvable bases, there may be a strongly coupled superconformal sector on the codimension-two (4,6) locus. 

In the 6D F-theory case, blowing up a codimension-two (4,6) point will give a non-zero v.e.v. to the scalar in the tensor multiplets, and the (1,0) SCFT will be deformed into the tensor branch. On the tensor branch, the low energy theory has a usual gauge theory description. If we shrink the exceptional divisors and go back to the superconformal point, then the gauge groups and matter on the exceptional divisors will become strongly coupled ``superconformal matter''~\cite{6dCM}. 

In 4D $\mc{N}=1$ theory, the tensor multiplets is replaced by a number of chiral multiplets. The situation is more subtle since the instanton effect from Euclidean D3 branes~\cite{4dCM} and $G_4$ flux may break the superconformal symmetry. However, one can generally expect a strongly coupled non-Lagrangian sector if there are (4,6) curves on the base threefold.}

\item{For the good bases, the low energy effective theory should be free of these SCFT sectors, and we have a 4D $\mc{N}=1$ supergravity coupled with a number of vector and chiral multiplets.}
\end{itemize}

In this paper, we generally accept all the resolvable bases and good bases. We will not consider other subtleties such as codimension-three (4,6) points~\cite{Candelas:2000nc,Baume:2015wia} or terminal singularities in the Weierstrass model that cannot be resolved~\cite{Terminal}. We will generally accept their appearance and leave their physical interpretation to future work.

\subsection{Generation of toric threefold bases}
\label{s:genbase}

We use the divisors on the good bases to train the classifier, and the bases are generated by two different methods. The first class of bases is the ``end point bases'' introduced in~\cite{skeleton}. 

We start with $\mathbb{P}^3$ and randomly blow up toric points or curves with the same probability. During the process, the base may contain toric curves where $(f,g)$ vanish to order $(4,6)$ or higher. However, it is always required to be resolvable, or equivalently the polytope $\mc{G}$ (\ref{gset}) should contain the origin $(0,0,0)$ in its interior. Finally, we will end up at a base without toric $(4,6)$ curves, which is called an end point base. It is impossible to blow up a toric curve or point on an end point base to get another resolvable base. The end point base may contain toric divisors with $E_8$ gauge group and non-toric $(4,6)$ curves on it, but we allow these to happen since we can easily blow up these $(4,6)$ curves. They are analogous to the $-9/-10/-11$ curves on 2D bases. In total, we have 2,000 end point bases generated in~\cite{skeleton}. 

The second class of bases is called ``intermediate base'' which is distinguished from the end point bases. Our method to generate these intermediate bases is similar to the random walk approach in \cite{MC}. We start from $\mathbb{P}^3$ and do a random toric blow up or blow down at each step with equal probability. In the whole process, it is required that no toric $(4,6)$ curve appears, but again the toric divisors with $E_8$ gauge group and non-toric $(4,6)$ curves on it are allowed. Each random walk sequence from $\mathbb{P}^2$ contains 10,000 bases $b_1,b_2,\dots,b_{10,000}$, however we only pick out the first 20 bases $b_1,b_2,\dots,b_{20}$ and the bases $b_{100n}, n\in\mathbb{Z}$. The reason is that bases related by a few blow up/downs have similar divisor structure, and we want to reduce repetitive samples in our data set. In total, we generate 1,500 of these random walk sequences and we take in total 180,000 bases out of them.

With the toric data, we can classify the toric divisors on these bases according to the $h^{1,1}(D)$ (equal to the number of their neighbor divisors minus 2) and compute the local triple intersection numbers and gauge groups. 

\subsection{Generation of the features}
\label{s:features}

Now we can generate the features for the machine learning program, which are local triple intersection numbers near a divisor $D$. For a divisor $D$ with Picard rank rk(Pic$(D))\equiv h^{1,1}(D)=p-2$, there are exactly $p$ toric divisors intersecting $D$. We relabel them by $D_1,\cdots,D_p$, and the toric curves on $D$ are $C_i=D_i\bigcap D$. The toric curves $C_i(i=1,\dots,p)$ are cyclic, such that the intersection numbers between two different curves are 
\be
C_i\cdot C_j=\left\{\begin{array}{rl} 1 & \mathrm{\ if\ }|i-j|=1\mathrm{\ or\ }p-1\\0 & \mathrm{\ otherwise}\end{array}\right.
\ee

Then we can define a $5p$-dimensional feature vector $V(D)$ with the triple intersection numbers near $D$. 

The first $p$ elements of $V(D)$ are $D_1^2 D, D_2^2 D,\dots, D_{p}^2 D$. Since $C_i^2=D_i^2 D(i=1,\dots,p)$, these numbers exactly fix the topological type of divisor $D$. The next $p$ elements of $V(D)$ are $D^2 D_1, D^2 D_2,\dots, D^2 D_{p}$. Since $D^2 D_i=N_D\cdot C_i(i=1,\dots,p)$, they fully determine the normal bundle $N_D$. 

The other $3p$ elements of $V(D)$ are $D_1^3,\dots, D_{p}^3$ and $D_1^2 D_2$, $D_2^2 D_1$, $D_2^2 D_3$, $D_3^2 D_2$,$\dots$, $D_{p}^2 D_1$, $D_1^2 D_{p}$. They are well defined numbers for each toric divisor $D$, and they encode the information of the neighboring divisors of $D$ in a subtle way.

These features are not entirely independent. For example, the Hirzebruch surfaces $\mathbb{F}_n$ divisors has $h^{1,1}(D)=2$ and four neighboring divisors. The four toric curves on $\mb{F}_n$ has self-intersection numbers $C_1^2=0$, $C_2^2=n$, $C_3^2=0$ and $C_4^2=-n$. Then we label the four neighboring divisors of $D$ by $D_1,\dots,D_4$, where $ D_1^2 D= D_3^2 D=0$, $D_2^2 D=n$, $D_4^2 D=-n$ and $v_1 v_2, v_2 v_3, v_3 v_4, v_4 v_1\in\Sigma(2)$. We denote the $(-n)$-curve on $\mathbb{F}_n$ by $S$ and the $0$-curve on $\mathbb{F}_n$ by $F$, which have intersection products $S\cdot S=-n$, $S\cdot F=1$, $F\cdot F=0$. Then we have $C_1=C_3=D\cdot D_1=D\cdot D_3=F$, $C_4=D\cdot D_4=S$, $C_2=D\cdot D_2=S+nF$. Now we can shorten $V(D)$ to the following 15-dimensional vector $\tilde{V}(D)$, whose components are also denoted as $f_0,\dots,f_{14}$.
\be
\bsp
\tilde{V}(D)=&(n, D^2 D_1, D^2 D_2, D_1^3, D_2^3, D_3^3, D_4^3, D_1^2 D_2, D_1 D_2^2, D_2^2 D_3, D_2 D_3^2, D_3^2 D_4, D_3 D_4^2, \\
&D_4^2 D_1, D_4 D_1^2)
\end{split}
\ee
The other quantities such as $D D_3^2, D D_4^2, D^2 D_3, D^2 D_4, D^3$ are all redundant, since there are linear relations between the curves $C_1, C_2, C_3, C_4$ on $\mathbb{F}_n$, see appendix~\ref{s:appFn}. We plot the features in figure~\ref{f:mlFn} with the diagrammatic presentation introduced in section~\ref{s:toric}. 

\begin{figure}
\centering
\includegraphics[height=7cm]{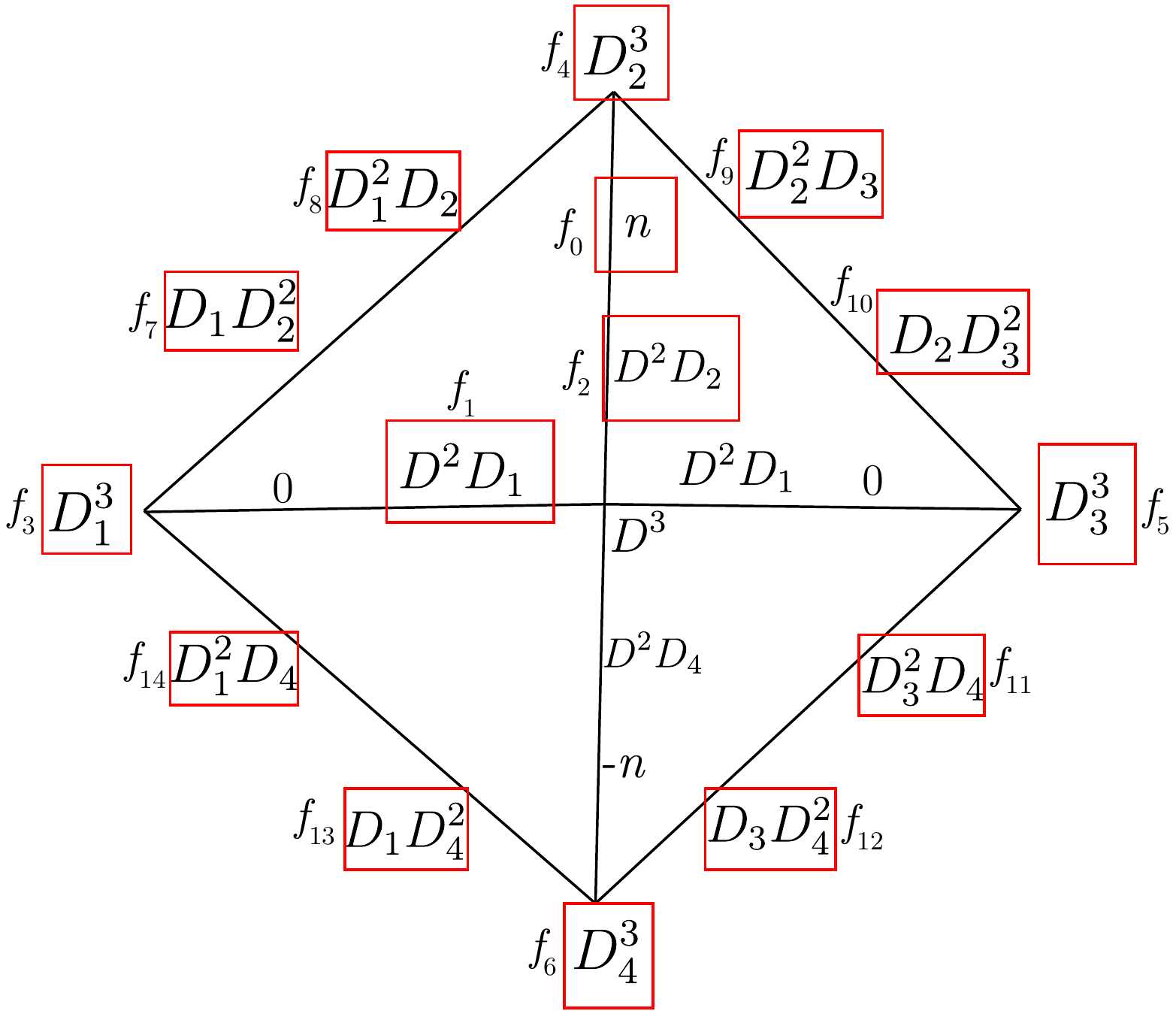}
\caption[E]{\footnotesize The features used in machine learning in the diagrammatic representation of triple intersection numbers near a $\mathbb{F}_n$ divisor $D$, labeled as $f_0,\dots,f_{14}$.}\label{f:mlFn}
\end{figure}

The normal bundle of $D$ can be written as $N_D=aS+bF$, and we have
\bea
f_1&\equiv& D^2 D_1=N_D\cdot F=a\\
f_2&\equiv& D^2 D_2=N_D\cdot (S+nF)=b.
\eea 

For divisors with other $h^{1,1}(D)$, we will present the shortened input vector $\tilde{V}(D)$ case by case in section~\ref{s:divisors}.

We will take the set of Hirzebruch surface on the end point bases as the sample set in section~\ref{s:method}, since the end point bases have similar structure~\cite{skeleton} and it provides a better test ground for various machine learning techniques. This data set is denoted as $S_{\rm end}(\mb{F}_n)$. In section~\ref{s:divisors}, we will provide more detailed results and include both the end point bases and intermediate bases in the training set. 

\section{A brief introduction of machine learning}
\label{s:ml}

In this section, we give a brief introduction of machine learning for the audience. First, we decribe the typical setup and procedure of a machine learning problem, including notation, training, and testing. Then we discuss the details and properties of a few most commonly used machine learning algorithms, including decision tree, feedforward neural network, logistic regression, random forest and support vector machine (SVM). 

There are two major categories of machine learning problems, namely, supervised learning and unsupervised learning. In supervised learning each input is associated with an output, and the objective of the algorithm is to predict the output for a given input.  Furthermore, when the output variable is from a set of categories (e.g.``cat",``dog") it is called a classification problem and the output is referred to as label, otherwise when the output takes continuous values it is called a regression problem. As opposed to supervised learning, in unsupervised learning input data do not have output associated, and the algorithm is tasked to classify the input data into different groups. In this paper, to fit the tasks described in previous sections, we will focus on supervised learning, and more specifically, classification algorithms.  

\subsection{Training and testing}
Following the description above, a machine learning classification problem involves a data set $(X,Y)$ where $X$ is an $N\times K$ matrix containing $N$ input $K$-dimensional variables, and $Y$ contains $N$ output variables. In general, training a machine learning algorithm can be summarized as an optimization problem which is to find
\bea
\hat{f}=\argmin \sum_{(x_i,y_i) \in (X,Y)} F\bigg[f(x_i) , y_i\bigg],\,\label{training}
\eea
where $F$ is the error function defined on the prediction $f(x_i)$ and actual output value $y_i$ for every pair $(x_i,y_i)\in (X,Y)$. An algorithm is specified by setting both $f$ and $F$. Note that many machine learning algorithms are semi/non-parametric, so sometimes $\hat{f}$ is not a closed-form function but rather a combination of operations. 

The optimization procedure that determines $\hat{f}$ is called training or fitting. Consequentially, the data used in the optimization, $(X,Y)$ is called training data, and any other data that is not part of training can be treated as test data. After the algorithm is trained, one can start making prediction on any input data $(x',y')$ via
\bea
\tilde{y}' = \hat{f}(x').
\eea 
By comparing the prediction $\tilde{y}'$ and the actual label $y'$, one can compute the performance of the algorithm. If $(x',y')\in (X,Y)$, this is called in-sample performance, as the test is done on the training data. Otherwise if $(x',y')\notin (X,Y)$ the performance is called out-of-sample. In-sample performance indicates goodness of the fit of the algorithm and out-of-sample performance shows the algorithm's real prediction power. In the case of classification, the accuracy can serve as a good performance measure, defined as 
\bea
ACC=\frac{\rm{Number \, of \, correct \, predictions}}{\rm{Number \, of \, all \,  predictions}}\,.
\eea

\subsection{Classification algorithms}\label{clfs}
Here we describe the details of the five major classification algorithms applied in this paper. More thorough description can be found in~\cite{Mehta2018}.

\begin{figure}
\centering
\includegraphics[height=5cm]{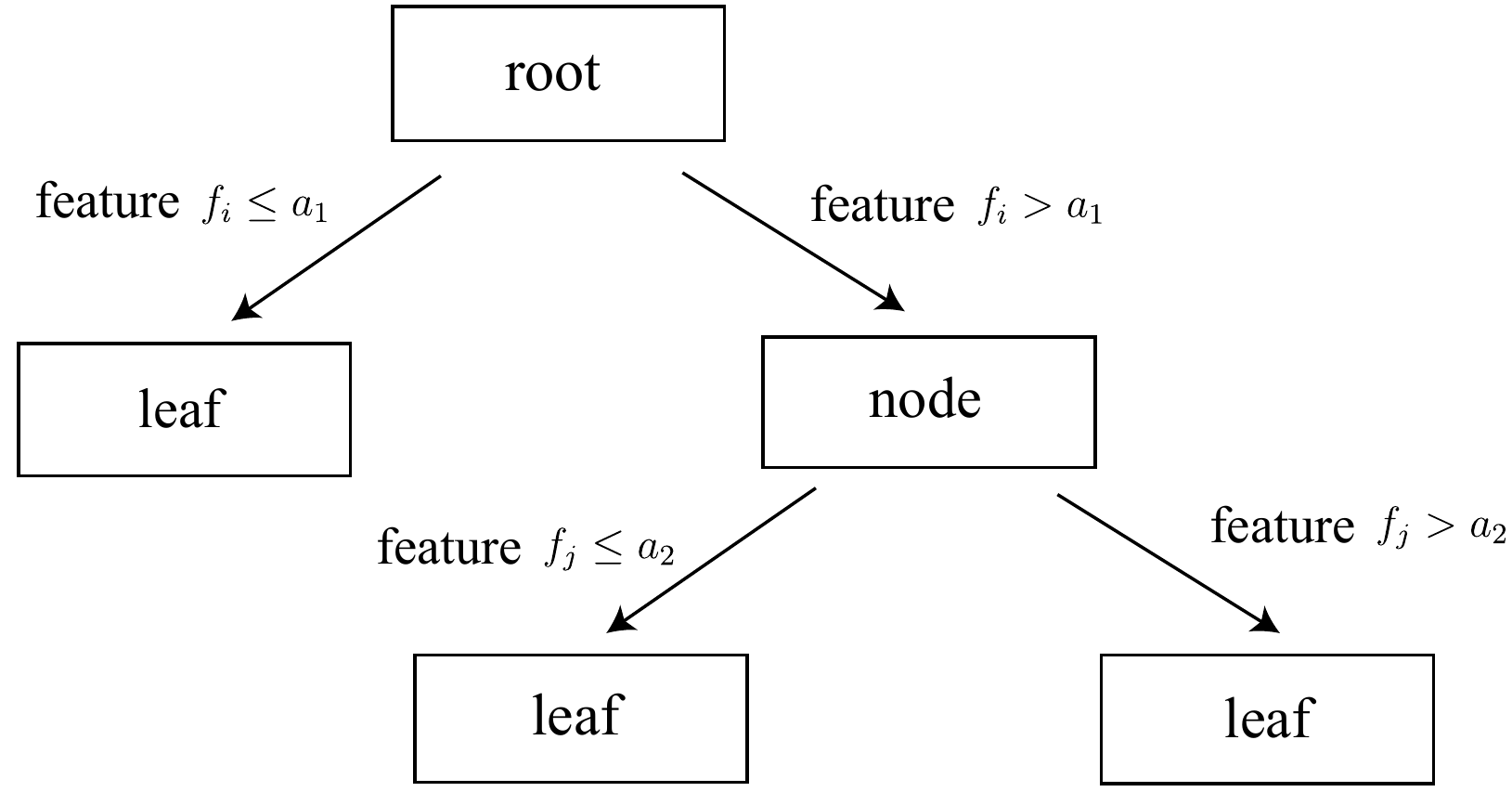}
\caption[E]{\footnotesize An illustration of splits of data in a decision tree. In the first split, the $i$-th feature splits at $a_1$ produces the highest information gain. In the second split, the $j$-th feature splits at $a_2$ produces the highest information gain.}\label{dt}
\end{figure}

\textbf{Decision tree} is a prototypical tree-based algorithm which consists of numerous sequencial ``splits". Each split divides the incoming data set into two non-overlapping subsets by partitioning on one feature.  
Succeeding splits take the previous split's output data sets as input, so the size of data sets for each split decreases. Once the stopping criteria are met for an output subset, it will no longer be split. We will refer to the input data sets as nodes if they are further split, otherwise they will be referred to as leaves. An illustration is shown in figure~\ref{dt}.

In the case of classification, a split on a data set is a binary division\footnote{Even though a dataset can be divided into multiple subsets at each split, it is common to restrict to binary divisions only. And we only consider binary divisions hereafter.}  that maximizes the information gain defined as
\be
{\rm{Information \,\, gain}} = I - \frac{N_{left}}{N} I_{left}  - \frac{N_{right}}{N} I_{right},
\ee\label{IG}
where $(I,N),(I_{left},N_{left}), (I_{right},N_{right})$ are respectively the (impurity, numbers of samples) in the input set, left output subset and right output subset. We choose the Gini index as the impurity measure, which is one of the most popular choices and is defined as follows:
\be
I=\sum_{i=1}^K p_i(1-p_i),
\ee\label{impurity}
where $i$ runs though all the $K$ classes and $p_i$ is the ratio of samples in class $i$ and all samples in the input set. The algorithm is recursive: it starts with the original data set as the parent node, computes the information gain of each possible split on every feature and applies the split that gives rise to the highest information gain evaluated by (\ref{IG}). Then the left and right nodes are regarded as parent nodes and the algorithm is repeated on them.

Since most problems studied in this paper are multiclass problems, i.e. the number of classes is greater than two, it is worth mentioning that some classification algorithms treat multiclass problems differently from binary cases. For some classifiers, multiclass prediction takes longer runtime and computation power than binary prediction. In the case of decision trees, as shown in (\ref{impurity}), the classifier is suitable for both binary and multiclass problems.  

\textbf{Random Forest} is an ensemble algorithm that consists of multiple decision trees. The prediction is given by majority voting/averaging the individual decision tree's prediction. Each tree is trained on different samples, which are drawn from the original training set with replacement (bootstrap). The randomness reduces potential overfitting in each decision tree, and often results in great enhancement in out-of-sample performance. 

\textbf{Logistic Regression} is the classification analog of linear regression, and is a special case of generalized linear models. In the case of binary class, logistic regression assigns $\{0,1\}$ to the two classes and models the prediction probability for the two classes $\{0,1\}$ as 
\bea
p(0|\vec{x})=y(\vec{x})=\frac 1{1+e^{-\vec{w}\cdot \vec{x}}}\,, \quad p(1|\vec{x})=1-p(0|\vec{x}),\label{lr}
\eea   
where $\vec{x}$ is the feature vector and $\vec{w}$ is the coefficient vector determined by maximizing the regularized log likelihood function:
\bea
\log P= \sum_{n=1}^N \{t_n\log y_n+(1-t_n)\log (1-y_n)\} - \lambda || w||_p\,,\label{lrerr}
\eea
where $\lambda$ is the regularization strength and $|| w||_p$ is the $L_p$ norm of $w$. The regularization strength is a hyperparameter that needs to be optimized by applying cross-validation. 

The function in (\ref{lr}) is only suited for binary classification. In the case of multiclass prediction, one can either modify (\ref{lr}) into multiple output functions, or apply generic multiclassification methods, such as "one-vs-rest" (OVR). When there are $K$ multiple labels ($K\geq3$), OVR is done by training $K$ binary classifiers, and the $i$-th classifier is to predict whether the output is label $i$ or not. We apply OVR for logistic regression in this paper.  

\textbf{Support Vector Machine} (SVM) is by nature a binary classifier. It is another generalization to linear models. By assigning $\{-1,1\}$ to the two classes, SVM makes prediction based on a generalized linear function
\bea
y=\pmb{w}\cdot \pmb \varphi(\vec{x}) + b \,,
\eea
where $\pmb w, b$ are determined by best dividing the two class samples in the feature space, and the prediction is $1/-1$ when $y$ is greater/smaller than 0. $\pmb \varphi$ is a transformation function on the original feature vectors, defined by a kernel function
\bea
k(\vec{x}, \vec{x}')=\pmb \varphi(\vec{x})\cdot \pmb \varphi(\vec{x}')\,.
\eea 
For all SVM classifiers considered in this section, we apply the rbf kernel defined as
\bea
k(\vec{x}, \vec{x}')=e^{-\gamma (\vec{x} - \vec{x}')^2}\,,
\eea
where $\gamma$ is a hyperparameter that can be optimized. As SVM is a binary classifier by design, we also apply OVR in our multiclassification problems.

\textbf{Feedforward Neural Network} (FNN) is one of the simplest neural network models. The graphic representation of a neural network is composed of a number of layers, including an input layer corresponding to the features and an output layer corresponding the labels. The rest is referred to as hidden layers. Each layer contains several neurons and different layers are connected by a certain topology. Mathematically, both neurons and connections between neurons correspond to variables of the prediction functions. A feedforward neural network with two hidden layers has the following function form
\bea
y_l(\pmb x, \pmb w)=O\left( \sum_k w^{(3)}_{lk} \cdot h\left(\sum_j w^{(2)}_{kj}\cdot h\left(\sum_i w^{(1)}_{ji} x_i +w^{(1)}_{j0} \right) +w^{(2)}_{k0}\right) +w^{(3)}_{l0} \right)\,,\label{fnn}
\eea
where $y_l$ is the prediction for the $l$-th label, $O$ is the output function, $h$ is the activation function, and $\pmb w = (
w^{(3)}, w^{(2)}, w^{(1)})$ are the coefficients determined by training. In addition to (\ref{fnn}), often it is better to include certain regularizations which can be non-parametric. Similar to decision tree, FNN can be designed in a multiclassification setting by choosing a proper output activation function.

\section{Machine learning algorithm comparison and selection}
\label{s:method}

In machine learning, a main question is what is the best algorithm for a specific problem. Since many ML algorithms are adaptive, this question in practice is often solved by empirically testing the performance of each algorithm and choosing the best one. In addition, the properties of a problem may call for a particular ML algorithm. For instance, neural network is usually preferred for image recognition tasks for its ability to handle large data sets in a parallelized fashion. In this section we take the same approach and compare the five ML algorithms introduced above on the data set $S_{\rm end}(\mb{F}_n)$ introduced in section~\ref{s:features}. Besides the prediction performance, we consider the ML algorithms' interpretability as another selection criterion. We show that out of the five algorithms, decision tree provides both high algorithm performance and good model interpretability, indicating that there is an inequality-based pattern in the data set.

\subsection{Class label imbalance and data resampling}
\label{s:resample}

The proportion of samples in each class of a data set are important to ML algorithms' training and evaluation. The disproportion of different classes, commonly referred to as class label imbalance, affects the classifier by making it biased towards the major classes over the minor ones. As a result, the regular classification measures are skewed to the major class label. For instance, a useless binary classifier that predicts only the major label can give $95\%$ accuracy if $95\%$ data is of the major label, albeit it does not provide any insight. 

Usually in the case of extremely unbalanced data sets, the minority classes are of higher interest. So it is important to analyze the imbalance before training a classifier and apply relevant techniques to deal with the imbalance. In our data sets, there tends to be a strong class label imbalance as gauge groups such as SU(3) have considerably smaller samples than others (yet they are of particular significance). For instance, the numbers of samples and class label imbalance of all gauge groups in the data set $S_{\rm end}(\mb{F}_n)$ are shown in table~\ref{imbal}. It is evident that the percentages of different gauge groups in the entire data set are extremely disproportional.
%\ref{imbal}

\begin{table}
\begin{center}
\begin{tabular}{|c |c |c|}
\hline
Gauge groups & number of samples & Fraction in the whole data set  \\
\hline
$\varnothing$ & 2053638 & 0.700337919752 \\
SU(2) & 520783 & 0.177599013488\\
$G_2$ & 286592 & 0.0977344814898\\
$F_4$ & 66934 & 0.0228260376564\\
$E_8$ & 4374 & 0.00149163487479\\ 
SU(3) & 24 & 8.1845534968e-06 \\
SO(8) & 8 & 2.72818449893e-06 \\
\hline
\end{tabular}
\end{center}
\caption[x]{\footnotesize  Class label imbalance in the data set $S_{\rm end}(\mb{F}_n)$.}\label{imbal}
\end{table}

\begin{table}
\begin{center}
\begin{tabular}{|c |c |c|}
\hline
Gauge groups & number of samples & Fraction in the whole data set  \\
\hline
SU(2) & 208313 & 0.145533490525\\
$\varnothing$ & 205364 & 0.143473233779 \\
$G_2$ & 200614 & 0.14015474631\\
$F_4$ & 200802 & 0.140286088551 \\
$E_8$ & 205578 & 0.143622740372 \\ 
SO(8) & 205360 & 0.143470439263 \\
SU(3) & 205344 & 0.1434592612 \\
\hline
\end{tabular}
\end{center}
\caption[x]{\footnotesize  Class label imbalance in the resampled data set $S'_{\rm end}(\mb{F}_n)$. Clearly resampling greatly reduces the imbalance.}\label{imbal1}
\end{table}

In order to train a classifer properly on an unbalanced data set, there are two common approaches: (1) resampling, which can be achieved by duplicating the minor class data and/or down-sampling the major class data; and (2) adding class weight to the training samples. One can define a class's weight as the normalized inverse of the sample's class percentage. When (2) is applied, it is important to note that the predictions will have a similar imbalance and using accuracy will still be biased. In this case, one can apply class label weighted accuracy instead of manipulating the data set. This measure is simply the orginal accuracy weighted by the class label's uniqueness on each sample. Intuitively it means that, when a correct/wrong prediction is done on a sample with a major label, the total weighted accuracy increases/decreases by a small amount; and a correct/wrong prediction is done on a sample with a minor label, the total weighted accuracy increases/decreases much more significantly. 

As shown in table~\ref{imbal}, the data set $S_{\rm end}(\mb{F}_n)$ is highly imbalanced. To overcome the imbalance and incorporate train-test split properly, we resample the data set by first duplicating samples in gauge group $SU(3), G_2, SO(8), F_4$, and $E_8$ to 10\% the size of the original data and then randomly drawing samples in gauge group $\varnothing$ and $SU(2)$ to 10\% the size of the original data. We call the resampled  data set $S'_{\rm end}(\mb{F}_n)$ (1431375 samples in total). The imbalance is completely resolved and is shown in table \ref{imbal1} for the resampled data set. To check the classifiers' scalability, we further down-sample the data set to 10\% and call this data set $S''_{\rm end}(\mb{F}_n)$ (143138 samples in total).   

\subsection{Prediction performance}
\label{s:performance}

We evaluate the performance of each classifier based on two measures: (1) weighted accuracy and (2) training time/efficiency. The core of most ML algorithms involves minimizing/maximizing an error/utility function, and this is typically done by numerical optimization methods. Difference in various algorithms' complexity results in different training time, and in some cases excessive complexity may lead to that a model not optimized, in addition to long training time.  

For each data set, we split it into non-overlapping training and test sets by randomly selecting (without replacement) $75\%$ original data as training and the rest as test. We train every ML model on the training set and use the trained model to make prediction on the test set. In the case of SVM, because the transformed feature space has a very high dimension, in general training more than 10000 samples is computationally unfeasible. This is solved by further down-sampling the training set to 10000 samples for SVM in both data sets. Nonetheless, the size of test set is the same for all classifiers to validate the performance comparability. The implementation and details for each classifer are listed below

\begin{itemize}
  \item LR: Scikit-learn \cite{sklearn}; hyperparameter $C$ optimized, regularization optimized between L1 and L2
  \item DT: Scikit-learn; untrimmed, no hyperparameter tuning
  \item RF: Scikit-learn; 10 trees, no hyperparameter tuning
  \item SVM: Scikit-learn; rbf kernel, hyperparameters $C, \gamma$ optimized
  \item FNN: Keras; 2 hidden layers (10, 10), output activation = softmax, dropout regularization added, epochs=5. The structure and hyperparameters have not been fully fine-tuned/optimized, so it can be expected that the accuracy may increase slightly upon further tuning. Yet given the apparent excessive runtime/complexity, we decide not to apply full tuning. 
\end{itemize}

The performance of all the classification algorithms described in Section~\ref{clfs} is presented in table~\ref{trainsetendFns1}. By comparing weighted accuracy, one finds that all non-linear algorithms (all but logistic regression) give considerably good results, and decision tree and random forest are the best performaners with only slight difference. The second criterion to consider is runtime, and the table shows that decision tree is much faster than other algorithms (which is clearly due to its simplicity)\footnote{The runtime presented for logistic regression and  SVM is based on the optimized hyperparameters. The optimization of hyperparameters involves multiple training/testing and the total runtime for both of them needs to be multiplied.}. Combining both performances, we conclude tentatively that decision tree is the best classification algorithm on our data set.

\begin{table}
\begin{center}
\begin{tabular}{|c |c |c|}
\hline
Classification method & Class-weighted Accuracy & RunTime (s) \\
\hline
Decision Tree &  0.995482019319 & 21.852001190185547\\
Feedforward Neural Network & 0.965769812798 &  11485 \\ 
Logistic Regression & 0.776428137819 & 324.3160767555237\\
Random Forest & 0.996106004368 & 52.732033252716064\\
Support Vector Machine & 0.973695351585 &  24.36295986175537\\
\hline
\end{tabular}
\end{center}
\caption[x]{\footnotesize  OOS weighted accuracy on resampled $S'_{\rm end}(\mb{F}_n)$.}\label{trainsetendFns1}

\end{table}

\begin{table}
\begin{center}
\begin{tabular}{|c |c |c|}
\hline
Classification method & Class-weighted Accuracy & RunTime (s) \\
\hline
Decision Tree &  0.991699490588 & 1.200284481048584\\
Feedforward Neural Network & 0.962707499358 &  279 \\ 
Logistic Regression & 0.773266180561 & 28.599193334579468\\
Random Forest & 0.993238403572 & 2.6778392791748047\\
Support Vector Machine & 0.972995952716 &  3.69943904876709\\
\hline
\end{tabular}
\end{center}
\caption[x]{\footnotesize  OOS weighted accuracy on re/down-sampled $S''_{\rm end}(\mb{F}_n)$.}\label{trainsetendFns2}

\end{table}

\subsection{Model interpretability}
\label{s:interpret}

Compared to traditional statistical models, one avantage of ML is its strong adaptability on data sets with complex patterns. However, this often leads to the fact that many ML algorithms lack interpretability, as one can hardly extract analytical results that explain the pattern in the data set. For instance, random forest can often provide high prediction performance, but its ensemble nature  makes it impractical to understand how the prediction is made based on simple rules. For our purposes, it is of particular interest to extract analytic understanding in addition to making prediction from the data sets. So we take interpretability as another algorithm selection criterion.

Among all the ML algorithms, decision tree is one of the most interpretable methods, given its simple algorithmic structure. Indeed, since the rule of each split is an inequality on a feature,  the decision function can be summarised as a collection of all the inequalities on the tree. For each input $\vec{x}$, the prediction rule is simply
\bea
\left\{L_i \le x_i \le U_i \right\},\, \text{for all $x_i$ involved in the decision}
\eea
where $L_i$ and $U_i$ are the lower/upper bound of all split inequalities involving $x_i$, i.e. the $i$-th feature of an input. Thus, by extracting the decision rules of certain-samples (e.g. geometries that have a SU(3) gauge group), we may gain insight about how features are related to the final prediction. Moreover, decision tree (and all the other tree-based algorithms) also has a built-in feature importance evaluation method. This is called Mean-Decreased-Impurity (MDI) importance measure. MDI computes the impurity-decrease (purity-increase in our langauage) weighted by number of samples on each node for every feature, and rank features' importance by their MDI values. This is helpful for interpreting how much contribution each feature provides to the whole prediction. 

With these considerations, we conclude that decision tree is the best ML algorithm to apply to our problem. In the rest of this paper, we will focus only on decision tree and we will apply the rule extraction and MDI feature importance on our data sets.  

\section{Detailed analysis of gauge group on divisors}
\label{s:divisors}

In this section, we present detailed results from the untrimmed decision tree and extract analytic rules for divisors with $h^{1,1}(D)=1,2,3$. For $h^{1,1}(D)>3$, we will only discuss the properties of the classifier and accuracies.

\subsection{$\mathbb{P}_2$}

For $h^{1,1}(D)=1$, the only possible topology of $D$ is $\mathbb{P}_2$. We can compress the vector $V(D)$ in section~\ref{s:features} into a 10D vector
\be
\tilde{V}(D)=(D^2 D_1, D_1^3, D_2^3, D_3^3,D_1^2 D_2, D_1 D_2^2,D_2^2 D_3,D_2 D_3^2, D_3^2 D_1, D_3 D_1^2),
\ee
whose components are denoted as $f_0,f_1,\dots,f_9$ in the following discussions. The information of the normal bundle $N_D=a H$ is explicitly given by $D^2 D_1=aH\cdot H=a$. 

As introduced in section~\ref{s:genbase}, we use a combination of $\mathbb{P}^2$ divisors on the end point bases and   intermediate bases as the initial data set $S(\mathbb{P}^2)$. 

In total, there are 113,219 samples with one of the following gauge groups: $\varnothing$, SU(2), SU(3), $G_2$, SO(8), $F_4$, $E_6$, $E_7$ and $E_8$. The total number of samples with each gauge group is listed in table~\ref{t:P2groups}.

\begin{table}
\begin{center}
\begin{tabular}{|c |c |c|c|c|c|c|c|c|}
\hline
\hline
 $\varnothing$ & SU(2) & SU(3) & $G_2$ & SO(8) & $F_4$ & $E_6$ & $E_7$ & $E_8$\\
\hline
81957 & 12807 & 940 & 10005 & 1403 & 4868 & 560 & 272 & 407\\
\hline
\end{tabular}
\end{center}
\caption[x]{\footnotesize  Total number of samples with each gauge group in the set $S(\mathbb{P}^2)$.}\label{t:P2groups}
\end{table}

The decision tree is trained on the set of samples $S'(\mathbb{P}^2)$ after up/down resampling, analogous to the resampling process in section~\ref{s:resample}. The number of samples in $S'(\mb{P}^2)$ with each label is balanced to $\sim 8200$. The (train set:test set) ratio is still (3:1). After the training, the IS and OOS accuracies are 0.912774 and 0.900694 respectively when tested on the resampled set. As another way to test the algorithm's predictability, we can use this decision tree on the data before resampling. When the decision tree is tested on the original set $S(\mathbb{P}^2)$ with 113,219 samples, the accuracy is $A=0.949111$. The maximal depth of the decision tree is $d_{max}=29$. 

\begin{figure}
\centering
\includegraphics[height=7cm]{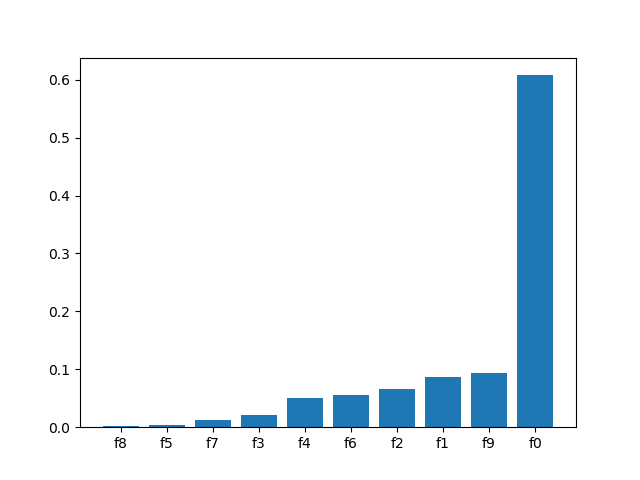}
\caption[E]{\footnotesize The feature importance of input vector elements $f_i$, for the $\mathbb{P}^2$ divisors on end point bases and intermediate bases in the resampled training set $S'(\mb{P}^2)$.}\label{f:featothP2_}
\end{figure}

We plot the feature importance of $f_i$ in figure~\ref{f:featothP2_}. The most important feature is $f_0=a$, which is expected since the normal bundle is the most direct information in the formula (\ref{normf}, \ref{normg}). 

The decision tree contains 2563 nodes and 1282 leaves. The structure is too complicated to be fully drawn. An efficient way to read out analytic conjectures from the decision tree is to sort the leaves $l$ according to the total number of samples $|S(l)|$ in $S(\mathbb{P}^2)$ which they apply to. We list a number of leaves with large $|S(l)|$ in table~\ref{t:rulesP2} (the tables are located at the end of this paper) . We only list the leaves that predict the existence of a non-Higgsable gauge group $G$ with probability higher than $80\%$ on a divisor $D$, which means that more than $80\%$ samples belong to the same gauge group based on the same rules.

\begin{table}
{\scriptsize
\resizebox{\linewidth} {\height}{%
    \setlength\tabcolsep{2pt}%
\begin{tabular}
{|c|c |c |c|c|c|c|c|c|c|c|c|c|}

\hline
\hline
$d$ & $|S(l)|$ & $f_0$ & $f_1$ & $f_2$ & $f_3$ & $f_4$ & $f_5$ & $f_6$ & $f_7$ & $f_8$ & $f_9$ & $G$\\
\hline
4 & 136 & $\leq -13$ & - & - & - & - & - & - & - & - & - & $E_8$\\
6 & 114 & $-11\sim -9$ & $\leq 2$ & $\geq 5$ & - & - & - & - & - & - & $\leq -1$ & $E_8$\\
10 & 163 & -9 & 3 & 5 & - & - & - & $\geq 0$ & - & - & $\geq 0$ & $57.1\% E_6$, $42.9\% E_7$\\
12 & 542 & -8 & $\geq 4$ & $\geq 5$ & - & $\geq 0$ & - & $\geq 0$ & - & - & $\geq 0$ & $95.6\% F_4$, $4.4\% E_6$\\
11 & 372 & -8 & 2 & $\geq 5$ & - & $\geq 0$ & - & $\geq 0$ & - & - & $\geq 0$ & $96.0\% F_4$, $4.0\% E_6$\\
12 & 287 & -8 & 3 & $\geq 5$ & - & $\geq 0$ & - & $\geq 0$ & - & - & $\geq 0$ & $95.1\% F_4$, $4.9\% E_6$\\ 
12 & 161 & -8 & 3 & 4 & - & $\geq 0$ & - & $\geq 0$ & - & - & $\geq 0$ & $94.4\% F_4$, $5.6\% E_6$\\
9 & 130 & -8 & $-5\sim -3$ & $\geq 3$ & - & $\geq -1$ & - & $\geq 0$ & - & - & - & $F_4$\\
6 & 1626 & -7 & - & - & - & $\geq 0$ & - & $\geq 0$ & - & - & $\geq 0$ & $F_4$\\
8 & 103 & -7 & $\geq -1$ & $\geq 4$ & - & $\geq 0$ & - & $\geq 0$ & - & - & $\leq -1$ & $F_4$\\
6 & 315 & -6 & - & - & - & $\leq -1$ & $\leq -1$ & - & - & - & - & $F_4$\\
11 & 280 & -6 & $\geq 2$ & $\geq 3$ & - & $\geq 0$ & - & $\geq 0$ & - & - & $\geq 0$ & $88.9\%$ SO(8), $10.1\% F_4$\\
6 & 234 & -6 & - & - & - & $\geq 0$ & - & - & - & - & $\leq -1$ & $F_4$\\
7 & 214 & -6 & - & - & - & $\geq 0$ & - & $\leq -1$ & - & - & $\geq 0$ & $F_4$\\
16 & 170 & -6 & 1 & $\geq 3$ & - & $\geq 0$ & - & $\geq 0$ & - & - & $\geq 0$ & $80.6\%$ SO(8), $19.4\% F_4$\\
7 & 2335 & -5 & - & - & - & $\geq 0$ & - & $\geq 0$ & - & - & $\geq 0$ & $G_2$\\
8 & 533 & -5 & $\geq 2$ & - & - & $\geq 0$ & - & $\geq 0$ & - & - & $\leq -1$ & $G_2$\\ 
8 & 508 & -5 & $\geq 2$ & - & - & $\leq -1$ & - & $\geq 0$ & - & - & $\geq 0$ & $G_2$\\
9 & 298 & -5 & $\leq 1$ & $\geq 3$ & - & $\leq -1$ & - & $\geq 0$ & - & - & $\geq 0$ & $G_2$\\
13 & 205 & -5 & 1 & 2 & - & $\leq -1$ & - & $\geq 0$ & - & - & $\geq 0$ & $87.3\% G_2$, $10.2\%$ SO(8), $2.5\% F_4$\\
8 & 172 & -5 & - & $\geq 3$ & $\geq 3$ & - & $\leq -2$ & $\leq -1$ & - & - & - & $G_2$\\
13 & 143 & -5 & $\geq 1$ & 2 & $\leq 9$ & $\geq 0$ & - & $\leq -1$ & - & - & $\geq 0$ & $88.8\% G_2$, $10.5\%$ SO(8), $0.7\% F_4$\\
9 & 127 & -5 & $\leq 1$ & - & $\geq 7$ & $\geq 0$ & - & $\geq 0$ & - & - & $\leq -1$ & $G_2$\\
14 & 116 & -5 & 0 & 2 & - & $\leq -1$ & - & $\geq 0$ & - & - & $\geq 0$ & $81.0\% G_2$, $18.1\%$ SO(8), $0.9\% F_4$\\ 
9 & 115 & -5 & - & $\geq 3$ & $\geq 9$ & - & $\geq -1$ & $\leq -1$ & - & - & - & $G_2$\\
11 & 105 & -5 & -1 & - & $\leq 6$ & $\geq 0$ & - & $\geq 0$ & - & - & $\leq -1$ & $G_2$\\
15 & 100 & -5 & 1 & $\geq 2 $ & $3\sim 6$ & $\geq 0$ & - & $\geq 0$ & - & - & $\leq -1$ & $82\% G_2$, $14\%$ SO(8), $4\% F_4$\\
6 & 980 & -4 & - & $\geq 3$ & - & - & - & - & - & - & $\leq -1$ & $G_2$\\
12 & 922 & -4 & $1\sim 7$ & $\geq 6$ & - & - & - & $\geq 0$ & $\geq -1$ & - & $\geq 0$ & $99.9\%$ SU(2), $0.1\% G_2$\\
11 & 643 & -4 & $\geq -7$ & $1\sim 2$ & - & $\geq 0$ & - & $\geq 0$ & $\geq -1$ & - & $\geq 0$ & SU(2)\\
13 & 631 & -4 & $-4\sim 5$ & $1\sim 2$ & $\geq 11$ & $\geq 0$ & $\geq -1$ & - & $\leq -2$ & - & - & SU(2)\\
8 & 475 & -4 & $\geq 2$ & $\leq 2$ & - & - & - & $\leq -1$ & $\geq -1$ &  - & - & $G_2$\\
13 & 346 & -4 & 0 & $\geq 6$ & - & $\geq 0$ & - & $\leq -1$ & $\geq -1$ & - & $\geq 0$ & $99.4\%$ SU(2), $0.6\% G_2$\\ 
11 & 301 & -4 & $\leq 1$ & $\leq 2$ & - & $\leq -1$ & - & $\leq -1$ & -1 & - & $\leq -1$ & $G_2$\\
7 & 285 & -4 & - & $\leq 2$ & - & $\leq -1$ & - & - & $\leq -2$ & - & - & $G_2$\\
11 & 214 & -4 & 0 & $\geq 6$ & - & - & - & - & $\leq -1$ & - & $\geq 0$ & $99\%$ SU(2), $1\% G_2$\\
11 & 209 & -4 & $\geq 5$ & $3\sim 5$ & - & - & - & $\geq 0$ & - & - & $\geq 0$ & SU(2)\\
10 & 160 & -4 & $0\sim 1$ & 4 & $\geq 7$ & - & - & - & - & - & $\geq 0$ & $99.4\%$ SU(2), $0.6\% G_2$\\
13 & 147 & -4 & 0 & $\geq 7$ & - & - & - & $\geq 0$ & $\geq -1$ & - & $\geq 0$ & SU(2)\\
13 & 137 & -4 & $-12\sim -1$ & $4\sim 5$ & $\leq 6$ & $\geq 0$ & - & $\geq -1$ & - & - & $\geq -1$ & $99.3\%$ SU(2), $0.7\% G_2$\\
20 & 133 & -4 & 0 & 1 & $\geq 2$ & $\leq -1$ & - & $\geq 0$ & $\geq -1$ & $\geq -1$ & $\leq -1$ & $72.2\% G_2$, $27.8\%$ SU(3)\\
15 & 124 & -4 & 6 & $1\sim 2$ & $\geq 11$ & $\geq 0$ & $\geq -1$ & - & $\leq -2$ & $\leq -2$ & - & SU(2)\\
14 & 118 & -4 & $\geq 7$ & $1\sim 2$ & $\geq 11$ & $\geq 0$ & $\geq -1$ & - & $\leq -2$ & - & - & SU(2)\\
9 & 105 & -4 & $\leq -1$ & $\geq 3$ & $\geq 6$ & - & $\geq -1$ & - & - & - & $\geq 0$ & $G_2$\\
10 & 617 & -3 & $\geq -1$ & $\leq 3$ & $3\sim 7$ & - & - & - & - & - & - & $97.4\%$ SU(2), $2.6\%\varnothing$\\
7 & 441 & -3 & - & $\leq 1$ & $\geq 2$ & - & - & - & - & - & $\leq -2$ & $99.8\%$ SU(2), $0.2\% G_2$\\
9 & 421 & -3 & - & $\geq 2$ & $\geq 8$ & - & $\leq -2$ & - & - & - & - &  $99\%$ SU(2), $1\% \varnothing$\\
12 & 390 & -3 & $\leq -4$ & $\geq 5$ & $3\sim 7$ & - & - & - & - & - & - & $97.7\%$ SU(2), $2.3\%\varnothing$\\
17 & 244 & -3 & $-4\sim 5$ & $\leq -2$ & 1 & $\leq -1$ & $\leq -2$ & $\geq -1$ & $\leq -1$ & - & $\geq -1$ & SU(2)\\
13 & 155 & -3 & $-12\sim 5$ & - & $\leq 0$ & - & $\geq -1$ & $\geq 0$ & - & - & $\geq -1$ & $99.4\%$ SU(2), $0.6\%\varnothing$\\
19 & 143 & -3 & -1 & 1 & 1 & $\leq -1$ & $\geq -1$ & $\geq -1$ & - & - & -1 & $93.7\%$ SU(2), $6.3\%\varnothing$\\  
10 & 128 & -3 & $\geq -3$ & $\geq 2$ & $\geq 8$ & - & $\geq -1$ & - & - & - & - & SU(2)\\
14 & 120 & -3 & $\leq 1$ & 0 & 1 & $\leq -2$ & - & $\geq -1$ & - & $\geq -1$ & $\geq -1$ & SU(2)\\
9 & 120 & -3 & $\leq 4$ & - & $\leq 1$ & - & - & $\geq 0$ & - & - & $\leq -2$ & SU(2)\\
10 & 118 & -3 & $\geq 0$ & $\leq 1$ & $\geq 2$ & $\leq -2$ & $\leq -1$ & - & - & - & $\geq -1$ & SU(2)\\
20 & 114 & -3 & -2 & 1 & 1 & $\leq -1$ & $\geq -1$ & $\geq 0$ & - & - & -1 & $80.7\%$ SU(2), $19.3\%\varnothing$\\ 
\hline
\end{tabular}
}}
\caption[x]{\footnotesize  The inequalities that predict the appearance of certain gauge group $G$ on a $\mathbb{P}^2$ divisor. $d$ is the depth of the leave in the tree. $|S(l)|$ denotes the number of samples in $S(\mathbb{P}^2)$ on which this rule will apply. The rules are sorted according to the normal bundle $N_D=f_0 H$ and the $|S(l)|$.}\label{t:rulesP2}
\end{table}

We can see that the analytic rules in table~\ref{t:rulesP2} mostly predict the common gauge groups SU(2), $G_2$ and $F_4$, except from the following five rules:

(1) If $f_0=a\leq -13$, the gauge group is $E_8$.

(2) If $f_0=a\in \{-9,-10,-11\}$, $f_1\leq 2$, $f_2\geq 5$, $f_9\leq -1$, the gauge group is $E_8$.

(3) If $f_0=a=-9$, $f_1=3$, $f_2=5$, $f_6\geq 0$ and $f_9\geq 0$, the gauge group is $E_6$ with a probability of $57.8\%$,   and $E_7$ with a probability of $42.2$.

(4) If $f_0=a=-6$, $f_1\geq 2$, $f_2\geq 3$, $f_4\geq 0$, $f_6\geq 0$, $f_9\geq 0$, then the gauge group is SO(8) with a probability of $89\%$, and $F_4$ with a probability of $11\%$.

(5) If $f_0=a=-6$, $f_1=1$, $f_2\geq 3$, $f_4\geq 0$, $f_6\geq 0$, $f_9\geq 0$, then the gauge group is SO(8) with a probability of $80.4\%$, and $F_4$ with a probability of $19.6\%$.

Another way to select informative leaves is listing the leaves with small depth, as in table~\ref{t:rulesdepthP2}. The reason is that the rules with small depth are generally simpler. However, there are leaves with small depth that apply to only few samples in $S(\mb{P}^2)$ or has low predictability, such as the rule giving 57\% SU(2) and 43\% $\varnothing$ in table~\ref{t:rulesdepthP2}. Hence we can not state that the shallow leaves give the best set of rules.

\begin{table}
\centering
{\scriptsize
    \setlength\tabcolsep{2pt}%
\begin{tabular}
{|c|c |c |c|c|c|c|c|c|c|c|c|c|}

\hline
\hline
$d$ & $|S(l)|$ & $f_0$ & $f_1$ & $f_2$ & $f_3$ & $f_4$ & $f_5$ & $f_6$ & $f_7$ & $f_8$ & $f_9$ & $G$\\
\hline
3 & 22 & $\geq -2$ & - & - & - & - & $\geq 6$ & - & - & - & - & $90.9\%$ SU(2), $8.1\%\varnothing$\\
4 & 136 & $\leq -13$ & - & - & - & - & - & - & - & - & - & $E_8$\\
4 & 17 & -8 & - & - & - & - & - & $\leq -1$ & - & - & - & $E_7$\\
5 & 14 & -7 & - & - & - & $\leq -1$ & - & $\leq -1$ & - & - & - & $E_6$\\
5 & 15 & $-8\sim -7$ & - & - & $\geq 6$ & $\leq -1$ & - & $\geq 0$ & - & - & - & $E_7$\\
6 & 1832 & $\geq -2$ & - & - & - & - & $\leq 5$ & $\geq -1$ & - & $\leq 0$ & $\leq -3$ & $\varnothing$\\ 
6 & 1626 & -7 & - & - & - & $\geq 0$ & - & $\geq 0$ & - & - & $\geq 0$ & $F_4$\\
6 & 980 & -4 & - & $\geq 3$ & - & - & - & - & - & - & $\leq -1$ & $G_2$\\
6 & 315 & -6 & - & - & - & $\leq -1$ & $\leq -1$ & - & - & - & - & $F_4$\\ 
6 & 234 & -6 & - & - & - & $\geq 0$ & - & - & - & - & $\leq -1$ & $F_4$\\
6 & 114 & $-11\sim -9$ & $\leq 2$ & $\geq 5$ & - & - & - & - & - & - & $\leq -1$ & $E_8$\\
6 & 99 & $-8\sim -7$ & - & $\leq 4$ & - & $\leq -1$ & - & $\geq 0$ & - & - & $\geq 0$ & $F_4$\\
6 & 81 & $-12$ & $\geq -2$ & $\leq 7$ & - & - & - & - & - & - & - & $E_8$\\
6 & 19 & -9 & - & - & - & - & - & $\leq -1$ & - & - & $\geq 0$ & $E_7$\\
6 & 9 & $-11\sim -10$ & - & - & - & - & - & $\leq -1$ & - & - & $\geq 0$ & $E_8$\\
6 & 4 & -7 & - & - & - & $\geq 0$ & - & $\leq -1$ & - & - & $\leq -1$ & $E_6$\\
6 & 4 & $-11\sim -9$ & - & $\leq 4$ & $\leq 7$ & - & - & - & - & - & $\leq -1$ & $E_7$\\
6 & 3 & $-8\sim -7$ & - & - & $\geq 6$ & $\leq -1$ & - & $\geq 0$ & - & - & $\leq -1$ & $E_6$\\
6 & 2 & -6 & - & - & - & $\leq -1$ & $\geq 0$ & - & - & - & - & SO(8)\\
6 & 1 & $-11\sim -9$ & - & $\leq 4$ & $\geq 8$ & - & - & - & - & - & $\leq -1$ & $E_8$\\ 
7 & 55878 & $\geq -2$ & - & - & - & $\geq -2$ & $-2\sim 5$ & - & - & $\geq -2$ & $\geq -2$ & $99.996\%\varnothing$, $0.004\%$ SU(2)\\
7 & 4183 & $\geq -2$ & - & - & - & $\geq 0$ & $\leq -3$ & - & - & $\geq -2$ & $\geq -2$ & $\varnothing$\\ 
7 & 2335 & -5 & - & - & - & $\geq 0$ & - & $\geq 0$ & - & - & $\geq 0$ & $G_2$\\
7 & 2300 & $\geq -2$ & - & $\leq 1$ & - & - & $\leq 5$ & - & $\leq -2$ & $\leq -3$ & $\geq -2$ & $99.96\%\varnothing$, $0.04\%$ SU(2)\\
7 & 441 & -3 & - & $\leq 1$ & $\geq 2$ & - & - & - & - & - & $\leq -2$ & $99.8\%$ SU(2), $0.2\% G_2$\\
7 & 285 & -4 & - & $\leq 2$ & - & $\leq -1$ & - & - & $\leq -2$ & - & - & $G_2$\\
7 & 214 & -6 & - & - & - & $\geq 0$ & - & $\leq -1$ & - & - & $\geq 0$ & $F_4$\\
7 & 28 & $\geq -2$ & - & - & - & - & $\leq 5$ & $\geq -1$ & $\geq -1$ & $\geq 1$ & $\leq -3$ & $\varnothing$\\
7 & 28 & $\geq -2$ & - & - & - & $\leq -2$ & $\leq -1$ & $\leq -2$ & - & - & $\leq -3$ & $\varnothing$\\
7 & 28 & $\geq -2$ & - & - & - & - & $\leq 5$ & $\geq -1$ & $\geq -2$ & $\geq 1$ & $\leq -3$ & $96.4\%\varnothing$, $3.6\%$ SU(2)\\
7 & 192 & $\geq -2$ & $\leq -1$ & $\geq 2$ & - & - & $\leq 5$ & - & - & $\leq -3$ & $\geq -2$ & $\varnothing$\\
7 & 21 & $\geq -2$ & - & - & - & - & $0\sim 5$ & $\leq -2$ & - & $\leq -1$ & $\leq -3$ & $57\%$ SU(2), $43\%\varnothing$\\
7 & 13 & -7 & $\leq 1$ & - & - & $\geq 0$ & - & $\geq 0$ & - & - & $\geq 0$ & $F_4$\\
7 & 7 & $\geq -2$ & - & - & - & - & $0\sim 5$ & $\leq -2$ & - & $\geq 0$ & $\leq -3$ & $\varnothing$\\ 
7 & 4 & -12 & $\leq -2$ & $5\sim 7$ & - & - & - & - & $\leq -2$ & - & - &$75\% E_8$, $25\% E_7$\\
7 & 4 & -12 & $\geq 5$ & $8$ & - & - & - & - & - & - & - & $E_7$\\
7 & 3 & -12 & $\leq 4$ & $\geq 9$ & - & - & - & - & - & - & - & $E_8$\\
7 & 3 & -12 & $\leq -2$ & $\leq 4$ & - & - & - & - & - & - & - & $E_8$\\
7 & 2 & $-11\sim -9$ & $\geq 3$ & $\geq 5$ & $\geq 8$ & - & - & - & - & - & $\leq -1$ & $E_8$\\
7 & 2 & -8 & $\geq 1$ & - & - & $\geq 0$ & - & $\geq 0$ & - & - & $\leq -1$ & $E_7$\\
7 & 2 & -5 & - & $\geq 3$ & - & - & - & $\leq -1$ & - & - & $\leq -1$ & $F_4$\\
7 & 1 & $-11\sim -9$ & $\geq 3$ & $\geq 5$ & $\leq 7$ & - & - & - & - & - & $\leq -1$ & $E_7$\\
\hline
\end{tabular}
}
\caption[x]{\footnotesize  Leaves in the decision tree with depth $d\leq 7$, applied to $\mathbb{P}^2$ divisors.}\label{t:rulesdepthP2}
\end{table}

Because the resampled training set in $S'(\mb{P}^2)$ has balanced labels, we have derived a large number of rules predicting rarer gauge groups such as SU(3), SO(8), $E_6$, $E_7$ and $E_8$. We list a number of these rules in table~\ref{t:rulesrareP2}.

\begin{table}
\centering
{\scriptsize
    \setlength\tabcolsep{2pt}%
\begin{tabular}{|c|c |c |c|c|c|c|c|c|c|c|c|c|}
\hline
\hline
$d$ & $|S(l)|$ & $f_0$ & $f_1$ & $f_2$ & $f_3$ & $f_4$ & $f_5$ & $f_6$ & $f_7$ & $f_8$ & $f_9$ & $G$\\
\hline
14 & 82 & -4 & -12 & -1$\sim$1 & - & $\leq -1$ & - & $\geq 0$ & $\geq -1$ & $\geq -1$ & $\leq -1$ & $91.5\%$SU(3), $8.5\% G_2$\\
18 & 88 & -4 & -1 & 1 & $2\sim 5$ & $\geq 0$ & $\geq -1$ & $\leq -1$ & $\geq -1$ & - & - & $76.1\%$ SU(3), $23.9\% G_2$\\
15 & 33 & -4 & -9 & $1\sim 2$ & - & $\leq -1$ & - & $\geq 0$ & $\geq -1$ & $\geq -1$ & $\leq -1$ & $81.8\% SU(3)$, $18.2\% G_2$\\
17 & 17 & -4 & $-9\sim -2$ & 1 & - & $\geq 0$ & - & -1 & $\geq -1$ & - & $\geq 0$ & $94.1\%$ SU(3), $5.9\%$ SU(2)\\
15 & 14 & -4 & -14 & 0$\sim$ 2 & - & $\leq -1$ & - & $\geq 0$ & $\geq -1$ & $\geq -1$ & $\geq 0$ & SU(3)\\
15 & 14 & -4 & $-5\sim 5$ & 2 & - & -1 & - & $\geq 0$ & $\geq -1$ & $\geq -1$ & $\geq 0$ & SU(3)\\
11 & 280 & -6 & $\geq 2$ & $\geq 3$ & - & $\geq 0$ & - & $\geq 0$ & - & - & $\geq 0$ & $88.9\%$ SO(8), $11.1\% F_4$\\
16 & 170 & -6 & 1 & $\geq 3$ & - & $\geq 0$ & - & $\geq 0$ & - & - & $\geq 0$ & $80.6\%$ SO(8), $19.4\% F_4$\\
16 & 97 & -6 & 0 & $\geq 3$ & - & $\geq 0$ & - & $\geq 0$ & - & - & $\geq 0$ & $89.7\%$ SO(8), $10.3\% F_4$\\
17 & 85 & -6 & 0 & 2 & - & $\geq 0$ & - & $\geq 0$ & - & - & $\geq 0$ & $71.8\%$ SO(8), $28.2\% F_4$\\
17 & 75 & -6 & 1 & 2 & - & $\geq 0$ & - & $\geq 0$ & - & - & $\geq 0$ & $81.3\%$ SO(8), $18.7\% F_4$\\
15 & 48 & -6 & $0\sim 1$ & $-2\sim 0$ & - & $\geq 0$ & - & $\geq 0$ & - & - & $\geq 0$ & $93.8\%$ SO(8), $6.2\% F_4$\\
15 & 42 & -6 & $-1$ & $\geq 3$ & - & $\geq 0$ & - & $\geq 0$ & - & - & $\geq 0$ & $73.8\%$ SO(8), $26.2\% F_4$\\
11 & 34 & -6 & $\geq 2$ & $-6\sim 2$ & - & $\geq 0$ & - & $\geq 0$ & - & - & $\geq 0$ & $93.8\%$ SO(8), $6.2\% F_4$\\
11 & 26 & -5 & $-1$ & - & - & $\leq -1$ & - & $\geq 0$ & - & - & $\leq -1$ & SO(8)\\
13 & 24 & -6 & -2 & $\geq 3$ & - & $\geq 0$ & - & $\geq 0$ & - & - & $\geq 0$ & SO(8)\\
10 & 163 & -9 & 3 & 5 & - & - & - & $\geq 0$ & - & - & $\geq 0$ & $57.0\% E_6$, $43.0\% E_7$\\
9 & 91 & -9 & $-1\sim 2$ & $\leq 5$ & - & - & - & $\geq 0$ & - & - & $\geq 0$ & $E_6$\\
9 & 64 & -9 & 2 & $\geq 6$ & - & - & - & $\geq 0$ & - & - & $\geq 0$ & $93.8\% E_6$, $6.2\% E_7$\\
11 & 52 & -9 & 4 & $\leq 5$ & - & - & - & $\geq 0$ & - & - & $\geq 0$ & $98.0\% E_6$, $2.0\% E_7$\\
11 & 46 & -9 & $\leq 1$ & $\geq 6$ & - & - & - & $\geq 0$ & - & - & $\geq 0$ & $80.4\% E_6$, $15.2\% E_7$, $4.3\% E_8$\\
11 & 21 & -9 & 4 & $\geq 6$ & - & $\geq 0$ & - & $\geq 0$ & - & - & $\geq 0$ & $E_6$\\
10 & 26 & -9 & $\geq 5$ & - & - & $\geq 0$ & - & $\geq 0$ & - & - & $\geq 0$ & $76.9\% E_6$, $23.1\% E_7$\\ 
9 & 17 & -9 & 3 & $\geq 6$ & - & - & - & $\geq 0$ & - & - & $\geq 0$ & $E_6$\\ 
8 & 32 & $-11\sim -10$ & - & $\leq 5$ & $\leq 8$ & - & - & $\geq 0$ & - & - & $\geq 0$ & $E_7$\\ 
13 & 29 & $-11\sim -10$ & 2 & $5\sim 6$ & $\geq 9$ & - & - & $\geq 0$ & - & - & $\geq 0$ & $72.4\% E_7$, $27.6\% E_8$\\ 
6 & 19 & -9 & - & - & - & - & - & $\leq -1$ & - & - & $\geq 0$ & $E_7$\\
4 & 17 & -8 & - & - & - & - & - & $\leq -1$ & - & - & - & $E_7$\\
5 & 15 & $-8\sim -7$ & - & - & $\geq 6$ & $\leq -1$ & - & $\geq 0$ & - & - & - & $E_7$\\
10 & 13 & $-10\sim -9$ & $\geq 6$ & $\geq 6$ & $\leq 8$ & - & - & $\geq 0$ & - & - & $\geq 0$ & $E_7$\\
4 & 136 & $\leq -13$ & - & - & - & - & - & - & - & - & - & $E_8$\\
6 & 114 & $-11\sim -9$ & $\leq 2$ & $\geq 5$ & - & - & - & - & - & - & $\leq -1$ & $E_8$\\
6 & 81 & $-12$ & $\geq -2$ & $\leq 7$ & - & - & - & - & - & - & - & $E_8$\\
\hline
\end{tabular}
}
\caption[x]{\footnotesize  The inequalities that predict the appearance of rarer gauge groups $G=$SU(3), SO(8), $E_6$, $E_7$ and $E_8$ on a $\mathbb{P}^2$ divisor. $|S(l)|$ denotes the number of samples among the 113,219 total samples which this rule will apply. The rules are sorted according to the gauge group $G$.}\label{t:rulesrareP2}
\end{table}

It is hard to check these rules analytically with (\ref{normf}, \ref{normg}), since the only directly relevant feature is the normal bundle coefficient $a\equiv f_0$. We summarize the possible gauge groups for different normal bundle $N_D=aH$ in table~\ref{t:normP2} using the data set $S(\mb{P}^2)$.

\begin{table}
\begin{center}
\begin{tabular}{|c |c |}
\hline
\hline
$a$ & possible gauge groups\\
\hline
$>-2$ & $\varnothing$\\
$-2$ & $\varnothing$, SU(2)\\
$-3$ & $\varnothing$, SU(2), SU(3), $G_2$\\
$-4$ & SU(2),SU(3),$G_2$, $SO(8)$, $F_4$\\
$-5$ & $G_2$, SO(8), $F_4$\\
$-6$ & SO(8), $F_4$\\
$-7$ & $F_4$, $E_6$\\
$-8$ & $F_4$, $E_6$, $E_7$\\
$-9$ & $E_6$, $E_7$, $E_8$\\
$-10$ & $E_7$, $E_8$\\
$-11$ & $E_7$, $E_8$\\
$-12$ & $E_7$, $E_8$\\
$<-12$ & $E_8$\\
\hline
\end{tabular}
\end{center}
\caption[x]{\footnotesize  Possible gauge groups on a $\mathbb{P}^2$ divisor with the given normal bundle $N_D=aH$.}\label{t:normP2}

\end{table}

For $D=\mathbb{P}^2$, $-K_D=3H$ and $N_D=aH$, hence the formula (\ref{normf}, \ref{normg}) becomes
\be
f_{k,D}\in\mc{O}[(12+(4-k)a-\sum_{D_i\bigcap D\neq\varnothing}\phi_i) H],\label{normfP2}
\ee
\be
g_{k,D}\in\mc{O}[(18+(6-k)a-\sum_{D_i\bigcap D\neq\varnothing}\gamma_i) H],\label{normgP2}
\ee
where $\phi_i$ and $\gamma_i$ are the order of vanishing of $f$ and $g$ on the divisors $D_i$ which intersect $D$.

If there is an $E_8$ on the divisor $D$, then $f_3$ and $g_4$ has to vanish, or equivalently we have
\be
12+a-\sum_{D_i\bigcap D\neq\varnothing}\phi_i<0\ ,\ 18+2a-\sum_{D_i\bigcap D\neq\varnothing}\gamma_i<0.\label{ineqP2E8}
\ee
Since $\phi_i,\gamma_i\geq 0$, if $a\leq -13$, then the inequalities (\ref{ineqP2E8}) are always satisfied. This condition exactly corresponds to the first rule in table~\ref{t:rulesP2}. If $-10\geq a\geq -12$, then the second inequality in (\ref{ineqP2E8}) is automatically satisfied, but the first inequality may not be satisfied. When $12+a-\sum_{D_i\bigcap D\neq\varnothing}\phi_i\geq 0$, the gauge group is expected to be $E_7$ instead of $E_8$ since $f_3$ is non-vanishing now. However, as we mentioned after (\ref{normg}), there may be non-local effects from other non-neighboring divisors which increases the order of vanishing. For example, we see from table~\ref{t:normP2} that the gauge group can be $E_8$ even if $a=-9$. However, the second inquality in (\ref{ineqP2E8}) cannot be satisfied since there cannot be another neighboring toric divisor $D_i$ with ord$_{D_i}(g)=1$. Otherwise, this will lead to a toric (4,6) curve which is not allowed in the set $S(\mb{F}_n)$ generated from good bases exclusively\footnote{$g$ cannot vanish on a non-toric divisor on $B$ since we are consider the generic fibration and the effective cone of $B$ is generated by the toric divisors.}. Thus we have found cases where the formula (\ref{normf}, \ref{normg}) cannot give us the correct non-Higgsable gauge group.

If we want an $F_4$ or $E_6$ gauge group on $D$, then $f_{2,D}$ and $g_{3,D}$ has to vanish but $g_{4,D}$ should not vanish. We have inequalities
\be
12+2a-\sum_{D_i\bigcap D\neq\varnothing}\phi_i<0\ ,\ 18+3a-\sum_{D_i\bigcap D\neq\varnothing}\gamma_i<0\ ,\ 18+2a-\sum_{D_i\bigcap D\neq\varnothing}\gamma_i\geq 0.\label{ineqP2F4}
\ee
Now the gauge group is $E_6$ if and only if $g_{4,D}$ is a locally complete square. In the case of generic fibration, this means $g_4$ is a single monomial which takes the form of a complete square locally. If $a=-9$ and the inequalities in (\ref{ineqP2F4}) are satisfied, then we can see that $g_{4,D}\in\mc{O}(0)$ and $\gamma_i=0$ for any $D_i$ intersects $D$. This exactly corresponds to the criterion of an $E_6$ gauge group as $g_{4,D}$ is a complex number in this case. Hence $F_4$ can only appear if $a\geq -8$, which is consistent with our observation in table~\ref{t:normP2}.

For the case of SO(8), the situation is similar. If we want an SO(8) or $G_2$ gauge group on $D$, $f_{1,D}$ and $g_{2,D}$ have to vanish but $g_{3,D}$ should not, we have
\be
12+3a-\sum_{D_i\bigcap D\neq\varnothing}\phi_i<0\ ,\ 18+4a-\sum_{D_i\bigcap D\neq\varnothing}\gamma_i<0\ ,\ 18+3a-\sum_{D_i\bigcap D\neq\varnothing}\gamma_i\geq 0.\label{ineqP2G2}
\ee
If $a=-6$, then the third inequality in (\ref{ineqP2G2}) means that $g_{3,D}$ is locally a complex number. Similarly, since
\be
f_2\in\mc{O}(12+2a-\sum_{D_i\bigcap D\neq\varnothing}\phi_i),
\ee
$f_2$ is also either locally a complex number or vanishes. Then the gauge group should be SO(8) if $a=-6$ and the conditions (\ref{ineqP2G2}) are satisfied. So we expect that $G_2$ can only appear if $a\geq -5$, which is consistent with table~\ref{t:normP2}.

\subsection{$\mathbb{F}_n$}
\label{s:Fn}

We apply the same method in section~\ref{s:features} to the data set $S(\mathbb{F}_n)$, which is a combination of $\mathbb{F}_n$ divisors on end point bases and intermediate bases. There are in total 6,300,170 samples in this set, and the total number of each gauge group is listed in table~\ref{t:Fngroups}.

\begin{table}
\begin{center}
\begin{tabular}{|c |c |c|c|c|c|c|c|c|c|}
\hline
\hline
 $\varnothing$ & SU(2) & SU(3) & $G_2$ & SO(7) & SO(8) & $F_4$ & $E_6$ & $E_7$ & $E_8$\\
\hline
5025302 & 696737 & 25816 & 411274 & 4 & 13583 & 113609 & 3670 & 2415 & 7760\\
\hline
\end{tabular}
\end{center}
\caption[x]{\footnotesize  Total number of samples with each gauge group in the set $S(\mathbb{F}_n)$.}\label{t:Fngroups}
\end{table}

We generate the resampled data set $S'(\mb{F}_n)$ similar to the procedure in section~\ref{s:resample}. After training the descision tree with 75\% of the data in $S'(\mb{F}_n)$, the IS and OOS accuracies are 0.982169 and 0.977909 respectively on the set $S'(\mb{F}_n)$. When the decision tree is tested on the original set $S(\mathbb{F}_n)$ with 6,300,170 samples, the accuracy is $A=0.978592$.

This decision tree has 66441 nodes and 33221 leaves, which is much larger than the decision tree for $\mathbb{P}^2$. This is due to the large total number of samples in the training set. The maximal depth of the decision tree is $d_{max}=49$.

\begin{figure}
\centering
\includegraphics[height=7cm]{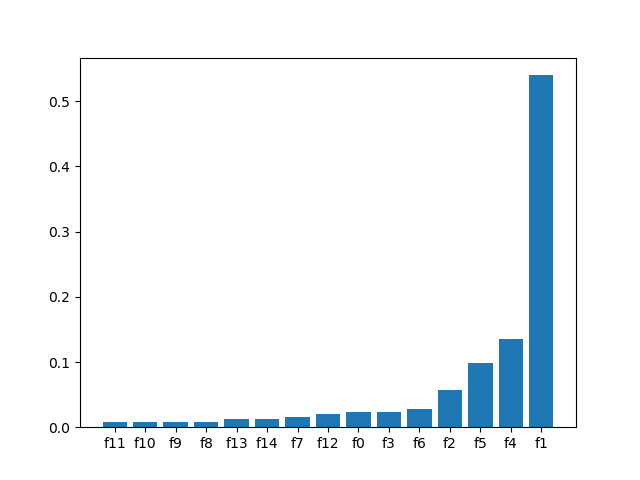}
\caption[E]{\footnotesize The feature importance of each input vector elements $f_i$, for the $\mathbb{F}_n$ divisors on end point bases and intermediate bases.}\label{f:featFn}
\end{figure}

We plot the feature importance of $f_i$ in figure~\ref{f:featFn}. The most important feature is $f_1=a$, the coefficient of $S$ in the normal bundle expression $N_D=aS+bF$. The next most important features are $f_4=D_2^3$, $f_5=D_3^3$, $f_2=b$, $f_3=D_1^3$, $f_6=D_4^3$, $f_0=n$ and $f_{13}=D_4^2 D_1$. It is interesting that the feature $f_0\equiv n$, which specifies the topology of $D$, has low importance. On the other hand, the canonical divisor of $\mb{F}_n$ is $-K(\mb{F}_n)=2S+(n+2)F$, which depends crucially on $n$. This counter-intuitive result may imply that the non-Higgsable gauge group on a divisor is not highly sensitive to its topology. 

\begin{table}
\centering
{\tiny

    \setlength\tabcolsep{2pt}%
\begin{tabular}{|c |c |c|c|c|c|c|c|c|c|c|c|}
\hline
\hline
$d$ & $|S(l)|$ & $f_0$ & $f_1$ & $f_2$ & $f_3$ & $f_4$ & $f_5$ & $f_6$ & $f_{13}$ & other $f_i$ & $G$\\
\hline
2 & 6831 & - & $\leq -9$ & - & - & - & - & - & - & - & $E_8$\\
7 & 4698 & - & -6 & - & - & - & $\leq 0$ & - & - & $f_7\geq 0$ & $F_4$\\
8 & 52528 & - & -5 & - & - & - & $\leq 3$ & $\leq 5$ & $\leq -1$ & $f_7\geq 0$, $f_{12}\leq -1$ & $F_4$\\
10 & 4478 & - & -5 & $-7\sim -4$ & $\geq 2$ & - & $\geq 4$ & $\leq 3$ & $\leq -1$ & - & $F_4$\\
9 & 4285 & 0 & -5 & - & - & - & $\leq 2$ & $\geq 6$ & $\leq -1$ & $f_7\geq 0$ & $F_4$\\
9 & 5680 & - & -4 & $\geq -12$ & - & $1\sim 3$ & $\geq 5$ & - & - & - & $G_2$\\
10 & 4249 & - & -4 & $\geq -12$ & $\geq 5$ & $1\sim 4$ & $\leq 4$ & - & - & $f_7\geq 0$ & $G_2$\\
12 & 4248 & $\geq 7$ & -4 & $\leq -13$ & - & 3 & - & - & - & $f_7\geq 0$, $f_{10}\geq 0$ & $G_2$\\
8 & 160214 & - & -3 & - & - & $\geq 6$ & $\geq -5$ & - & - & $f_{12}\leq -1$ & $G_2$\\
9 & 42972 & - & -3 & - & - & $\leq 2$ & $\geq -5$ & - & - & $f_{11}=0$, $f_{12}\leq -2$ & $G_2$\\
9 & 42871 & - & -3 & - & - & $3\sim 5$ & $\geq -5$ & $\geq 7$ & - & $f_{12}\leq -1$ & $G_2$\\
16 & 17985 & - & -3 & -2 & $\geq 6$ & $0\sim 2$ & $\geq 12$ & $-4\sim 2$ & - & $f_{12}\geq 0$, $f_{14}\leq -1$ & SU(2)\\
12 & 11200 & - & -3 & $\leq -3$ & - & $\geq 7$ & $\geq -5$ & - & - & $f_{12}\geq 0$, $f_{14}\geq -2$ & $G_2$\\ 
14 & 8483 & - & -3 & $\geq -12$ & - & $\leq 1$ & $-5\sim 11$ & - & $\leq -2$ & $f_{12}\geq -1$, $f_{14}=0$ & $G_2$\\
13 & 3751 & $\geq 1$ & -3 & $\geq -5$ & - & $3\sim 5$ & $\geq -5$ & $\leq 6$ & - & $f_7\geq 0$, $f_{10}\geq 0$, $f_{12}\leq -1$ & $G_2$\\
16 & 3536 & - & -3 & $\leq -13$ & $\leq 6$ & $-7\sim 2$ & $-5\sim 3$ & $\geq 10$ & - & $f_{11}\geq 0$, $f_{12}\geq -1$ & SU(2)\\
16 & 3355 & $\geq 2$ & -3 & $-12\sim -7$ & $\geq -11$ & 2 & $-5\sim 11$ & - & $\leq -1$ & $f_7\geq 0$, $f_{12}\geq -1$ & $G_2$\\
20 & 3230 & - & -3 & $\geq -4$ & $\geq 3$ & 2 & $3\sim 11$ & - & $\leq -1$ & $f_7\geq 0$, $f_{10}\geq 0$, $f_{11}\geq 0$, $f_{12}\geq -1$, $f_{14}\geq 0$ & SU(2)\\
12 & 245756 & - & -2 & $\leq -3$ & - & $\geq 16$ & - & - & $\leq -1$ & $f_7\geq -1$, $f_8\leq -2$ & SU(2)*\\
14 & 11713 & $\geq 3$ & -2 & $\leq -3$ & - & $14\sim 51$ & - & - & $\leq -1$ & $f_7\geq -1$, $f_8\geq -1$ & SU(2)*\\
12 & 9063 & - & -2 & $\geq -11$ & $\leq 0$ & $5\sim 12$ & $\leq 2$ & - & $\leq -3$ & $f_{12}\leq -1$, $f_{14}\leq 0$ & SU(2)\\
16 & 8989 & $\geq 1$ & -2 & $\geq -11$ & $\geq 1$ & $10\sim 12$ & $\geq 0$ & $\geq 13$ & - & $f_9\leq -2$, $f_{12}\leq -1$ & SU(2)\\
13 & 8710 & - & -2 & $\geq -11$ & $\geq -18$ & $\leq 4$ & - & - & $\leq -3$ & $f_{10}\geq 0$, $f_{12}\leq -1$, $f_{14}=0$ & SU(2)\\
13 & 6218 & $\geq 2$ & -2 & $\leq -2$ & - & $14\sim 15$ & - & - & $\leq -1$ & $f_7\geq -1$, $f_8\leq -2$ & SU(2)\\
19 & 3017 & $\geq 1$ & -2 & $\geq -11$ & $\geq 1$ & $10\sim 12$ & $0\sim 2$ & $8\sim 12$ & - & $f_9\leq -2$, $f_{12}\leq -1$, $f_{14}\leq -1$ & SU(2)\\
\hline
\end{tabular}
}
\caption[x]{\footnotesize  The inequalities that predict the appearance of certain gauge group $G$ on a $\mathbb{F}_n$ divisor. $|S(l)|$ denotes the number of samples among the 6,300,170 total samples which this rule will apply. The rules are sorted according to the normal bundle coefficient $a$ in $N_D=aS+bF$ and $|S(l)|$. We only list the rules that apply to more than 3,000 samples. The rules with $G=$SU(2)* predicts the existence of SU(2) with at least 99.94\% possibility.}\label{t:rulesFn}
\end{table}

\begin{table}
\centering
{\tiny

    \setlength\tabcolsep{2pt}%
\begin{tabular}{|c |c |c|c|c|c|c|c|c|c|c|c|}
\hline
\hline
$d$ & $|S(l)|$ & $f_0$ & $f_1$ & $f_2$ & $f_3$ & $f_4$ & $f_5$ & $f_6$ & $f_{13}$ & other $f_i$ & $G$\\
\hline
2 & 6831 & - & $\leq -9$ & - & - & - & - & - & - & - & $E_8$\\
4 & 199 & $\leq 1$ & $-8\sim -7$ & $\leq -13$ & - & - & - & - & - & - & $E_8$\\
5 & 8 & $\geq 2$ & $-8\sim -7$ & $\leq -16$ & - & - & - & - & - & - & $E_8$\\
6 & 18 & - & $-8\sim -7$ & $\geq -12$ & - & - & - & $\geq 2$ & - & $f_7\leq -1$, $f_{10}\geq 0$ & $E_8$\\
6 & 17 & $\geq 2$ & $-8\sim -7$ & $-15\sim -13$ & $\geq 3$ & - & - & - & - & - & $E_7$\\
6 & 12 & $\geq 2$ & -6 & - & - & - & $\leq 3$ & - & - & $f_7\leq -1$ & $F_4$\\
6 & 8 & - & -6 & - & - & $\leq 7$ & $\geq 4$ & $\geq 7$ & - & - & $E_7$\\ 
7 & 4698 & - & -6 & - & - & - & $\leq 0$ & - & - & $f_7\geq 0$ & $F_4$\\
7 & 2921 & - & -2 & $\leq -12$ & - & $\leq 12$ & - & - & $\leq -3$ & - & SU(2)\\ 
7 & 1951 & - & -3 & - & - & $\geq 3$ & $\leq -6$ & - & $\leq -1$ & - & $G_2$\\
7 & 63 & - & -2 & - & - & $13\sim 32$ & - & - & - & $f_7\leq -16$ & $96.8\%\varnothing$, $3.2\%$ SU(2)\\
7 & 19 & - & -5 & - & $\leq 0$ & - & $\leq 2$ & - & $\geq 0$ & - & $F_4$\\
7 & 12 & $-8\sim -7$ & $\geq -12$ & - & $\leq 0$ & - & $\geq 2$ & - & - & $f_7\leq -1$, $f_{10}\leq -1$ & $E_8$\\
7 & 9 & - & -6 & - & $\geq 3$ & $\geq 8$ & $\geq 4$ & $\geq 7$ & - & - & $66.7\% E_8$, $33.3\% E_7$\\
7 & 9 & $\leq 1$ & -6 & - & - & $\leq 0$ & $\leq 3$ & - & - & $f_7\leq -1$ & $F_4$\\
7 & 7 & - & -8 & $\geq -12$ & - & $\geq 10$ & $\leq 7$ & - & - & $f_7\geq 0$ & $E_8$\\
7 & 6 & - & $-8\sim -7$ & $\geq -12$ & - & $\leq 3$ & - & $\leq 1$ & - & $f_7\leq -1$, $f_9\geq 0$ & $E_7$\\
7 & 5 & $\geq 2$ & $-8\sim -7$ & -15 & $\leq 2$ & - & - & - & - & - & $E_8$\\
7 & 2 & 0 & $-8\sim -7$ & $\geq -12$ & $\geq 4$ & $\geq 8$ & - & - & - & $f_7\geq 0$ & $E_7$\\
7 & 1 & - & -6 & - & $\leq 2$ & $\geq 8$ & $\geq 4$ & $\geq 7$ & - & - & $E_8$\\
7 & 1 & - & -2 & - & - & $\geq 32$ & - & - & - & $f_7\leq -16$ & SU(2)\\
\hline
\end{tabular}
}
\caption[x]{\footnotesize  Leaves in the decision tree with depth $d\leq 7$, applied to $\mathbb{F}_n$ divisors.}\label{t:rulesdepthFn}
\end{table}

We make a similar selected list of rules in table~\ref{t:rulesFn}. We also list the leaves with small depth in table~\ref{t:rulesdepthFn}. Note that the rules with small depth often apply to few samples since they correspond to extremal cases. The rules in table~\ref{t:rulesFn} only give SU(2), $G_2$, $F_4$ or $E_8$ gauge group. We list a number of rules for SU(3), SO(8), $E_6$, $E_7$ and $E_8$ in table~\ref{t:rulesrareFn}\footnote{The number of samples for SO(7) is too small.}. One can see that the leaves giving SU(3) or SO(8) typically have large depth and the rules are highly complicated, except for the following two rules giving mainly SU(3):

(1) $d=13$, $f_1=-3$, $f_2=-11\sim -3$, $f_4\leq -5$, $f_5\geq 12$, $f_7\geq 0$, $f_{11}\leq -1$, $f_{13}\geq -1$;

(2) $d=9$, $f_1\geq -1$, $f_3\leq -3$, $f_4\leq -35$, $f_6=2$, $f_{14}\geq 0$.

\begin{table}
{\tiny
\resizebox{\linewidth} {\height}{%
    \setlength\tabcolsep{1pt}%
\begin{tabular}{|c|c |c |c|c|c|c|}
\hline
\hline
$d$ & $|S(l)|$ & $f_0$ & $f_1$ & $f_2$ &  other $f_i$ & $G$\\
\hline
27 & 358 & 3 & -2 & $-11\sim -4$ & $f_3=0$, $f_4=6\sim 9$, $f_5=-1\sim 2$, $f_6=-6\sim 9$, $f_7\leq -1$, $f_9=-1$, $f_{12}\geq 0$, $f_{14}\leq 0$ & $98.6\%$ SU(3), $0.8\%\varnothing$, $0.4\%$ SU(2), $0.4\% G_2$\\
34 & 388 & $\geq 2$ & -3 & -6 & $f_3=1$, $f_4=1$, $f_5=-3\sim 11$, $f_6=3\sim 4$, $f_7\geq 0$, $f_8\leq -1$, $f_9\geq 0$, $f_{10}=-1$, $f_{12}\geq 0$, $f_{13}=-1$, $f_{14}\geq 0$ & $91.0\%$ SU(3), $9.0\% G_2$\\
36 & 189 & $\leq 1$ & $-3$ & $-4$ & $f_3=6\sim 9$, $f_4=1$, $f_5=-5\sim 11$, $f_6=3$, $f_7\leq -1$, $f_8\leq -1$, $f_9\geq 0$, $f_{10}=-1$, $f_{12}\geq 0$, $f_{13}=-1$, $f_{14}\leq -1$ & SU(3)\\
29 & 180 & $\geq 1$ & -3 & $\geq -4$ & $f_3=0\sim 3$, $f_4=2$, $f_5=4\sim 8$, $f_6=0\sim 4$, $f_7\geq 0$, $f_8\geq -1$, $f_{10}\leq -1$, $f_{11}\leq -1$, $f_{12}=-1$, $f_{13}=-1$, $f_{14}\geq 0$ & SU(3)\\
29 & 131 & $\leq 2$ & -3 & -4 & $f_3=-11\sim -3$, $f_4=8\sim 12$, $f_5=-3\sim -2$, $f_6=6\sim 8$, $f_7\geq 0$, $f_9\leq -2$, $f_{12}=-2\sim -1$, $f_{13}\geq -2$, $f_{14}=-2\sim -1$ & $96.2\%$ SU(3), $3.1\%\varnothing$, $0.7\%$ SU(2)\\
25 & 129 & - & -2 & $-11\sim -5$ & $f_3=-1$, $f_4=9\sim 12$, $f_5=-2\sim -1$, $f_6=10\sim 14$, $f_9\leq -2$, $f_{12}=-2\sim -1$, $f_{13}=-2\sim -1$, $f_{14}\geq 0$ & $97.8\%$ SU(3), $2.2\%$ SU(2)\\
27 & 115 & - & -3 & -7 & $f_3\leq -1$, $f_4=0\sim 1$, $f_5=-5\sim 0$, $f_7\leq -1$, $f_8\geq -1$, $f_{10}\geq 0$, $f_{11}\geq 0$, $f_{12}\geq -1$, $f_{13}=-1$, $f_{14}\leq -1$ & SU(3)\\
26 & 110 & $\leq 1$ & -3 & $\geq -4$ & $f_3=1$, $f_4=2$, $f_5=2\sim 11$, $f_6\leq 3$, $f_7=0$, $f_9\geq 0$, $f_{10}=-1$, $f_{11}\geq 0$, $f_{12}\geq -1$, $f_{13}\leq -1$, $f_{14}\geq 0$ & $99.1\%$ SU(3), $0.9\% G_2$\\ 
38 & 109 & 1 & -2 & $-11\sim -3$ & $f_3=-1\sim 1$, $f_4=-18\sim -6$, $f_5=2\sim 5$, $f_6=3$, $f_7\geq -1$, $f_8=-3\sim, -2$, $f_9=-1$, $f_{10}=-1$, $f_{11}\leq -1$, $f_{12}\leq -1$, $f_{13}\geq -2$, $f_{14}=-1$ & SU(3)\\
26 & 100 & - & -2 & -5 & $f_3=-1$, $f_4=5\sim 12$, $f_5=-3\sim 2$, $f_6=10$, $f_{10}\geq 0$, $f_{12}=-1$, $f_{13}=-1$, $f_{14}\leq -1$ & SU(3)\\
39 & 100 & $\leq 1$ & -3 & -4 & $f_3=1$, $f_4=1$, $f_5=-5\sim 11$, $f_6=3$, $f_7\leq -1$, $f_8\leq -1$, $f_9\geq 0$, $f_{10}=-1$, $f_{12}\geq 0$, $f_{13}=-1$, $f_{14}\leq -1$ & SU(3)\\
22 & 193 & - & -4 & $\geq -7$ & $f_3=0$, $f_4=1\sim 4$, $f_5\leq 4$, $f_6=3$, $f_7\geq 0$, $f_{10}\geq 0$, $f_{11}\leq -1$, $f_{12}\geq 0$, $f_{14}=0$ & SO(8)\\
25 & 221 & $\leq 1$ & -4 & $\geq -9$ & $f_3=-1$, $f_4=1\sim 4$, $f_5=3\sim 4$, $f_6=3$, $f_7\geq 0$, $f_{10}\geq 0$, $f_{11}\leq -1$, $f_{14}=0$ & $86.4\%$ SO(8), $13.6\% F_4$\\
24 & 183 & - & -4 & $\geq -7$ & $f_3=2\sim 3$, $f_4=1\sim 4$, $f_5=3\sim 4$, $f_6=3$, $f_7\geq 0$, $f_{10}\geq 0$, $f_{11}\leq -1$, $f_{12}\geq 0$, $f_{14}=0$ & $89.1\%$ SO(8), $10.9\% F_4$\\
20 & 152 & - & -4 & $\geq -9$ & $f_3\leq -5$, $f_4=1\sim 4$, $f_5\leq 4$, $f_6=3$, $f_7\geq 0$, $f_{10}\geq 0$, $f_{11}\leq -1$, $f_{14}=0$ & SO(8)\\
20 & 141 & - & -4 & -8 & $f_3\leq 4$, $f_4=2$, $f_5\leq 4$, $f_6=2\sim 3$, $f_7\leq -1$, $f_8\geq 0$, $f_9\geq 0$, $f_{11}\leq -1$, $f_{13}\geq 0$ & $92.2\%$ SO(8), $7.8\%F_4$\\
27 & 142 & $\geq 1$ & -4 & -6 & $f_3=3$, $f_4=1\sim 4$, $f_5\leq 4$, $f_6=2$, $f_7\geq 0$, $f_9\geq 0$, $f_{10}\leq -1$, $f_{11}\leq -1$, $f_{12}\geq 0$, $f_{13}\geq 0$, $f_{14}\leq -1$ & $88.7\%$ SO(8), $11.3\% F_4$\\
26 & 132 & $\geq 1$ & -4 & -6 & $f_3=3$, $f_4=1\sim 4$, $f_5\leq 4$, $f_6=2$, $f_7\geq 0$, $f_{10}\geq 0$, $f_{11}\leq -1$, $f_{12}\geq 0$, $f_{13}\geq 0$, $f_{14}\leq -1$ & $95.5\%$ SO(8), $4.5\% F_4$\\
19 & 150 & - & -4 & -8 & $f_3\leq 4$, $f_4=1$, $f_5\leq 4$, $f_6\leq 1$, $f_7\leq -1$, $f_8\geq 0$, $f_9\geq 0$, $f_{11}\leq -1$ & $83.3\%$ SO(8), $16.7\% F_4$\\
21 & 111 & - & -4 & -6 & $f_3\geq 3$, $f_4=-6$, $f_5\leq 3$, $f_6\geq 0$, $f_7\geq -1$, $f_8\geq 0$, $f_9\geq 0$, $f_{14}\geq 0$ & $99.1\%$ SO(8), $0.9\% F_4$\\
25 & 115 & - & -4 & $\geq -4$ & $f_3=1$, $f_4=3$, $f_5\leq 4$, $f_6=3$, $f_7\geq 0$, $f_9\geq 0$, $f_{10}\leq -1$, $f_{11}\leq -1$, $f_{12}\geq 0$, $f_{14}\geq 0$ & $91.3\%$ SO(8), $8.7\% F_4$\\
29 & 103 & $\leq 1$ & -4 & -6 & $f_3=-3\sim -2$, $f_4=1\sim 3$, $f_5\leq 4$, $f_6=3$, $f_7\geq 0$, $f_9\geq 0$, $f_{10}\leq -1$, $f_{11}\leq -1$, $f_{12}\geq 0$, $f_{14}=0$ & $99\%$ SO(8), $1\% F_4$\\
24 & 116 & - & -4 & $\geq -7$ & $f_3=1$, $f_4=1\sim 4$, $f_5=3\sim 4$, $f_6=3$, $f_7\geq 0$, $f_{10}\geq 0$, $f_{11}\leq -1$, $f_{12}\geq 0$, $f_{14}=0$ & $87.1\%$ SO(8), $12.9\% F_4$\\
11 & 708 & $\leq 1$ & -6 & -9 & $f_4\geq 6$, $f_5\geq 4$, $f_6\leq 6$, $f_7\geq 0$, $f_{14}\geq 0$ & $E_6$\\
11 & 519 & $\leq 1$ & -6 & -9 & $f_5\geq 4$, $f_6=3\sim 6$, $f_7\leq -1$, $f_8\geq 0$ & $E_6$\\
9 & 277 & 0 & -6 & -6 & $f_5\geq 4$, $f_6\leq 6$ & $E_6$\\
13 & 191 & $\leq 1$ & -6 & -9 & $f_3\geq 3$, $f_4\leq 5$, $f_5\geq 4$, $f_6\leq 6$, $f_7\geq 0$, $f_9\geq 0$, $f_{14}\geq 0$ & $E_6$\\
10 & 142 & $\leq 1$ & -5 & -9 & $f_3\leq 4$, $f_5\geq 4$, $f_9\geq 0$ & $E_6$\\
14 & 125 & 2 & -6 & $\leq -12$ & $f_3\geq -1$, $f_4\geq 3$, $f_5\geq 4$, $f_6\leq 6$, $f_{11}\leq -1$ & $E_6$\\
12 & 114 & $\leq 1$ & -6 & -9 & $f_5\geq 4$, $f_6\leq 6$, $f_7\geq 0$, $f_9\geq 0$, $f_{14}\leq -1$ & $E_6$\\
10 & 540 & - & $-8\sim -7$ & $\geq -10$ & $f_4\leq 9$, $f_5\leq 6$, $f_6\leq 2$, $f_7\geq 0$ & $E_7$\\
12 & 176 & - & $-8\sim -7$ & $-10\sim -4$ & $f_4\leq 9$, $f_5=7$, $f_6\leq 2$, $f_7\geq 0$, $f_{11}\leq -1$ & $E_7$\\
15 & 172 & - & $-8\sim -7$ & $\geq -10$ & $f_3\geq 8$, $f_4=1\sim 4$, $f_6=2$, $f_7\leq -1$, $f_8\geq 0$, $f_9\geq 0$, $f_{10}\leq -1$, $f_{11}\leq -1$ & $97.1\% E_7$, $2.9\% E_8$\\
13 & 126 & - & $-8\sim -7$ & $\geq -10$ & $f_3=-8\sim 4$, $f_4=1\sim 7$, $f_5\leq 7$, $f_6=3\sim 4$, $f_7\geq 0$ & $E_7$\\
2 & 6831 & - & $\leq -9$ & - & - & $E_8$\\
4 & 199 & $\leq 1$ & $-8\sim -7$ & $\leq -13$ & - & $E_8$\\
11 & 110 & - & $-8\sim -7$ & $\geq -8$ & $f_3\geq 1$, $f_6=1$, $f_7\leq -1$, $f_9\leq -1$, $f_{13}\geq 0$ & $88.9\% E_8$, $11.1\% E_7$\\
\hline
\end{tabular}
}}
\caption[x]{\footnotesize  The inequalities that predict the appearance of certain gauge group $G=$SU(3), SO(8), $E_6$, $E_7$ or $E_8$ on a $\mathbb{F}_n$ divisor. $|S(l)|$ denotes the number of samples among the 6,300,170 total samples which this rule will apply. The rules are sorted according to the gauge group and $|S(l)|$. We only list the rules that apply to more than 100 samples and giving rare gauge groups more than $80\%$ of the times.}\label{t:rulesrareFn}
\end{table}

Now we use (\ref{normf}, \ref{normg}) to analyze some of the rules. For $D=\mathbb{F}_n$, $-K_D=2S+(n+2)F$ and $N_D=aS+bF$, hence the formula (\ref{normf}, \ref{normg}) becomes
\be
f_{k,D}\in\mc{O}[(8+(4-k)a)S+(4(n+2)+(4-k)b)F-\sum_{D_i\bigcap D\neq\varnothing}\phi_i D_i\bigcap D],\label{normfFn}
\ee
\be
g_{k,D}\in\mc{O}[(12+(6-k)a)S+(6(n+2)+(6-k)b)F-\sum_{D_i\bigcap D\neq\varnothing}\gamma_i D_i\bigcap D],\label{normgFn}
\ee
where $\phi_i$ and $\gamma_i$ are the order of vanishing of $f$ and $g$ on the divisors $D_i$ which intersect $D$.

If $a\leq -9$, it is clear that $f_{3,D}$ in (\ref{normfFn}) vanishes, since the coefficient $8+(4-k)a$ becomes negative. Similarly, $g_{4,D}$ in (\ref{normgFn}) vanishes. Hence the gauge group has to be $E_8$, as given by the first rule in table~\ref{t:rulesFn}.

If $a=-7$ or $-8$, $n\leq 1$ and $b\leq -13$, then the coeffcient $4(n+2)+(4-k)b$ in (\ref{normfFn}) becomes negative for $k=3$, hence $f_{3,D}$ vanishes. One can also see that $g_{4,D}$ vanishes since the coefficient $6(n+2)+2b$ is negative in (\ref{normgFn}), hence the gauge group is $E_8$. This is exactly the second last rule in table~\ref{t:rulesrareFn}.

If $a=-6$, we can see that $f_{2,D}$ in (\ref{normfFn}) vanishes since $8+(4-k)a<0$. Similarly $g_{3,D}$ vanishes hence $(f,g)$ vanishes to at least order $(3,4)$ on $D$. Hence the gauge group is minimally $F_4$. The gauge group is $E_6$ if $g_{4,D}\in\mc{O}(0)$ and $g_{4,D}$ is locally a complete square. This can happen when $6(n+2)+2b=0$ and $\sum_{D_i\bigcap D\neq\varnothing}\gamma_i D_i\bigcap D=0$. The rules in table~\ref{t:rulesrareFn} that predict $E_6$ gauge group roughly all belong to this class. For example, the rule with $|S(l)|=277$ states that if $n=0$, $a=b=-6$, $f_5\geq 4$, $f_6\leq 6$, then the gauge group is $E_6$. If $n=0$, $a=b=-6$, then the gauge group cannot be $F_4$ since $g_{4,D}\in\mc{O}(0)$ already. The additional rules $f_5\geq 4$, $f_6\leq 6$ help to make sure that the gauge group is not larger than $E_6$ in a subtle way.

\subsection{Toric surfaces with $h^{1,1}=3$}
\label{s:S3n}

The toric surfaces with $h^{1,1}(D)=3$ are generated by blowing of $\mathbb{F}_n$ at the intersection points of toric curves. They form a simple one parameter family $S_{3,n} (n\in\mathbb{Z},n\geq 0)$. $D$ has five neighboring divisors $D_1,\cdots,D_5$, where $D_i\bigcap D=C_i$ gives the toric curve on the divisor $D$. The five corresponding toric curves $C_1,C_2,\dots C_5$ on $D$ has the following selfiintersection numbers:
\be
C_1^2=0, C_2^2=n, C_3^2=-1, C_4^2=-1, C_5^2= -(n+1).
\ee

The linear equivalence conditions of $C_i(i=1,\dots, 5)$ are
\be
C_1=C_3+C_4\ ,\ C_2=C_5+nC_3+(n+1)C_5.
\ee
The input vector we use is 19-dimensional, with the following form:
\be
\bsp
V=&(n,D^2 D_1,D^2 D_2, D^2 D_3, D_1^3, D_2^3, D_3^3, D_4^3, D_5^3, D_1^2 D_2, D_2^2 D_1, D_2^2 D_3, D_3^2 D_2, D_3^2 D_4,\\
& D_4^2 D_3, D_4^2 D_5, D_5^2 D_4, D_5^2 D_1, D_1^2 D_5).
\end{split}
\ee
We call these 19 entries $f_0,\dots,f_{18}$. The numbers $f_1,f_2,f_3$ encodes the information of the normal bundle of $D$. Suppose that the normal bundle of $D$ has the form
\be
N_D=a C_3+b C_4+c C_5,
\ee
Then we have equations
\be
\bsp
&f_1=N_D\cdot C_1=c\\
&f_2=N_D\cdot C_2=a\\
&f_3=N_D\cdot C_3=b-a
\end{split}
\ee

There are in total 3,832,969 divisors with $h^{1,1}(D)=3$ on the end point bases and intermediate bases we have generated, which we called the set $S(3)$. We list the number of divisors with each gauge group in table~\ref{t:S3ngroups}. After training a decision tree on the up/down resampled data $S'(3)$, the decision tree has 110,732 nodes and 55,367 leaves. The maximal depth of the tree is $d_{max}=47$. The in-sample and out-of-sample accuracies on the up/down resampled data are 0.976481 and 0.962241, while the accuracy on the original data set $S(3)$ is $A=0.945464$.

\begin{table}
\begin{center}
\begin{tabular}{|c |c |c|c|c|c|c|c|c|c|}
\hline
\hline
 $\varnothing$ & SU(2) & SU(3) & $G_2$ & SO(7) & SO(8) & $F_4$ & $E_6$ & $E_7$ & $E_8$\\
\hline
2629209 & 544104 & 56710 & 397029 & 15 & 32592 & 145676 & 11802 & 4644 & 11188\\
\hline
\end{tabular}
\end{center}
\caption[x]{\footnotesize  Total number of sample divisors with each gauge group in the set $S(3)$, which is the set of divisor with $S_{3,n}$ topology.}\label{t:S3ngroups}
\end{table}

\begin{figure}
\centering
\includegraphics[height=8cm]{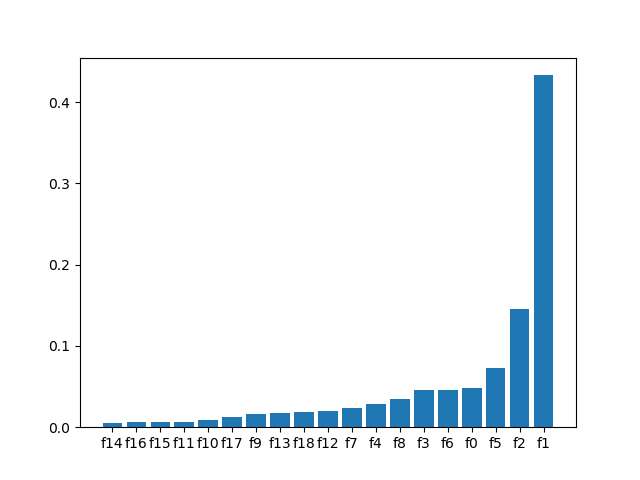}
\caption[E]{\footnotesize The feature importance of input vector elements $f_i$, for the $S_{3,n}$ divisors with $h^{1,1}(D)=3$ on end point bases and intermediate bases.}\label{f:featS3n}
\end{figure}

We plot the feature importance in figure~\ref{f:featS3n}. It seems that $f_1$ is the most important feature to determine the gauge group. $f_2$ and $f_5$ also has significantly higher importance. We list a number of leaves with large number of applied samples and small depth in table~\ref{t:rulesS3n}.

\begin{table}
\begin{center}
{\tiny

\begin{tabular}{|c |c |c|c|c|c|}
\hline
\hline
$d$ & $|S(l)|$ & $f_1$ & $f_2$ & other $f_i$ & $G$\\
\hline
3 & 8493 & $\leq -9$ & - & - & $E_8$\\
5 & 973 & $-8\sim -7$ & $\leq -9$ & $f_0=0$ & $E_8$\\
6 & 208 & -6 & $\leq -9$ & $f_0=0$ & $E_8$\\
7 & 194 & -5 & $\geq -6$ & $f_3\leq -1$, $f_8\geq 6$, $f_{17}\geq 0$ & $F_4$\\
7 & 3976 & -4 & $\geq -3$ & $f_4\geq 4$, $f_9\geq 0$, $f_{13}\geq 0$ & $G_2$\\
7 & 134 & -4 & $\leq -9$ & $f_5\geq 7$ & $E_8$\\
7 & 1456 & -4 & $\geq -3$ & $f_4\leq 3$, $f_7\geq 6$, $f_{13}\geq 0$ & $G_2$\\
8 & 1349 & -4 & $\geq -3$ & $f_4\geq 5$, $f_8\leq 2$, $f_{13}\leq -1$ & $G_2$\\
8 & 148 & $-8\sim -7$ & $\geq -7$ & $f_3\leq -5$, $f_6\leq 3$, $f_{14}\geq 0$ & $E_7$\\
8 & 1691 & $\geq -2$ & $\geq -1$ & $f_3\geq -1$, $f_4\geq 3$, $f_7\geq 5$, $f_{13}\geq 4$, $f_{15}\leq -1$ & $99.8\%\varnothing$, $0.2\%$ SU(2)\\
8 & 314 & $\geq -2$ & $\geq -1$ & $f_3=-1\sim 1$, $f_4\geq 3$, $f_7\geq 3$, $f_{13}\geq 2$, $f_{15}\geq 0$ & SU(2)\\
8 & 286 & $\geq -1$ & $-7\sim -4$ & $f_8\leq 16$, $f_{10}\geq -1$ & $99.3\% F_4$, $0.7\%\varnothing$\\
8 & 3736 & -2 & $\geq -1$ & $f_3\leq -2$, $f_4\leq 2$, $f_{11}\leq -1$, $f_{12}\leq -2$ & $\varnothing$\\
8 & 4063 & -3 & - & $f_5\leq 0$, $f_6\geq 2$, $f_{17}\leq -2$, $f_{18}=0$ & $G_2$\\
8 & 2604 & -4 & $\leq -4$ & $f_4\geq 5$, $f_5\leq 4$, $f_6\leq 4$, $f_9\geq 0$ & $G_2$\\
8 & 960 & -3 & $\leq -15$ & $f_4\geq -1$, $f_5\leq 0$, $f_6\leq 1$, $f_8\leq 3$ & $G_2$\\
8 & 1093 & -5 & $\geq -5$ & $f_3\leq -10$, $f_8=2\sim 5$ & $F_4$\\
9 & 2050 & $\geq -2$ & $\geq -1$ & $f_3\leq -2$, $f_4\geq 3$, $f_7\leq 2$, $f_{12}\leq -2$, $f_{13}\geq 2$, $f_{15}\geq 0$ & $\varnothing$\\
9 & 3736 & $\geq -1$ & $\leq -2$ & $f_0=0$, $f_3\leq -1$, $f_{10}\leq -3$, $f_{17}\leq -2$ & $99.95\%\varnothing$, $0.05\%$ SU(2)\\
9 & 223 & -6 & -6 & $f_0\geq 1$, $f_3\geq -3$, $f_4\geq 2$, $f_9\geq 0$ & $F_4$\\
9 & 16462 & -2 & $\leq -2$ & $f_4\leq 2$, $f_5\leq 11$, $f_6\leq 3$, $f_{17}\leq -3$, $f_{18}\leq 0$ & $99.99\%\varnothing$, $0.01\%$ SU(2)\\
9 & 1411 & $\geq -1$ & $-4\sim -3$ & $f_0\geq 2$, $f_{10}\geq -1$ & $\varnothing$\\
9 & 15513 & -3 & - & $f_0=0$, $f_5\geq 3$, $f_9\leq 1$, $f_{10}\leq -2$, $f_{18}\geq 0$ & $G_2$\\
9 & 147 & -5 & $\geq -5$ & $f_3\leq -11$, $f_8\leq 1$ & $F_4$\\
9 & 294 & $\geq -2$ & $\geq -1$ & $f_3\leq -2$, $f_4\geq 3$, $f_{11}\geq 0$, $f_{12}\leq -4$, $f_{15}\leq -1$ & $\varnothing$\\
9 & 14338 & -3 & $\leq -3$ & $f_5=1\sim 2$, $f_8\leq 5$, $f_{17}\leq -2$, $f_{18}\leq 0$ & $G_2$\\
9 & 410 & $\geq -2$ & $\geq -1$ & $f_3\geq -1$, $f_4\geq 3$, $f_7\leq 4$, $f_{13}\geq 4$, $f_{15}\leq -1$, $f_{18}\geq 1$ & $\varnothing$\\
9 & 12110 & $\geq -2$ & $\geq -1$ & $f_3\leq -2$, $f_4\geq 3$, $f_{10}\leq 1$, $f_{11}\leq -1$, $f_{12}\leq -2$, $f_{15}\leq -1$ & $\varnothing$\\
9 & 11659 & -3 & - & $f_3\geq -4$, $f_5=1\sim 2$, $f_8\geq 6$, $f_{10}\leq 0$, $f_{17}\geq -4$ & $99.99\% G_2$, $0.01\% F_4$\\
9 & 529 & -2 & -2 & $f_4\leq -3$, $f_5\geq 12$, $f_{10}\leq 4$, $f_{18}\geq 1$ & $97.7\%$ SU(2), $2.3\%\varnothing$\\
9 & 166200 & $\geq -1$ & $\leq -2$ & $f_0\geq 1$, $f_3=-1$, $f_{10}=-2$ & $99.999\%\varnothing$, $0.001\%$ SU(2)\\
9 & 612 & -3 & - & $f_5=1\sim 2$, $f_6\geq 2$, $f_7\geq 1$, $f_8\geq 6$, $f_{10}\geq 1$ & $G_2$\\
9 & 278 & -4 & $-8\sim -4$ & $f_3=-3$, $f_8\leq 5$ & $E_6$\\
9 & 6141 & -4 & $\leq -4$ & $f_3\geq -2$, $f_5\leq 3$, $f_6\geq 5$, $f_{12}\geq 0$ & $G_2$\\
9 & 2535 & -6 & $\leq -14$ & $f_0\geq 1$, $f_8\leq 0$, $f_{15}\geq 0$ & $F_4$\\
9 & 5585 & -2 & $\geq -1$ & $f_3\geq -1$, $f_4\leq 2$, $f_5\geq 4$, $f_9\leq -2$, $f_{13}\leq 1$ & $\varnothing$\\
9 & 45945 & $\geq -1$ & $\leq -2$ & $f_0\geq 1$, $f_3\leq -1$, $f_9\geq 0$, $f_{10}\leq -3$ & $\varnothing$\\
 
\hline
\end{tabular}
}
\end{center}
\caption[x]{\footnotesize  The inequalities that predict the appearance of certain gauge group $G$ on a $S_{3,n}$ divisor. $|S(l)|$ denotes the number of samples among the 3,832,969 total samples which this rule will apply. We only list the rules that apply  to more than 100 samples and has depth $d\leq 9$.}\label{t:rulesS3n}
\end{table}

Most of the rules are predicting SU(2), $G_2$, $F_4$ or $E_8$ gauge group. However, there is also one rule predicting $E_7$ and another rule predicting $E_6$. Similar to the cases of $D=\mb{F}_n$, we can see that the number $f_0\equiv n$ specifying the topology of $D$ is not very important in these rules.

\subsection{Toric surfaces with $h^{1,1}>3$}

For toric divisors with $h^{1,1}(D)>3$, we use the original $5(h^{1,1}(D)+2)$-dimensional vector described in Section~\ref{s:features}, which contains some redundant information. The labeling of the $p=h^{1,1}(D)+2$ neighbor divisors is chosen such that the curve $C_1=D_1\bigcap D$ has the lowest self-intersection number among $C_i(i=1,\dots,p)$. Then $C_1,\dots,C_p$ curves form a cyclic toric diagram of $D$.

We only list some general information about the sample divisors and the decision tree. We list the number of divisors with each gauge group in the original data sets $S(h^{1,1}(D))$ in table~\ref{t:othergroups}.

\begin{table}
\begin{center}
{\scriptsize
\begin{tabular}{|c|c|c |c |c|c|c|c|c|c|c|c|}
\hline
\hline
$h^{1,1}(D)$ & $N$ & $\varnothing$ & SU(2) & SU(3) & $G_2$ & SO(7) & SO(8) & $F_4$ & $E_6$ & $E_7$ & $E_8$\\
\hline
4 & 4,557,007 & 2997632 & 742695 & 90854 & 483568 & 2 & 43140 & 168607 & 12651 & 5343 & 12516\\
5 & 1,792,867 & 823253 & 457698 & 68018 & 312898 & 1 & 23913 & 90786 & 5927 & 1562 & 8811\\
6 & 1,008,600 & 498699 & 238615 & 38492 & 166968 & 0 & 10312 & 48119 & 1771 & 438 & 5186\\
7 & 578953 & 298197 & 132599 & 20181 & 91764 & 0 & 5282 & 25768 & 617 & 30 & 4515\\
8 & 365346 & 210308 & 71062 & 9705 & 49757 & 1 & 2641 & 16996 & 346 & 0 & 4530\\
9 & 240144 & 143847 & 44129 & 5566 & 30293 & 0 & 1325 & 10782 & 100 & 61 & 4041\\
10 & 237500 & 174572 & 29131 & 3562 & 17943 & 0 & 833 & 7610 & 19 & 0 & 3830\\
11 & 87873 & 45537 & 18810 & 2596 & 12118 & 0 & 417 & 4997 & 12 & 0 & 3386\\
12 & 70992 & 42723 & 12372 & 1526 & 6997 & 0 & 377 & 3887 & 9 & 0 & 3101\\
13 & 45032 & 25462 & 8500 & 1021 & 4353 & 0 & 146 & 2689 & 0 & 0 & 2861\\
14 & 44646 & 30499 & 5536 & 767 & 3258 & 0 & 71 & 1985 & 0 & 0 & 2530\\
15 & 33323 & 22822 & 4107 & 504 & 2034 & 0 & 12 & 1548 & 0 & 0 & 2296\\
16 & 62617 & 55410 & 2781 & 228 & 968 & 0 & 3 & 1164 & 0 & 0 & 2063\\
17 & 27902 & 22140 & 1996 & 199 & 781 & 0 & 10 & 891 & 0 & 0 & 1885\\
18 & 10805 & 6458 & 1364 & 204 & 467 & 0 & 4 & 709 & 0 & 0 & 1619\\
19 & 9579 & 6384 & 761 & 47 & 254 & 0 & 0 & 574 & 0 & 0 & 1559\\ 
20 & 7000 & 4486 & 531 & 20 & 157 & 0 & 0 & 420 & 0 & 0& 1386\\
\hline
\end{tabular}
}
\end{center}
\caption[x]{\footnotesize  Total number of sample divisors with each gauge group for each $h^{1,1}(D)$. $N$ is the total number of divisors with each $h^{1,1}(D)$.}\label{t:othergroups}
\end{table}

\begin{figure}
\centering
\includegraphics[height=6cm]{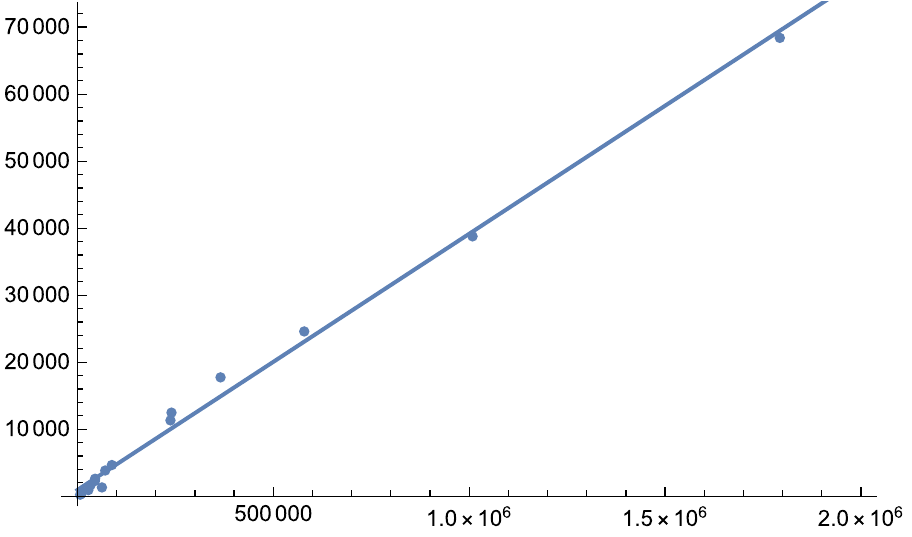}
\caption[E]{\footnotesize The linear relation between the number of nodes of the decision tree and the number of data  samples in $S(h^{1,1}(D))$ for different $h^{1,1}(D)>4$.}\label{f:sampnodes}
\end{figure}

The decision tree is trained on 75\% of the resampled data set $S'(h^{1,1}(D))$ and we list the accuracy, total number of nodes and maximal depth of the decision tree in table~\ref{t:otherinfo}, including the cases for $h^{1,1}(D)=1,2,3$ as well. As we can see from table~\ref{t:othergroups} and \ref{t:otherinfo}, for the cases of $h^{1,1}(D)>4$, the number of nodes and leaves in the decision tree is roughly proportional to the number of data samples in $S(h^{1,1}(D))$. We plot the linear model and data points in figure~\ref{f:sampnodes}. The linear relation is
\be
N_{nodes}=0.038218N_{samples}+937.59,
\ee
with $R^2=0.994635$. This indicates that the decision tree approach on divisors with larger $h^{1,1}(D)$ has a universality. On the other hand, the maximal depth of the decision tree is not significantly correlated to the total number of nodes.

We can see that the in-sample accuracy roughly increases as $h^{1,1}(D)$ becomes bigger. For $h^{1,1}(D)>7$, the in-sample accuracy becomes very high ($>99.98\%$). In principle, if there is not a case where two samples with different labels share the identical features, then an untrimmed decision tree should give perfect in-sample accuracy, as samples with different labels can always be split into different nodes. The low in-sample accuracy for $h^{1,1}(D)=1$ implies that there are many samples where the features are not enough to distinguish the gauge group. On the other hand, for larger $h^{1,1}(D)$, there are more features and this problem is less significant, since it is less likely to find two samples with exactly the same features.

On the other hand, the out-of-sample accuracy and the actual accuracy on the original data set are not clearly correlated with $h^{1,1}(D)$. Nonetheless, the accuracies are always between $85\%\sim 99\%$.

\begin{table}
\begin{center}
\begin{tabular}{|c|c|c |c |c|c|c|}
\hline
\hline
$h^{1,1}(D)$ & $N_{\rm nodes}$ & $N_{\rm leaves}$ & IS acc. & OOS acc. & $A$ & $d_{max}$\\
\hline
1 & 2563 & 1282 & 0.912774 & 0.900694 & 0.949111 & 29\\
2 & 66441 & 33221 & 0.982169 & 0.977909 & 0.978592 & 49\\
3 & 110732 & 55367 & 0.976481 & 0.962241 & 0.945464 & 47\\
4 & 173485 & 86743 & 0.981032 & 0.959135 & 0.937576 & 49\\
5 & 68393 & 34197 & 0.990597 & 0.949702 & 0.926666 & 53\\
6 & 38779 & 19390 & 0.995579 & 0.945967 & 0.924599 & 52\\
7 & 24587 & 12294 & 0.997105 & 0.937040 & 0.909661 & 51\\
8 & 17737 & 8869 & 0.998167 & 0.944062 & 0.911947 & 50\\
9 & 12475 & 6238 & 0.998337 & 0.937586 & 0.903046 & 50\\
10 & 11315 & 5658 & 0.998713 & 0.954421 & 0.940406 & 53\\
11 & 4655 & 2328 & 0.998970 & 0.886774 & 0.853343 & 43\\
12 & 3839 & 1920 & 0.998433 & 0.919833 & 0.887834 & 43\\
13 & 2607 & 1304 & 0.999923 & 0.892036 & 0.873670 & 40\\
14 & 2271 & 1136 & 0.999866 & 0.927583 & 0.917897 & 44\\
15 & 1639 & 820 & 0.999553 & 0.941823 & 0.916046 & 50\\
16 & 1313 & 657 & 0.998731 & 0.990263 & 0.967306 & 60\\
17 & 915 & 458 & 0.999711 & 0.966474 & 0.940061 & 44\\
18 & 569 & 285 & 1.0 & 0.900461 & 0.867998 & 31\\
19 & 363 & 182 & 1.0 & 0.923681 & 0.899510 & 18\\
20 & 233 & 117 & 1.0 & 0.915584 & 0.895196 & 22\\ 
\hline
\end{tabular}
\end{center}
\caption[x]{\footnotesize  The information of the decision tree for each $h^{1,1}(D)$. $N_{\rm nodes}$ and $N_{\rm leaves}$ are the total number of nodes and leaves in the decision tree. IS acc. and OOS acc. are the in-sample and out-of-sample accuracy tested on the up/down resampled data set $S'$ as described in section~\ref{s:resample}. $A$ is the accuracy on the original dataset without resampling. $d_{max}$ is the maximal depth of the decision tree.}\label{t:otherinfo}
\end{table}

\section{Checking whether a curve is a (4,6)-curve}
\label{s:46curve}

\begin{figure}
\centering
\includegraphics[height=6cm]{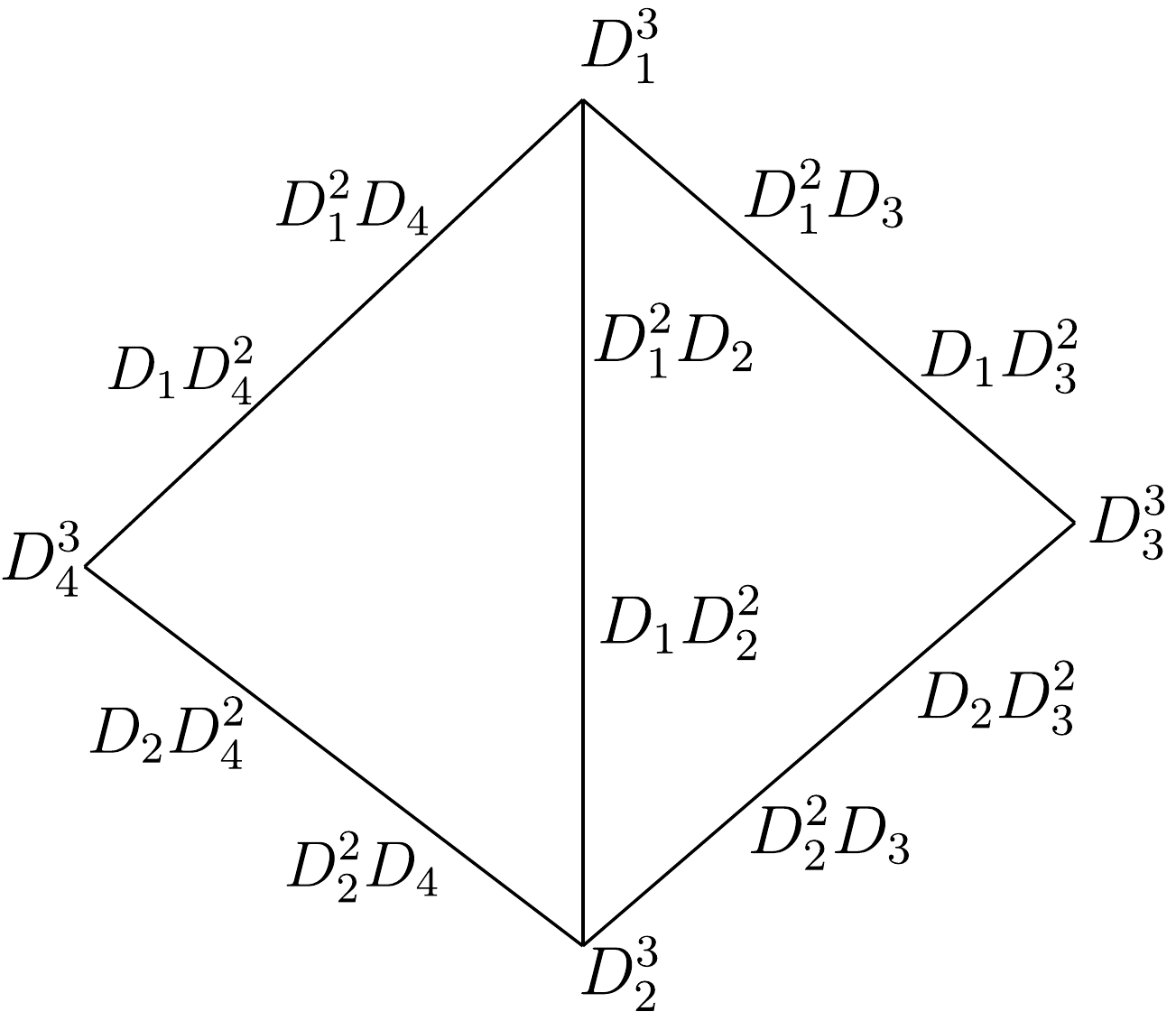}
\caption[E]{\footnotesize The 14 triple intersection numbers used as the features in machine learning to determine whether the curve $D_1 D_2$ is a (4,6) curve or not.}\label{f:curve}
\end{figure}

Besides the decision of non-Higgsable gauge groups, we also attempt to use machine learning to decide whether a toric curve $v_i v_j$ on a general resolvable base is a (4,6)-curve or not. We use the 14 local triple intersection numbers shown in figure~\ref{f:curve} as the features. The input vector is
\be
(D_1^3, D_1^2 D_2, D_1 D_2^2, D_2^3, D_1^2 D_3, D_1 D_3^2, D_1^2 D_4, D_1 D_4^2, D_2^2 D_3, D_2 D_3^2, D_2^2 D_4, D_2 D_4^2, D_3^3, D_4^3).
\ee
We label them by $f_0\sim f_{13}$. The output label is binary, 0 for curves without (4,6) singularity and 1 for (4,6) curves.

The toric threefold bases are generated from a similar approach as the intermediate bases in Section~\ref{s:genbase}. We start from base $b_1=\mathbb{P}^3$ and randomly blow up/down once in each step, generating 10,000 bases in the sequence. The difference is that we allow all the resolvable bases with (4,6) curves to appear in this sequence. To reduce repetition, we only pick $b_1,\dots,b_{20}$ and $b_{100k} (k\in\mathbb{Z})$. Then we use every toric curves on these bases to generate the training data set. In total, we have performed 25 random walk sequences and generated 3,000 bases.

In total there are 12,125,945 sample curves, among which 1,342,652 of them has (4,6) singularity. After processing the original data set by resampling, the decision tree has 193,121 nodes and 96,561 leaves. The maximal depth is $d_{max}=51$. The in-sample and out-of-sample accuracy on the resampling data set is 0.997106 and 0.953865 respectively. The accuracy on the original data set is $A=0.957505$.

\begin{figure}
\centering
\includegraphics[height=6cm]{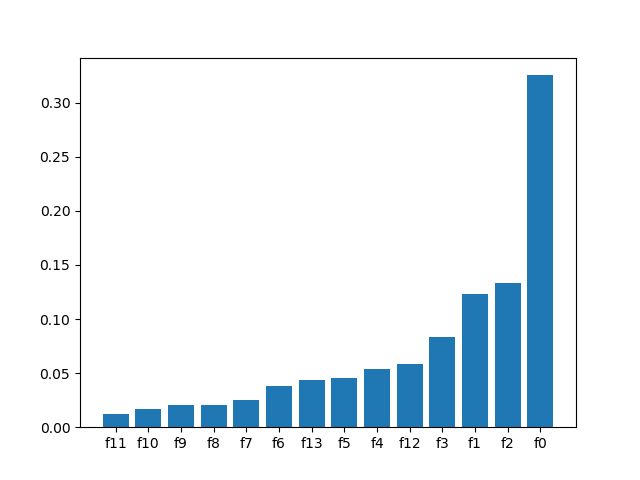}
\caption[E]{\footnotesize The feature importance of input vector elements $f_i$, for the curves on resolvable bases.}\label{f:feat46}
\end{figure}

The feature importance of $f_i$ is plotted in figure~\ref{f:feat46}. We can see that $f_0$, $f_1$, $f_2$ and $f_3$ are the most important features, which is expected since they sit closer to the curve $D_1 D_2$.

\begin{table}
\begin{center}
{\tiny
\resizebox{\linewidth} {\height}{%
    \setlength\tabcolsep{2pt}%
\begin{tabular}{|c |c |c|c|c|c|c|c|}
\hline
\hline
$d$ & $|S(l)|$ & $f_0$ & $f_1$ & $f_2$ & $f_3$ & other $f_i$ & $p_{(4,6)}$\\
\hline
4 & 128613 & $\leq -8$ & $\leq -2$ & $\leq -2$ & - & - & $100\%$\\
5 & 37363 & $-7\sim 16$ & $\leq -2$ & -2 & - & - & 100\%\\
26 & 35673 & $28\sim 143$ & $-13\sim 0$ & $\leq 5$ & $\geq 11$ & $f_4\geq -4$, $f_5=-1\sim 26$, $f_6\geq -4$, $f_7=-3\sim 17$, $f_9=-7\sim 0$, $f_{11}\leq 1$, $f_{12}\leq 4$, $f_{13}\leq 14$ & $99.97\%$\\
13 & 22600 & $\leq -197$ & $0\sim 5$ & $\leq -3$ & $\geq 5$ & $f_4\geq -5$, $f_5\leq -1$, $f_6\geq -5$, $f_{12}\geq -6$ & $100\%$\\
11 & 18744 & $-7\sim 16$ & $\leq -2$ & $\leq -3$ & - & $f_8\leq 11$, $f_{10}\leq 3$, $f_{13}\leq 2$ & $100\%$\\
9 & 14823 & $\geq 17$ & $\leq -2$ & $\leq -2$ & $5\sim 7$ & $f_5\leq -1$, $f_9\leq 1$ & $100\%$\\
6 & 13715 & $\geq 17$ & $\leq -2$ & $\leq -2$ & $\leq 4$ & - & $100\%$\\
15 & 12257 & $\leq -197$ & $0\sim 4$ & $\leq -3$ & $f_3\leq 4$ & $f_4=-5\sim -4$, $f_5\leq -1$, $f_6\geq -5$, $f_{12}\geq -6$ & $99.93\%$\\
20 & 11290 & $22\sim 27$ & $-15\sim -2$ & $\leq 5$ & $\geq 8$ & $f_5=-2\sim 0$, $f_6=-2\sim 0$, $f_7\geq -3$, $f_9=-5\sim 1$, $f_{11}\leq 1$ & $99.8\%$\\
16 & 10515 & $-196\sim -104$ & $0\sim 1$ & $\leq -3$ & - & $f_4\geq -5$, $f_5=-3\sim -1$, $f_6\geq -5$, $f_7\geq -6$, $f_8\geq -7$, $f_{10}\leq 0$ & $99.8\%$\\
23 & 10331 & $-103\sim -25$ & 0 & $\leq -3$ & $\geq -31$ & $f_4\geq -5$, $f_5\leq -1$, $f_6\geq -5$, $f_7\leq -1$, $f_8=-1\sim 8$, $f_{10}\leq 0$, $f_{11}\leq 12$, $f_{12}\leq 25$, $f_{13}=-8\sim 4$ & $99.6\%$\\
28 & 10113 & $28\sim 143$ & $-8\sim 3$ & $\leq 5$ & $8\sim 10$ & $f_4\geq -4$, $f_5=-1\sim 26$, $f_6\geq -3$, $f_7=-3\sim 17$, $f_9=-7\sim 0$, $f_{11}\leq 1$, $f_{12}\leq 3$, $f_{13}\leq 14$ & $98.7\%$\\
19 & 9756 & $\leq -238$ & $6\sim 15$ & $-22\sim -4$ & $\geq -3$ & $f_4=-5\sim -4$, $f_5\leq -1$, $f_6\geq -5$, $f_{7}\geq -2$ & $99.6\%$\\
21 & 9675 & $\leq -194$ & $6\sim 23$ & $-22\sim -6$ & $\leq -5$ & $f_4\geq -5$, $f_5\leq -1$, $f_6\geq -5$, $f_{11}\geq -4$, $f_{13}\leq 7$ & $99.99\%$\\
6 & 9547 & $\leq -8$ & $\leq -7$ & -1 & - & $f_5\leq 1$ & $100\%$\\
23 & 8105 & $-7\sim 16$ & -1 & $\leq -3$ & $-1\sim 15$ & $f_5=-2\sim 0$, $f_7\leq 0$, $f_8\leq 11$, $f_9\geq -2$, $f_{10}\leq 0$, $f_{11}=-1\sim 1$, $f_{13}\leq 2$ & $98.9\%$\\
17 & 7054 & $\geq 21$ & $\geq -1$ & $\leq -3$ & $5\sim 7$ & $f_4\leq -2$, $f_5\leq -1$, $f_7\leq -1$, $f_9\leq 0$, $f_{12}=-23\sim 13$ & $99.93\%$\\
14 & 7044 & $\geq 61$ & $\geq 0$ & $-48\sim -4$ & $\leq 4$ & $f_4\geq -4$, $f_6\geq -5$, $f_7\geq -1$ & $99.5\%$\\
27 & 6954 & $28\sim 143$ & $0\sim 3$ & $\leq -4$ & $\geq 11$ & $f_4\geq -4$, $f_5=-1\sim 26$, $f_6\geq -4$, $f_7=-3\sim 17$, $f_9=-7\sim 0$, $f_{11}\leq 1$, $f_{12}\leq 4$, $f_{13}\leq 14$ & $100\%$\\
8 & 6888 & $-164\sim -29$ & -1 & $\leq -3$ & - & $f_{10}\geq -1$ & $99.9\%$\\
9 & 6630 & $-28\sim -8$ & -1 & $\leq -3$ & $-76\sim 20$ & $f_{11}\leq 13$ & $99.6\%$\\
11 & 6586 & $-8\sim 16$ & $\leq -4$ & $-1\sim 2$ & $\geq 17$ & - & $99.8\%$\\
12 & 6198 & $\leq -491$ & -1 & $\leq -2$ & - & $f_7\geq -7$, $f_8\geq -3$, $f_9\leq 2$ & $100\%$\\
16 & 6089 & $17\sim 21$ & $\geq -22$ & $\leq 2$ & $\geq 13$ & $f_5\leq -1$, $f_6\geq -4$, $f_7\leq -1$, $f_8\leq 5$, $f_9\leq 1$, $f_{10}\geq -2$, $f_{12}\leq 4$ & $99.75\%$\\
15 & 5255 & $\leq -197$ & $0\sim 1$ & $\leq -3$ & $\leq 4$ & $f_4\geq -3$, $f_5\leq -1$, $f_6\geq -5$, $f_{12}\geq -6$ & $100\%$\\
11 & 4285 & $\geq 28$ & - & $\leq -2$ & $6\sim 7$ & $f_5\geq 0$, $f_9\leq 1$ & $99.3\%$\\ 
\hline
\end{tabular}
}}
\end{center}
\caption[E]{\footnotesize The inequalities that predict whether a curve is a (4,6)-curve, where we use $p_{4,6}$ to denote the probability. $|S(l)|$ denotes the number of samples among the 12,125,945 total samples which this rule will apply. We only list the rules that apply to more than 4000 samples and predict the appearance of (4,6) curve with $p_{(4,6)}>80\%$.}\label{t:rules46}
\end{table}

\begin{figure}
\centering
\includegraphics[height=5cm]{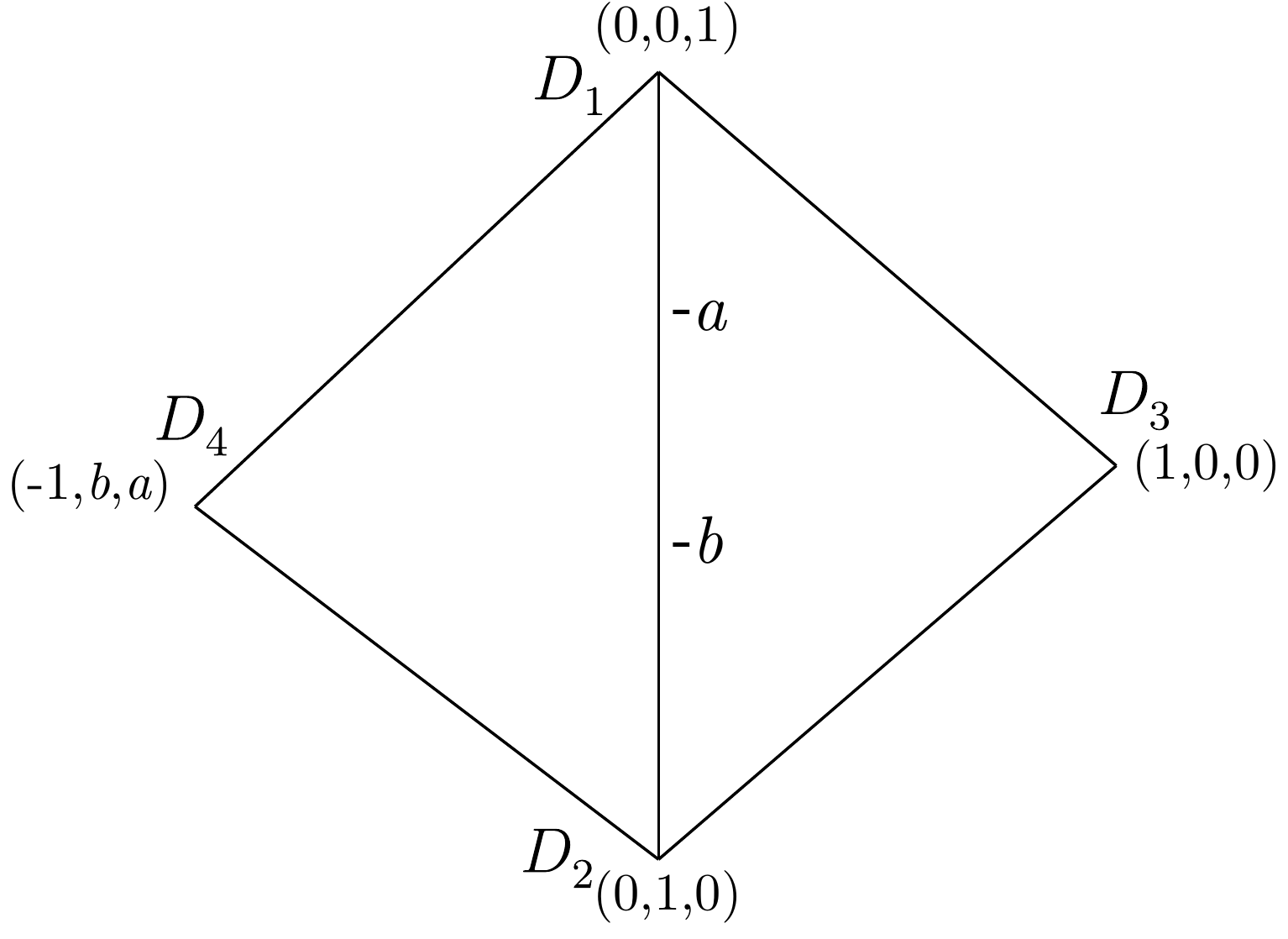}
\caption[E]{\footnotesize The universal configuration of local toric geometry near a toric curve $D_1\bigcap D_2$.}\label{f:46curve}
\end{figure}

We list a number of leaves with big $|S(l)|$ in table~\ref{t:rules46}. From table~\ref{t:rules46}, it seems that the curve is usually a (4,6)-curve whenever $f_1\leq -2$ and $f_2\leq -2$. Actually this rule always holds for any toric curve. We will now derive this analytically.

Suppose that $f_1=-a$, $f_2=-b$, then the local toric geometry near the toric curve is always described by figure~\ref{f:46curve} up to an SL$(3,\mathbb{Z})$ transformation on the toric rays\footnote{In this paper, SL$(3,\mb{Z})$ always include the matrices with determinant $\pm 1$.}. The reason is that since the two 3D cones have unit volume, we can always transform $v_1$, $v_2$ and $v_3$ to (0,0,1), (0,1,0) and (1,0,0). Then the relations $D_1^2 D_2=-a$ and $D_1 D_2^2=-b$ fix the toric ray $v_4$ to be $(-1,b,a)$. With the toric rays in figure~\ref{f:46curve}, any monomial $(x,y,z)\in\mc{F}$ satisfies
\be
x,y,z\geq -4\ ,\ -x+by+az\geq -4,
\ee 
which implies that $by+az\geq -8$. Now if $a,b\geq 2$, this means $y+z\geq -4$ for any $(x,y,z)\in\mc{F}$. Since the order of vanishing of $f$ on this toric curve $v_1 v_2$ is given by
\be
\mathrm{ord}_{v_1 v_2}(f)=\min_{(x,y,z)\in\mc{F}}(y+z+8),
\ee
$f$ vanishes to order 4 or higher on the curve $v_1 v_2$.

Similarly, any monomial $(x,y,z)\in\mc{G}$ satisfies
\be
x,y,z\geq -6\ ,\ -x+by+az\geq -6,
\ee 
which implies that $by+az\geq -12$. If $a,b\geq 2$, then $y+z\geq -6$ and $g$ vanishes to order 6 or higher on $v_1 v_2$.

Hence we have proved that if $D_1^2 D_2\leq -2$, $D_2^2 D_1\leq -2$, then the toric curve $D_1\bigcap D_2$ is a (4,6) curve. One can also prove in the same fashion if $D_1^2 D_2=-a (a\leq 1)$ and $D_2^2 D_1< (6a-12)$, $D_1\bigcap D_2$ is a (4,6) curve.

\section{Applications}

\subsection{Applying the rules on bases with toric (4,6) curves}
\label{s:applyres}

\begin{figure}
\centering
\includegraphics[height=6cm]{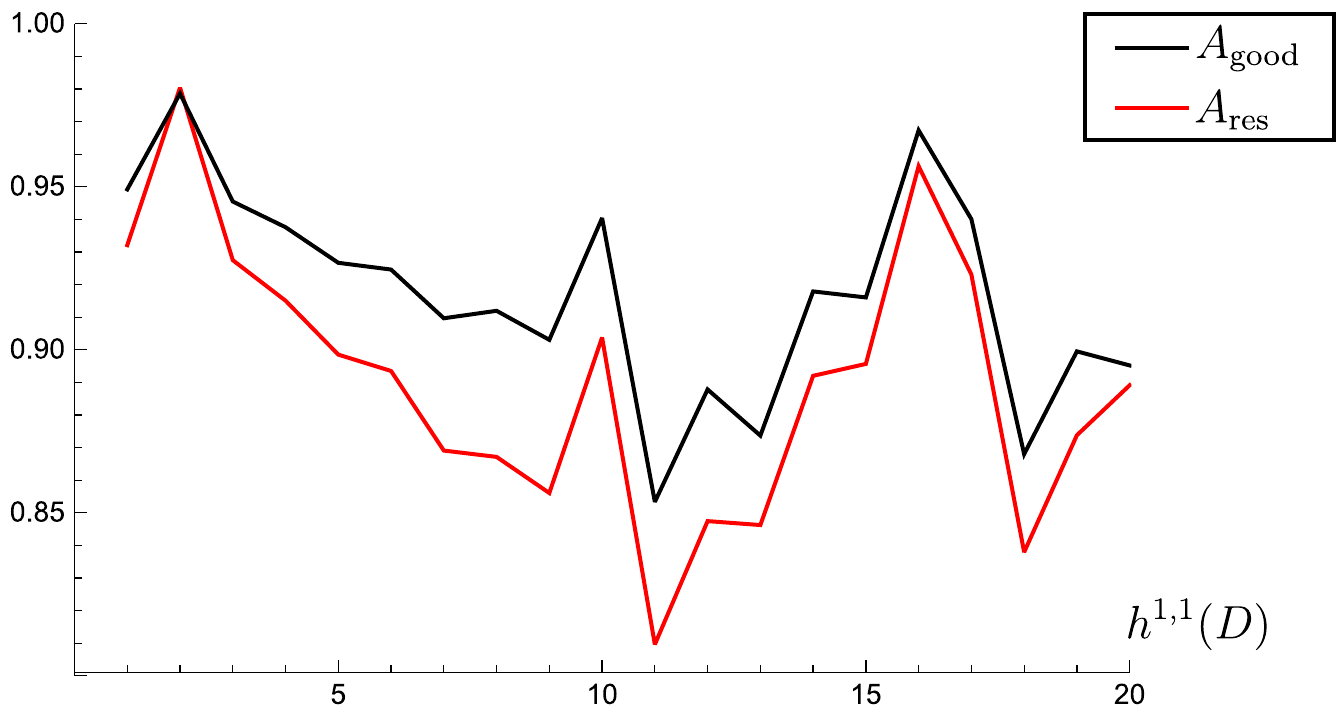}
\caption[E]{\footnotesize The comparison of the accuracies on good bases and resolvable bases for different $h^{1,1}(D)$. The black line is the accuracies $A_{\rm good}$ on good bases and the red line is the accuracies $A_{\rm res}$ on resolvable bases. Although the classifiers for each $h^{1,1}(D)$ are trained on the good bases, $A_{\rm good}$ does not represent in-sample accuracy as the train set is a resampled subset.}\label{f:accuracy}
\end{figure}

In our train set, we do not use the data from resolvable bases with toric (4,6) curves. Now we want to know if the rules derived from the good bases can apply to resolvable bases as well. We have applied the classifiers trained from the good bases in section~\ref{s:divisors} to the divisors on the resolvable bases generated in section~\ref{s:46curve}, and we list the accuracies for each $h^{1,1}(D)$ in table~\ref{t:resdivisors}. We plot the comparison of the accuracy on the resolvable bases and good bases in figure~\ref{f:accuracy}.

\begin{table}
\begin{center}
\begin{tabular}{|c|c|c |c |}
\hline
\hline
$h^{1,1}(D)$ & $N_{\rm res}$ & $A_{\rm res}$ & $A_{\rm good}$\\
\hline
1 & 105994 & 0.932084 & 0.949191\\
2 & 2935137 & 0.980387 & 0.978592 \\
3 & 3786373 & 0.927479 & 0.945464 \\
4 & 4546551 & 0.915048 & 0.937576 \\
5 & 1744718 & 0.898510 & 0.926666 \\
6 & 984305 & 0.893490 & 0.924599 \\
7 & 561131 & 0.869083 & 0.909661 \\
8 & 364615 &  0.867141 & 0.911947 \\
9 & 238652 & 0.856069 & 0.903046 \\
10 & 235184 & 0.903769 & 0.940406\\
11 & 88308 & 0.809622 & 0.853343 \\
12 & 71467 & 0.847447 & 0.887834 \\
13 & 44476 & 0.846242 & 0.873670 \\
14 & 44791 & 0.892008 & 0.917897 \\
15 & 33362 & 0.895658 & 0.916046 \\
16 & 62744 & 0.956308 & 0.967306 \\
17 & 27595 & 0.923074 & 0.940061 \\
18 & 10407 & 0.837934 & 0.867998 \\
19 & 8773 & 0.873788 & 0.899510 \\
20 & 6458 & 0.889113 & 0.895196 \\ 
\hline
\end{tabular}
\end{center}
\caption[x]{\footnotesize  The testing results of classifier trained in section~\ref{s:divisors} on the resolvable bases. $N_{\rm res}$ is the total number of sample divisors on the resolvable bases with a certain $h^{1,1}(D)$ and $A_{\rm res}$ is the accuracy. We have listed the accuracy $A_{\rm good}$ on the good bases for comparison.}\label{t:resdivisors}
\end{table}

As we can see, the accuracies on this set of resolvable bases are usually a bit lower than the accuracies on the set of good bases. Nonetheless, the accuracies are still always higher than 80\%. For the case of Hirzebruch surfaces with $h^{1,1}(D)=2$, the accuracy on resolvable bases is 0.980387, which is even higher than the accuracy on good bases! This implies that the rules of non-Higgsable gauge groups we have derived in section~\ref{s:divisors} universally apply to the good bases and resolvable bases.

Another interesting feature in figure~\ref{f:accuracy} is the peaks for both $A_{\rm res}$ and $A_{\rm good}$, for example  at $h^{1,1}(D)=10$, 16 and 17. For some reason, the rules of non-Higgsable gauge groups for these $h^{1,1}(D)$ are more organized, and we can get a high accuracy despite of the lacking of training samples. It may be interesting to investigate this phenomenon in future work.

\subsection{An SU(3) chain}

In this section, we present some local constructions of non-Higgsable clusters using the analytic rules we have derived in section~\ref{s:divisors}.

In 6D F-theory, the only possible appearance of a non-Higgsable SU(3) gauge group is on an isolated $(-3)$-curve with no charged matter~\cite{clusters}. However, we will construct an infinite chain of non-Higgsable SU(3) gauge groups on a 3D base using the analytic rules we have discovered in table~\ref{t:rulesrareFn}.

The rule with $d=25$, $S(l)=129$ states that if $f_1=D^2 D_1=-2$, $f_2=D^2 D_2=-11\sim -5$, $f_3=D_1^3=-1$, $f_4=D_2^3=9\sim 12$, $f_5=D_3^3=-2\sim -1$, $f_6=D_4^3=10\sim 14$, $f_9=D_2^2 D_3\leq -2$, $f_{12}=D_4^2 D_3=-2\sim -1$, $f_{13}=D_4^2 D_1=-2\sim -1$, $f_{14}=D_1^2 D_4\geq 0$, then the gauge group on $D$ is most likely SU(3). The divisors $D$, $D_1\sim D_4$ are locally assigned as in figure~\ref{f:mlFn}.

\begin{figure}
\centering
\includegraphics[height=8cm]{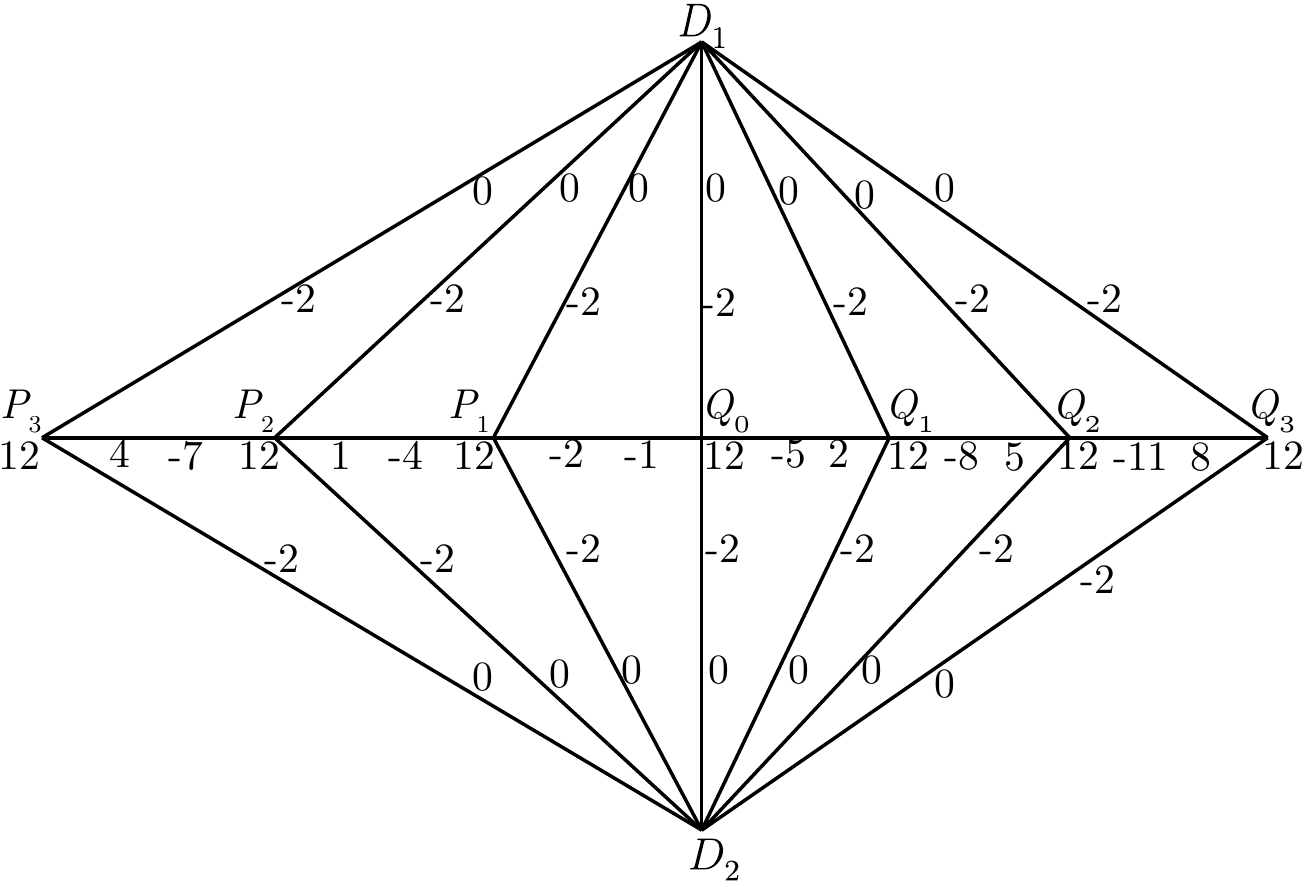}
\caption[E]{\footnotesize The configuration of an SU(3) chain. There are non-Higgsable SU(3) gauge groups on divisors $P_n$ and $Q_n$. The figure can be extended on the left and right to $P_n$ and $Q_n(n>3)$.}\label{f:SU3chain}
\end{figure}

Then we can use this to construct a chain configuration as in figure~\ref{f:SU3chain}. In this particular example, the divisors $P_n$s are $\mathbb{F}_{3n-2}$ and $Q_n$s are $\mathbb{F}_{3n+2}$. The non-vanishing triple intersection numbers between $Q_n$ and $P_n$ are
\be
\bsp
&Q_n^2 Q_{n+1}=-3n-5\ ,\ Q_{n+1}^2 Q_n=3n+2\ ,\ P_n^2 P_{n+1}=-3n-1\ ,\ P_{n+1}^2 P_n=3n-2\ ,\\
& Q_0^2 P_1=-1\ ,\ P_1^2 Q_0=-2.
\end{split}
\ee
Notice that the sum of two numbers on each edge between $Q_n$ and $P_n$ is always 3.

Using the formula (\ref{D3Fn}), one can compute that the self-triple intersection numbers of $P_n$ and $Q_n$ are all 12. Hence the conditions $D_2^3=9\sim 12$ and $D_4^3=10\sim 14$ in the rule we derived from machine learning are satisfied.

Now we check the gauge group analytically using the formula (\ref{normf}) and (\ref{normg}), assuming $f$ and $g$ does not vanish on $D_1$ and $D_2$. For $Q_n(n>0)$, $-K_{Q_n}=2S+(3n+4)F$ and $N_{Q_n}=-2S-(3n+5)F$. Then $f_1$ on $Q_n$ is given by
\be
\bsp
f_{Q_n,1}&\in\mc{O}(4(2S+(3n+4)F)-3(2S+(3n+5)F)-\sum\phi_j C_{ij})\\
&=\mc{O}(2S+(3n+1)F-\mathrm{ord}_{Q_{n-1}}(f)S-\mathrm{ord}_{Q_{n+1}}(f)(S+(3n+2)F).
\end{split}
\ee
Hence if $f$ vanishes to at least order 1 on $Q_{n-1}$ and $Q_{n+1}$, $f_1$ always vanishes. $g_2$ on $Q_n$ is given by
\be
g_{Q_n,2}\in\mc{O}(6(2S+(3n+4)F)-4(2S+(3n+5)F)-\mathrm{ord}_{Q_{n-1}}(g)S-\mathrm{ord}_{Q_{n+1}}(g)(S+(3n+2)F)).
\ee
Hence if $g$ vanish to order 2 on $Q_{n-1}$ and $Q_{n+1}$, we will exactly get $g_2\in\mc{O}(0)$, which is the condition for the gauge group to be SU(3). The situation for $P_n$ and $Q_0$ is analogous.

We can also check this by assigning toric rays to each of these divisors. $D_1$ and $D_2$ are given by $(1,0,0)$ and $(-1,-2,-5)$. $Q_n$s are given by $(0,n,n-1)$ and $P_n$s are given by $(0,-n,-n-1)$. Under these conditions, the only monomial in $g_{Q_0,2}$ is $(-6,-4,4)$ for the configuration in figure~\ref{f:SU3chain} or any extended version of it. There are no (4,6) curves as well.

\begin{figure}
\centering
\includegraphics[height=8cm]{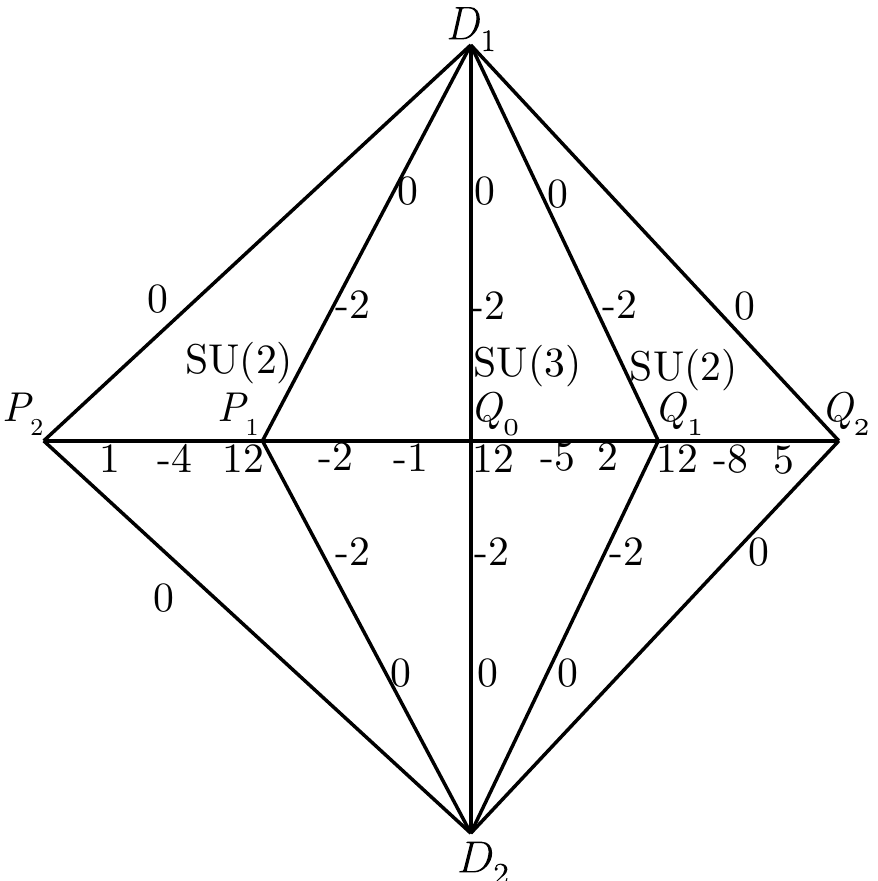}
\caption[E]{\footnotesize An SU(2) $\times$ SU(3)$\times$ SU(2) configuration.}\label{f:SU3SU2}
\end{figure}

Similarly, we can slightly modify the SU(3) chain structure in the last section to get a local SU(2)$\times$ SU(3)$\times$ SU(2) configuration, as in figure~\ref{f:SU3SU2}.

In this case $g$ still vanishes to order 2 on $P_1$, $Q_0$ and $Q_1$, hence
\be
g_{Q_0,2}\in\mc{O}(6(2S+4F)-4(2S+5F)-\mathrm{ord}_{P_1}(g)S-\mathrm{ord}_{Q_{1}}(g)(S+2F))=\mc{O}(0),
\ee
and the gauge group on $Q_0$ is SU(3).

For $Q_1$ and $P_1$, since $\mathrm{ord}_{P_2}(g),\mathrm{ord}_{Q_2}(g)\leq 1$, $g_{Q_1,2},g_{P_1,2}\neq\mc{O}(0)$. Hence the gauge groups on $Q_1$ and $P_1$ are type IV SU(2) or type III SU(2) depending on the order of vanishing of $f$. 

Since the constructions are all independent of the global structure of the compact base threefold, this can be applied to non-GUT type model building using non-Higgsable gauge groups~\cite{ghst} or 4D $\mc{N}=1$ SCFT~\cite{4dCM}.

\section{Conclusion and future directions}
\label{s:con}

In this paper, we have partially solved the problem of reading out the non-Higgsable gauge group on a toric divisor $D$ in 4D F-theory. Using decision tree classification algorithm, we achieved 85\%-98\% out-of-sample accuracies on divisors with different $h^{1,1}(D)$, see table~\ref{t:otherinfo} for details. For the divisors with $h^{1,1}(D)\leq 3$, this methodology is limited by the insufficiency of the features. This is because there exist many samples with the same features but different labels. In the physical language, it means that the set of local triple intersection numbers near $D$ cannot uniquely determine the non-Higgsable gauge group. This problem cannot be resolved by machine learning techniques, and we can only add more local geometric information and increase the number of features. However, we expect the decision tree's structure and rules to be more complicated, which is a trade off. For the divisors with $h^{1,1}(D)\geq 4$, it turns out that the in-sample accuracy is significantly higher than the out-of-sample accuracy. Hence we should modify the machine learning method to improve the predictability.

Besides the predictability, the machine learning algorithm's interpretability also has crucial importance for our purpose. It will be useful if we can simplify the decision tree's structure. For example, if we have two features $f_i$ and $f_j$, then a linear combination $f_i+f_j$ may be a better variable than $f_i$ and $f_j$ such that the decision tree will have a smaller depth with the feature. In our decision trees for $\mb{F}_n$ and $S_{3,n}$ divisors, it turns out that the number $n$ specifying the topology type of $D$ is not very important. It is possible that a combination of $n$ and the normal bundle coefficients may act as a better feature in the decision tree approach. We will leave this exploration to future work.

We have generated various analytic rules from the decision trees through out section~\ref{s:divisors}. But it is worth noting that these rules are derived empirically and not necessarily rigorous. It is hard to prove these rules apart from a small number of simple ones. However, we expect that a particular gauge group will appear for most of the times ($>99\%$) on a generic base. Of course, it is useful to test these rules on other set of bases as well.

We have applied the trained decision tree to divisors on resolvable bases, and the accuracies are 80\%-98\% for different $h^{1,1}(D)$, see table~\ref{t:resdivisors} and figure~\ref{f:accuracy}. We see that the rules trained from the good bases can be applied to resolvable bases as well. 

In section~\ref{s:46curve}, we presented a simple analysis of the criteria for (4,6) curve. In the future, it is worth investigating the blow up sequences of different (4,6) curves, which will lead to a set of 4D conformal matter. Machine learning techniques may be useful in this problem as well since there are many classes of these (4,6) curves. Similarly, it is interesting to study the blow up of a point where $(f,g)$ vanishes to order $(8,12)$ or higher~\cite{4dCM}, since they are common on a general resolvable bases constructed in section~\ref{s:applyres}.

Of course, another interesting direction is to apply our results to toric divisors on non-toric threefolds. If the divisor $D$ still have $p=h^{1,1}(D)+2$ neighboring divisors, then the local geometric structure is similar to our samples and the analytic rules should apply. Because many of the analytic rules are insensitive to the topology of the divisor $D$, as we have mentioned in section~\ref{s:Fn}, they may be applicable to non-toric divisors as well. However, we currently do not have such a non-toric threefold database and the non-Higgsable gauge groups information to check these rules.

\acknowledgments

We would like to thank Thomas Grimm, Jim Halverson, Cody Long and Washington Taylor for useful discussions. We would also like to thank the organizers of the String Data workshops at Northeastern university and LMU Munich for their hospitality. This research was supported by the DOE under contract 
\#DE-SC00012567.

\appendix
\section{Constraints on triple intersection numbers near a divisor}

The triple intersection numbers near a toric divisor $D$ on a toric threefold is constrained. Denote the $p$ neighboring toric divisors of $D$ by $D_i(i=1,\dots,p)$, where $p=h^{1,1}(D)+2$, then there is a simple constraint on $D_i^2 D$:
\be
\sum_{i=1}^p D_i^2 D=3(4-p)=3(2-h^{1,1}(D)).
\ee
This is due to the fact that $D_i^2 D$ equals to the self-intersection number of $C_i$ on the complex surface $D$. The sum of these self-intersection numbers $\sum C_i^2=3(2-h^{1,1}(D))$, since $\sum C_i^2$ is equal to 3 for $\mathbb{P}^2$, 0 for $\mathbb{F}_n$, and each toric blow up reduces this number by 3.

$D^2 D_i$ and $D^3$ are also subject to some constraints. We will analyze them explicitly for divisors $D$ with $h^{1,1}(D)=1,2,3$.

\subsection{$\mathbb{P}^2$}
\label{s:appP2}

For $D=\mathbb{P}^2$, it has three neighbor divisors $D_1,D_2,D_3$. Denote the toric ray of $(D, D_1, D_2, D_3)$ by $(v,  v_1, v_2, v_3)$, we can do a SL$(3,\mathbb{Z})$ transformation on the set of toric rays to transform $v$, $v_1$ and $v_2$ to $(0,0,-1)$, $(1,0,0)$ and $(0,1,0)$. Since the 3D cones $v v_1 v_3$ and $v v_2 v_3$ have unit volume, we can only set $v_3$ to be $v_3=(-1,-1,a), (a\in\mathbb{Z})$. Now we can use the linear equivalence equations (\ref{tripint1}):
\be
D\cdot D_1\cdot (-D+a D_3)=D\cdot D_2\cdot(-D+a D_3)=D\cdot D_3\cdot(-D+a D_3)=0
\ee
to compute
\be
D^2 D_1=D^2 D_2=D^2 D_3=a.
\ee
Then we can use (\ref{tripint2}):
\be
D^2 (-D+a D_3)=0
\ee
to compute  
\be
D^3=a^2.
\ee
Hence we have a relation
\be
D^3=(D^2 D_1)^2=(D^2 D_2)^2=(D^2 D_3)^2.
\ee
Denote the curves on $D$ by $C_i=D\bigcap D_i$, then we can observe this linear equivalence relation on $D$:
\be
C_1=C_2=C_3.
\ee

\subsection{$\mathbb{F}_n$}
\label{s:appFn}

For $D=\mathbb{F}_n$, it has four neighbor divisors $D_1,D_2,D_3,D_4$ with toric rays $v_i (i=1,\dots,4)$. We take these these toric rays to be
\be
v=(0,0,-1)\ ,\ v_1=(1,0,0)\ ,\ v_2=(0,1,0)\ ,\ v_3=(-1,-n,b)\ ,\ v_4=(0,-1,a).
\ee
We can explicitly see that these 3D rays can be projected to a 2D subspace and they $v_1,v_2,v_3,v_4$ explicitly give the toric rays of $\mathbb{F}_n$. Any toric divisor $\mathbb{F}_n$ and its four neighbors can be transformed into the above form with a SL$(3,\mb{Z})$ transformation and a permutation. We have the following triple intersection numbers which equal to the self-intersection numbers of the curve on $D$.
\be
D_1^2 D=D_3^2 D=0\ ,\ D_2^2 D=n\ ,\ D_4^2 D=-n.
\ee
Using (\ref{tripint1}), we have
\be
\bsp
&D\cdot D_1\cdot(a D_4+b D_3-D)=D\cdot D_2\cdot(a D_4+b D_3-D)=0,\\
&D\cdot D_3\cdot(a D_4+b D_3-D)=D\cdot D_4\cdot(a D_4+b D_3-D)=0.
\end{split}
\ee
Hence we can read off
\be
D^2 D_1=D^2 D_3=a\ ,\ D^2 D_2=b\ ,\ D^2 D_4=b-na.\label{divrelFn}
\ee
Then we can use
\be
D^2(a D_4+b D_3-D)=0
\ee
to compute
\be
D^3=2ab-na^2.\label{D3Fn}
\ee
The relations (\ref{divrelFn}) can be checked with the linear relations of the curves on $D$:
\be
C_1=C_3\ ,\ C_4=C_2-n C_1.
\ee

\subsection{$S_{3,n}$}
\label{s:appS3n}

For $D=S_{3,n}$, it has five neighbor divisors $D_1,D_2,D_3,D_4,D_5$ with toric rays $v_i (i=1,\dots,5)$. We take these these toric rays to be
\be
\bsp
&v=(0,0,-1)\ ,\ v_1=(0,1,0)\ ,\ v_2=(1,0,0)\ ,\ v_3=(-n,-1,a)\ ,\ v_4=(-n-1,-1,b)\ ,\ \\
&v_5=(-1,0,c).
\end{split}
\ee
We have
\be
D_1^2 D=0\ ,\ D_2^2 D=n\ ,\ D_3^2 D=D_4^2 D=-1\ ,\ D_5^2 D=-(n+1).
\ee
The linear relations of the curves on $D$ are
\be
C_1=C_4+C_5\ ,\ C_2=nC_3+(n+1)C_4+C_5.\label{linearS3n}
\ee
With (\ref{tripint1}), we can compute
\be
\bsp
&D^2 D_1=c\ ,\ D^2 D_2=a\ ,\ D^2 D_3=b-a\ ,\ D^2 D_4=\frac{(n+1)a-nb-c}{n}\ ,\ \\&D^2 D_5=\frac{(n-1)a+nb+(n+1)c}{n}.
\end{split}
\ee
Then with (\ref{tripint2}), we can compute
\be
D^3=\frac{ab-(a-b)^2 n+((b-a)c+c^2)(1+n)}{n}.
\ee
Using the variables $D^2 D_1=f_1$, $D^2 D_2=f_2$, $D^2 D_3=f_3$, we have
\be
D^3=\frac{f_2(f_2+f_3)-f_3^2 n+f_1^2(1+n)+f_1 f_3 (1+n)}{n}
\ee

\end{document}